\documentclass[twocolumn,traditabstract,longauth]{aa}
\usepackage{graphicx,amsmath}
\usepackage{epsf}
\usepackage{txfonts}
\usepackage{url}
\usepackage[breaklinks, colorlinks, citecolor=blue]{hyperref}
\usepackage{fixltx2e}
\usepackage{natbib}
\usepackage{upgreek}
\usepackage{ifthen}

\def\gtsim{~\rlap{$>$}{\lower 1.0ex\hbox{$\sim$}}}
\bibpunct{(}{)}{;}{a}{}{,} 
\def\setsymbol#1#2{\expandafter\def\csname #1\endcsname{#2}}
\def\getsymbol#1{\csname #1\endcsname}

\def\Planck{\textit{Planck}}

\def\allearlypapers{\nocite{planck2011-1.1, planck2011-1.3, planck2011-1.4, planck2011-1.5, planck2011-1.6, planck2011-1.7, planck2011-1.10, planck2011-1.10sup, planck2011-5.1a, planck2011-5.1b, planck2011-5.2a, planck2011-5.2b, planck2011-5.2c, planck2011-6.1, planck2011-6.2, planck2011-6.3a, planck2011-6.4a, planck2011-6.4b, planck2011-6.6, planck2011-7.0, planck2011-7.2, planck2011-7.3, planck2011-7.7a, planck2011-7.7b, planck2011-7.12, planck2011-7.13}}

\newbox\tablebox    \newdimen\tablewidth
\def\leaderfil{\leaders\hbox to 5pt{\hss.\hss}\hfil}
\def\endPlancktable{\tablewidth=\columnwidth 
    $$\hss\copy\tablebox\hss$$
    \vskip-\lastskip\vskip -2pt}
\def\endPlancktablewide{\tablewidth=\textwidth 
    $$\hss\copy\tablebox\hss$$
    \vskip-\lastskip\vskip -2pt}
\def\tablenote#1 #2\par{\begingroup \parindent=0.8em
    \abovedisplayshortskip=0pt\belowdisplayshortskip=0pt
    \noindent
    $$\hss\vbox{\hsize\tablewidth \hangindent=\parindent \hangafter=1 \noindent
    \hbox to \parindent{$^#1$\hss}\strut#2\strut\par}\hss$$
    \endgroup}
\def\doubleline{\vskip 3pt\hrule \vskip 1.5pt \hrule \vskip 5pt}

\def\L2{\ifmmode L_2\else $L_2$\fi}

\def\DeltaT{\ifmmode \Delta T\else $\Delta T$\fi}
\def\deltat{\ifmmode \Delta t\else $\Delta t$\fi}
\def\fknee{\ifmmode f_{\rm knee}\else $f_{\rm knee}$\fi}
\def\Fmax{\ifmmode F_{\rm max}\else $F_{\rm max}$\fi}
\def\solar{\ifmmode{\rm M}_{\mathord\odot}\else${\rm M}_{\mathord\odot}$\fi}
\def\Msolar{\ifmmode{\rm M}_{\mathord\odot}\else${\rm M}_{\mathord\odot}$\fi}
\def\Lsolar{\ifmmode{\rm L}_{\mathord\odot}\else${\rm L}_{\mathord\odot}$\fi}
\def\inv{\ifmmode^{-1}\else$^{-1}$\fi}
\def\mo{\ifmmode^{-1}\else$^{-1}$\fi}
\def\sup#1{\ifmmode ^{\rm #1}\else $^{\rm #1}$\fi}
\def\expo#1{\ifmmode \times 10^{#1}\else $\times 10^{#1}$\fi}
\def\,{\thinspace}
\def\lsim{\mathrel{\raise .4ex\hbox{\rlap{$<$}\lower 1.2ex\hbox{$\sim$}}}}
\def\gsim{\mathrel{\raise .4ex\hbox{\rlap{$>$}\lower 1.2ex\hbox{$\sim$}}}}

\def\simprop{\mathrel{\raise .4ex\hbox{\rlap{$\propto$}\lower 1.2ex\hbox{$\sim$}}}}
\def\deg{\ifmmode^\circ\else$^\circ$\fi}
\def\pdeg{\ifmmode $\setbox0=\hbox{$^{\circ}$}\rlap{\hskip.11\wd0 .}$^{\circ}
          \else \setbox0=\hbox{$^{\circ}$}\rlap{\hskip.11\wd0 .}$^{\circ}$\fi}
\def\arcs{\ifmmode {^{\scriptstyle\prime\prime}}
          \else $^{\scriptstyle\prime\prime}$\fi}
\def\arcm{\ifmmode {^{\scriptstyle\prime}}
          \else $^{\scriptstyle\prime}$\fi}
\newdimen\sa  \newdimen\sba
\def\parcs{\sa=.07em \sba=.03em
     \ifmmode \hbox{\rlap{.}}^{\scriptstyle\prime\kern -\sba\prime}\hbox{\kern -\sa}
     \else \rlap{.}$^{\scriptstyle\prime\kern -\sba\prime}$\kern -\sa\fi}
\def\parcm{\sa=.08em \sba=.03em
     \ifmmode \hbox{\rlap{.}\kern\sa}^{\scriptstyle\prime}\hbox{\kern-\sba}
     \else \rlap{.}\kern\sa$^{\scriptstyle\prime}$\kern-\sba\fi}
\def\ra[#1 #2 #3.#4]{#1\sup{h}#2\sup{m}#3\sup{s}\llap.#4}
\def\dec[#1 #2 #3.#4]{#1\deg#2\arcm#3\arcs\llap.#4}
\def\deco[#1 #2 #3]{#1\deg#2\arcm#3\arcs}
\def\rra[#1 #2]{#1\sup{h}#2\sup{m}}

\def\dots{\relax\ifmmode \ldots\else $\ldots$\fi}
\def\WHzsr{\ifmmode $W\,Hz\mo\,sr\mo$\else W\,Hz\mo\,sr\mo\fi}
\def\mHz{\ifmmode $\,mHz$\else \,mHz\fi}
\def\GHz{\ifmmode $\,GHz$\else \,GHz\fi}
\def\mKs{\ifmmode $\,mK\,s$^{1/2}\else \,mK\,s$^{1/2}$\fi}
\def\muKs{\ifmmode \,\mu$K\,s$^{1/2}\else \,$\mu$K\,s$^{1/2}$\fi}
\def\muKRJs{\ifmmode \,\mu$K$_{\rm RJ}$\,s$^{1/2}\else \,$\mu$K$_{\rm RJ}$\,s$^{1/2}$\fi}
\def\muKHz{\ifmmode \,\mu$K\,Hz$^{-1/2}\else \,$\mu$K\,Hz$^{-1/2}$\fi}
\def\MJysr{\ifmmode \,$MJy\,sr\mo$\else \,MJy\,sr\mo\fi}
\def\MJysrmK{\ifmmode \,$MJy\,sr\mo$\,mK$_{\rm CMB}\mo\else \,MJy\,sr\mo\,mK$_{\rm CMB}\mo$\fi}
\def\microns{\ifmmode \,\mu$m$\else \,$\mu$m\fi}

\def\muK{\ifmmode \,\mu$K$\else \,$\mu$\hbox{K}\fi}
\def\microK{\ifmmode \,\mu$K$\else \,$\mu$\hbox{K}\fi}
\def\muW{\ifmmode \,\mu$W$\else \,$\mu$\hbox{W}\fi}
\def\kms{\ifmmode $\,km\,s$^{-1}\else \,km\,s$^{-1}$\fi}
\def\kmsMpc{\ifmmode $\,\kms\,Mpc\mo$\else \,\kms\,Mpc\mo\fi}
\providecommand{\sorthelp}[1]{}

\newcommand{\planck}{\Planck} 
\newcommand{\herschel}{\Herschel}
\newcommand{\Herschel}{\textit{Herschel\/}}
\newcommand{\IRAS}{IRAS}
\newcommand{\iras}{\IRAS}
\newcommand{\WMAP}{\textit{WMAP\/}}
\newcommand{\COBE}{COBE}
\newcommand{\DIRBE}{\COBE-DIRBE}
\newcommand{\Spitzer}{\textit{Spitzer\/}}
\newcommand{\ISO}{ISO}
\newcommand{\ghz}{\,GHz}

\newcommand{\MHz}{\,MHz}
\newcommand{\kpc}{\,kpc}
\newcommand{\hi}{\ion{H}{i}}
\newcommand{\um}{\,$\upmu$m}

\newcommand{\healpix}{{\tt HEALPix}}
\newcommand{\mdotyr}{\,M$_\odot$\,yr$^{-1}$}
\newcommand{\mkpc}{$\Msolar$\,kpc$^{-2}$}
\newcommand{\ha}{H$\alpha$}

\newcommand{\changed}[1]{#1}

\setsymbol{dusttau}{$(1.2\pm0.2)\times10^{-5}$}
\setsymbol{dustbeta}{$1.62\pm0.11$}
\setsymbol{tdust}{$18.2\pm1.0$}
\setsymbol{freefree}{$1.8\pm1.3$}
\setsymbol{cmbamp}{$(1.7\pm1.0)\times10^{-6}$}
\setsymbol{Async}{$9.5\pm1.1$}
\setsymbol{alphasync}{$-0.92\pm0.16$}
\setsymbol{AMEamp}{$(7.7\pm3.3)\times10^{16}$}
\setsymbol{residual100}{$-0.08\pm1.16$}
\setsymbol{residual100percent}{$1.3$\,\%}
\setsymbol{residual217}{$-2.2\pm8.2$}
\setsymbol{residual217percent}{$3.5$\,\%}
\setsymbol{chi2}{15.4}
\setsymbol{ndof}{14}

\begin{document}

\title{\Planck\ intermediate results. XXV.\\The Andromeda Galaxy as seen by \Planck}

\titlerunning{Andromeda as seen by \Planck}
\author{\small
Planck Collaboration: P.~A.~R.~Ade\inst{82}
\and
N.~Aghanim\inst{55}
\and
M.~Arnaud\inst{69}
\and
M.~Ashdown\inst{65, 6}
\and
J.~Aumont\inst{55}
\and
C.~Baccigalupi\inst{81}
\and
A.~J.~Banday\inst{90, 10}
\and
R.~B.~Barreiro\inst{61}
\and
N.~Bartolo\inst{28, 62}
\and
E.~Battaner\inst{91, 92}
% ...
}
\institute{\small
APC, AstroParticule et Cosmologie, Universit\'{e} Paris Diderot, CNRS/IN2P3, CEA/lrfu, Observatoire de Paris, Sorbonne Paris Cit\'{e}, 10, rue Alice Domon et L\'{e}onie Duquet, 75205 Paris Cedex 13, France\goodbreak
\and
Aalto University Mets\"{a}hovi Radio Observatory and Dept of Radio Science and Engineering, P.O. Box 13000, FI-00076 AALTO, Finland\goodbreak
\and
African Institute for Mathematical Sciences, 6-8 Melrose Road, Muizenberg, Cape Town, South Africa\goodbreak
\and
Agenzia Spaziale Italiana Science Data Center, Via del Politecnico snc, 00133, Roma, Italy\goodbreak
\and
Aix Marseille Universit\'{e}, CNRS, LAM (Laboratoire d'Astrophysique de Marseille) UMR 7326, 13388, Marseille, France\goodbreak
\and
Astrophysics Group, Cavendish Laboratory, University of Cambridge, J J Thomson Avenue, Cambridge CB3 0HE, U.K.\goodbreak
\and
Astrophysics \& Cosmology Research Unit, School of Mathematics, Statistics \& Computer Science, University of KwaZulu-Natal, Westville Campus, Private Bag X54001, Durban 4000, South Africa\goodbreak
\and
Atacama Large Millimeter/submillimeter Array, ALMA Santiago Central Offices, Alonso de Cordova 3107, Vitacura, Casilla 763 0355, Santiago, Chile\goodbreak
\and
CITA, University of Toronto, 60 St. George St., Toronto, ON M5S 3H8, Canada\goodbreak
\and
CNRS, IRAP, 9 Av. colonel Roche, BP 44346, F-31028 Toulouse cedex 4, France\goodbreak
\and
California Institute of Technology, Pasadena, California, U.S.A.\goodbreak
\and
Centro de Estudios de F\'{i}sica del Cosmos de Arag\'{o}n (CEFCA), Plaza San Juan, 1, planta 2, E-44001, Teruel, Spain\goodbreak
\and
Computational Cosmology Center, Lawrence Berkeley National Laboratory, Berkeley, California, U.S.A.\goodbreak
\and
DSM/Irfu/SPP, CEA-Saclay, F-91191 Gif-sur-Yvette Cedex, France\goodbreak
\and
DTU Space, National Space Institute, Technical University of Denmark, Elektrovej 327, DK-2800 Kgs. Lyngby, Denmark\goodbreak
\and
D\'{e}partement de Physique Th\'{e}orique, Universit\'{e} de Gen\`{e}ve, 24, Quai E. Ansermet,1211 Gen\`{e}ve 4, Switzerland\goodbreak
\and
Departamento de Astrof\'{i}sica, Universidad de La Laguna (ULL), E-38206 La Laguna, Tenerife, Spain\goodbreak
\and
Departamento de F\'{\i}sica, Universidad de Oviedo, Avda. Calvo Sotelo s/n, Oviedo, Spain\goodbreak
\and
Department of Astrophysics/IMAPP, Radboud University Nijmegen, P.O. Box 9010, 6500 GL Nijmegen, The Netherlands\goodbreak
\and
Department of Physics \& Astronomy, University of British Columbia, 6224 Agricultural Road, Vancouver, British Columbia, Canada\goodbreak
\and
Department of Physics and Astronomy, Dana and David Dornsife College of Letter, Arts and Sciences, University of Southern California, Los Angeles, CA 90089, U.S.A.\goodbreak
\and
Department of Physics and Astronomy, University College London, London WC1E 6BT, U.K.\goodbreak
\and
Department of Physics, Florida State University, Keen Physics Building, 77 Chieftan Way, Tallahassee, Florida, U.S.A.\goodbreak
\and
Department of Physics, Gustaf H\"{a}llstr\"{o}min katu 2a, University of Helsinki, Helsinki, Finland\goodbreak
\and
Department of Physics, Princeton University, Princeton, New Jersey, U.S.A.\goodbreak
\and
Department of Physics, University of California, Santa Barbara, California, U.S.A.\goodbreak
\and
Department of Physics, University of Illinois at Urbana-Champaign, 1110 West Green Street, Urbana, Illinois, U.S.A.\goodbreak
\and
Dipartimento di Fisica e Astronomia G. Galilei, Universit\`{a} degli Studi di Padova, via Marzolo 8, 35131 Padova, Italy\goodbreak
\and
Dipartimento di Fisica e Scienze della Terra, Universit\`{a} di Ferrara, Via Saragat 1, 44122 Ferrara, Italy\goodbreak
\and
Dipartimento di Fisica, Universit\`{a} La Sapienza, P. le A. Moro 2, Roma, Italy\goodbreak
\and
Dipartimento di Fisica, Universit\`{a} degli Studi di Milano, Via Celoria, 16, Milano, Italy\goodbreak
\and
Dipartimento di Fisica, Universit\`{a} degli Studi di Trieste, via A. Valerio 2, Trieste, Italy\goodbreak
\and
Dipartimento di Fisica, Universit\`{a} di Roma Tor Vergata, Via della Ricerca Scientifica, 1, Roma, Italy\goodbreak
\and
Discovery Center, Niels Bohr Institute, Blegdamsvej 17, Copenhagen, Denmark\goodbreak
\and
European Southern Observatory, ESO Vitacura, Alonso de Cordova 3107, Vitacura, Casilla 19001, Santiago, Chile\goodbreak
\and
European Space Agency, ESAC, Planck Science Office, Camino bajo del Castillo, s/n, Urbanizaci\'{o}n Villafranca del Castillo, Villanueva de la Ca\~{n}ada, Madrid, Spain\goodbreak
\and
European Space Agency, ESTEC, Keplerlaan 1, 2201 AZ Noordwijk, The Netherlands\goodbreak
\and
Facolt\`{a} di Ingegneria, Universit\`{a} degli Studi e-Campus, Via Isimbardi 10, Novedrate (CO), 22060, Italy\goodbreak
\and
Gran Sasso Science Institute, INFN, viale F. Crispi 7, 67100 L'Aquila, Italy\goodbreak
\and
HGSFP and University of Heidelberg, Theoretical Physics Department, Philosophenweg 16, 69120, Heidelberg, Germany\goodbreak
\and
Haverford College Astronomy Department, 370 Lancaster Avenue, Haverford, Pennsylvania, U.S.A.\goodbreak
\and
Helsinki Institute of Physics, Gustaf H\"{a}llstr\"{o}min katu 2, University of Helsinki, Helsinki, Finland\goodbreak
\and
INAF - Osservatorio Astrofisico di Catania, Via S. Sofia 78, Catania, Italy\goodbreak
\and
INAF - Osservatorio Astronomico di Padova, Vicolo dell'Osservatorio 5, Padova, Italy\goodbreak
\and
INAF - Osservatorio Astronomico di Roma, via di Frascati 33, Monte Porzio Catone, Italy\goodbreak
\and
INAF - Osservatorio Astronomico di Trieste, Via G.B. Tiepolo 11, Trieste, Italy\goodbreak
\and
INAF/IASF Bologna, Via Gobetti 101, Bologna, Italy\goodbreak
\and
INAF/IASF Milano, Via E. Bassini 15, Milano, Italy\goodbreak
\and
INFN, Sezione di Bologna, Via Irnerio 46, I-40126, Bologna, Italy\goodbreak
\and
INFN, Sezione di Roma 1, Universit\`{a} di Roma Sapienza, Piazzale Aldo Moro 2, 00185, Roma, Italy\goodbreak
\and
INFN/National Institute for Nuclear Physics, Via Valerio 2, I-34127 Trieste, Italy\goodbreak
\and
Imperial College London, Astrophysics group, Blackett Laboratory, Prince Consort Road, London, SW7 2AZ, U.K.\goodbreak
\and
Infrared Processing and Analysis Center, California Institute of Technology, Pasadena, CA 91125, U.S.A.\goodbreak
\and
Institut Universitaire de France, 103, bd Saint-Michel, 75005, Paris, France\goodbreak
\and
Institut d'Astrophysique Spatiale, CNRS (UMR8617) Universit\'{e} Paris-Sud 11, B\^{a}timent 121, Orsay, France\goodbreak
\and
Institut d'Astrophysique de Paris, CNRS (UMR7095), 98 bis Boulevard Arago, F-75014, Paris, France\goodbreak
\and
Institute for Space Sciences, Bucharest-Magurale, Romania\goodbreak
\and
Institute of Astronomy, University of Cambridge, Madingley Road, Cambridge CB3 0HA, U.K.\goodbreak
\and
Institute of Theoretical Astrophysics, University of Oslo, Blindern, Oslo, Norway\goodbreak
\and
Instituto de Astrof\'{\i}sica de Canarias, C/V\'{\i}a L\'{a}ctea s/n, La Laguna, Tenerife, Spain\goodbreak
\and
Instituto de F\'{\i}sica de Cantabria (CSIC-Universidad de Cantabria), Avda. de los Castros s/n, Santander, Spain\goodbreak
\and
Istituto Nazionale di Fisica Nucleare, Sezione di Padova, via Marzolo 8, I-35131 Padova, Italy\goodbreak
\and
Jet Propulsion Laboratory, California Institute of Technology, 4800 Oak Grove Drive, Pasadena, California, U.S.A.\goodbreak
\and
Jodrell Bank Centre for Astrophysics, Alan Turing Building, School of Physics and Astronomy, The University of Manchester, Oxford Road, Manchester, M13 9PL, U.K.\goodbreak
\and
Kavli Institute for Cosmology Cambridge, Madingley Road, Cambridge, CB3 0HA, U.K.\goodbreak
\and
Kazan Federal University, 18 Kremlyovskaya St., Kazan, 420008, Russia\goodbreak
\and
LAL, Universit\'{e} Paris-Sud, CNRS/IN2P3, Orsay, France\goodbreak
\and
LERMA, CNRS, Observatoire de Paris, 61 Avenue de l'Observatoire, Paris, France\goodbreak
\and
Laboratoire AIM, IRFU/Service d'Astrophysique - CEA/DSM - CNRS - Universit\'{e} Paris Diderot, B\^{a}t. 709, CEA-Saclay, F-91191 Gif-sur-Yvette Cedex, France\goodbreak
\and
Laboratoire Traitement et Communication de l'Information, CNRS (UMR 5141) and T\'{e}l\'{e}com ParisTech, 46 rue Barrault F-75634 Paris Cedex 13, France\goodbreak
\and
Laboratoire de Physique Subatomique et Cosmologie, Universit\'{e} Grenoble-Alpes, CNRS/IN2P3, 53, rue des Martyrs, 38026 Grenoble Cedex, France\goodbreak
\and
Laboratoire de Physique Th\'{e}orique, Universit\'{e} Paris-Sud 11 \& CNRS, B\^{a}timent 210, 91405 Orsay, France\goodbreak
\and
Lawrence Berkeley National Laboratory, Berkeley, California, U.S.A.\goodbreak
\and
Lebedev Physical Institute of the Russian Academy of Sciences, Astro Space Centre, 84/32 Profsoyuznaya st., Moscow, GSP-7, 117997, Russia\goodbreak
\and
Max-Planck-Institut f\"{u}r Astrophysik, Karl-Schwarzschild-Str. 1, 85741 Garching, Germany\goodbreak
\and
McGill Physics, Ernest Rutherford Physics Building, McGill University, 3600 rue University, Montr\'{e}al, QC, H3A 2T8, Canada\goodbreak
\and
National University of Ireland, Department of Experimental Physics, Maynooth, Co. Kildare, Ireland\goodbreak
\and
Nicolaus Copernicus Astronomical Center, Bartycka 18, 00-716 Warsaw, Poland\goodbreak
\and
Niels Bohr Institute, Blegdamsvej 17, Copenhagen, Denmark\goodbreak
\and
Optical Science Laboratory, University College London, Gower Street, London, U.K.\goodbreak
\and
SISSA, Astrophysics Sector, via Bonomea 265, 34136, Trieste, Italy\goodbreak
\and
School of Physics and Astronomy, Cardiff University, Queens Buildings, The Parade, Cardiff, CF24 3AA, U.K.\goodbreak
\and
Sorbonne Universit\'{e}-UPMC, UMR7095, Institut d'Astrophysique de Paris, 98 bis Boulevard Arago, F-75014, Paris, France\goodbreak
\and
Space Sciences Laboratory, University of California, Berkeley, California, U.S.A.\goodbreak
\and
Special Astrophysical Observatory, Russian Academy of Sciences, Nizhnij Arkhyz, Zelenchukskiy region, Karachai-Cherkessian Republic, 369167, Russia\goodbreak
\and
Sterrewacht Leiden, Leiden University, PO Box 9513, 2300 RA Leiden, The Netherlands\goodbreak
\and
Sub-Department of Astrophysics, University of Oxford, Keble Road, Oxford OX1 3RH, U.K.\goodbreak
\and
UK ALMA Regional Centre Node, Jodrell Bank Centre for Astrophysics, Alan Turing Building, School of Physics and Astronomy, The University of Manchester, Oxford Road, Manchester, M13 9PL, U.K.\goodbreak
\and
UPMC Univ Paris 06, UMR7095, 98 bis Boulevard Arago, F-75014, Paris, France\goodbreak
\and
Universit\'{e} de Toulouse, UPS-OMP, IRAP, F-31028 Toulouse cedex 4, France\goodbreak
\and
University of Granada, Departamento de F\'{\i}sica Te\'{o}rica y del Cosmos, Facultad de Ciencias, Granada, Spain\goodbreak
\and
University of Granada, Instituto Carlos I de F\'{\i}sica Te\'{o}rica y Computacional, Granada, Spain\goodbreak
% ...
}

\authorrunning{Planck Collaboration}

\abstract{The Andromeda Galaxy (M31) is one of a few galaxies that has sufficient angular size on the sky to be resolved by the \Planck\ satellite. \Planck\ has detected M31 in all of its frequency bands, and has mapped out the dust emission with the High Frequency Instrument, clearly resolving multiple spiral arms and sub-features. We examine the morphology of this long-wavelength dust emission as seen by \Planck, including a study of its outermost spiral arms, and investigate the dust heating mechanism across M31. We find that dust dominating the longer wavelength emission ($\gtrsim 0.3$\,mm) is heated by the diffuse stellar population (as traced by 3.6\um\ emission), with the dust dominating the shorter wavelength emission heated by a mix of the old stellar population and star-forming regions (as traced by 24\um\ emission). We also fit spectral energy distributions (SEDs) for individual 5\arcm\ pixels and quantify the dust properties across the galaxy, taking into account these different heating mechanisms, finding that there is a linear decrease in temperature with galactocentric distance for dust heated by the old stellar population, as would be expected, with temperatures ranging from around 22\,K in the nucleus to 14\,K outside of the 10\,kpc ring. Finally, we measure the integrated spectrum of the whole galaxy, which we find to be well-fitted with a global dust temperature of (\getsymbol{tdust})\,K with a spectral index of \getsymbol{dustbeta} (assuming a single modified blackbody), and a significant amount of free-free emission at intermediate frequencies of 20--60\GHz, which \changed{corresponds to a star formation rate of around 0.12\mdotyr}. We find a $2.3\,\sigma$ detection of the presence of spinning dust emission, with a 30\GHz\ amplitude of $0.7\pm0.3$\,Jy, which is in line with expectations from our Galaxy.}


\maketitle   

\allearlypapers

\section{Introduction} \label{sec:intro}

The infrared (IR) and submillimetre (submm) wavelength domain is particularly useful for understanding the processes driving star formation in various galactic environments, since dust grains re-emit in this frequency window the energy that has been absorbed from the UV-optical starlight.  Our view of the global inner and outer disk star formation and ISM properties of the spiral galaxy in which we live is limited by our position inside the Galaxy, but our nearest spiral neighbour, the Andromeda Galaxy (also known as Messier~31), offers the best view of the environmental effects within an entire galaxy, particularly because of its large angular extent on the sky.

M31 has been extensively studied at IR/submm wavelengths with data from the {\it Infrared Astronomical Satellite} \citep[\IRAS,][]{Habing1984,Walterbos1987}, the {\it Infrared Space Observatory} \citep[\ISO,][]{Haas1998}, the {\it Spitzer Space Telescope} \citep{Barmby2006,Gordon2006,Tabatabaei2010}, and, most recently, the {\it Herschel Space Observatory} \citep{Fritz2012, Groves2012, Smith2012,Ford2013,Draine2013,Viaene2014,Kirk2015}. Except for the \Herschel\ data, these IR observations have been restricted to observing the peak of the dust emission in the far infrared (FIR) as well as mid-infrared (MIR) emission from $\ga\!100$~K dust and polycyclic aromatic hydrocarbons.  In contrast, due to the large angular size of M31 it has been particularly difficult to map the entire galaxy at wavelengths longer than $500$\um, which is needed to constrain the Rayleigh-Jeans side of the dust emission and the contribution of non-thermal emission sources to the spectral energy distribution (SED).  In fact, the submm data for nearby spiral galaxies that have been published \citep[e.g.,][]{Dunne2000, Stevens2005, Dale2007} have had low signal-to-noise levels, have been biased towards infrared-bright sources, or have only covered the centres of galaxies. 

Data from \Planck\ \citep{tauber2010a}\footnote{\Planck\ (\url{http://www.esa.int/Planck}) is a project of the European Space Agency -- ESA -- with instruments provided by two scientific Consortia funded by ESA member states (in particular the lead countries: France and Italy) with contributions from NASA (USA), and telescope reflectors provided in a collaboration between ESA and a scientific Consortium led and funded by Denmark.}, which range from 28.4 to 857\GHz\ (10.5 to 0.35\,mm) with angular resolution between 31\arcm\ and 5\arcm, allow us to examine the Rayleigh-Jeans tail of the dust SED and the transition into free-free and synchrotron emission at longer wavelengths.  Moreover, since \Planck\ observed the entire sky at high sensitivity, its High Frequency Instrument \citep[HFI,][]{Lamarre2010} provides high signal-to-noise maps of M31 that extend to the outermost edges of the galaxy.

\Planck's large-scale map of the region provides the opportunity to study dust heating in M31 out to the optical radius of the galaxy.  Prior investigations with \IRAS\ of dust heating in our Galaxy and others had produced seemingly contradictory results, with some studies indicating that dust seen at \mbox{60--100\um} was heated primarily by star-forming regions \citep{Devereux1990,Buat1996} and others demonstrating that evolved stellar populations could partially contribute to heating the dust seen by \IRAS\ \citep[e.g.,][]{Lonsdale1987, Walterbos1987,Sauvage1992,Walterbos1996}.  More recent observations with \Herschel\ of a number of galaxies, including M81, M83, NGC2403 \citep{Bendo2010, Bendo2012} and M33 \citep{Boquien2011}, demonstrated that dust-dominating emission at wavelengths shorter than $160$\um\ was primarily heated by star-forming regions (henceforth abbreviated as ``SFR dust''), while dust-dominating emission at wavelengths over $250$\um\ may be primarily heated by the total stellar populations, including evolved stars in the discs and bulges of the galaxies (henceforth abbreviated as interstellar radiation field dust, or ``ISRF dust.''  \Planck\ observations of the SED of M31 allow us to examine dust heating within the galaxy at frequencies much lower than was possible with \Herschel. Once the dust heating sources are identified empirically and the SED is separated into different thermal components based on the heating sources, it will be possible to more accurately measure the temperature of the coldest dust within M31, which is also critically important to properly estimate the dust mass.

Non-thermal emission from M31 can also be measured at the lowest frequencies covered by \Planck. Synchrotron emission from M31 was discovered in the early days of radio astronomy \citep{Brown1950,Brown1951}.  It has subsequently been mapped at low frequencies \citep{Beck1998,Berkhuijsen2003}, and associated emission has even been detected at gamma-ray frequencies \citep{Abdo2010}. However, the synchrotron emission has not been studied at higher frequency. Free-free emission is also expected from M31.  This emission may be used to measure star formation rates in a way that is not affected by dust extinction or reliant upon assumptions about the dust heating sources \citep[e.g., see][]{Murphy2011}. Free-free emission from M31 was first seen by \citet{Hoernes1998}, as well as \citet{Berkhuijsen2003} and \citet{Tabatabaei2013}. \Planck\ data provide the opportunity to characterize the high-frequency radio emission for the first time.

Section \ref{sec:planck} of this paper describes the \Planck\ data, and Sect. \ref{sec:ancillary} the ancillary data that we use here. We discuss the morphology of the dust as seen by \Planck\ in Sect. \ref{sec:morphology}, the colour ratios and the implications they have on the dust heating mechanism in Sect. \ref{sec:colourratios}, and the spectral energy distributions on 5\arcm\ scales in Sect. \ref{sec:seds} and for the whole of M31 in Sect. \ref{sec:lowressed}. We conclude in Sect. \ref{sec:conclusions}.

\section{\Planck\ data} \label{sec:planck}
\Planck\ \citep{tauber2010a, planck2011-1.1} is the third generation space mission to measure the anisotropy of the cosmic microwave background (CMB).  It observes the sky in nine frequency bands covering 30--857\,GHz (10.5 to 0.35\,mm) with high sensitivity and angular resolution from 31\arcm\ to 5\arcm.  The Low Frequency Instrument \citep[LFI;][]{Mandolesi2010, Bersanelli2010, planck2011-1.4} covers the 30, 44, and 70\,GHz (10.5, 6.8, and 4.3\,mm) bands with amplifiers cooled to 20\,\hbox{K}.  The High Frequency Instrument (HFI; \citealt{Lamarre2010, planck2011-1.5}) covers the 100, 143, 217, 353, 545, and 857\,GHz (3.0, 2.1, 1.38, 0.85, 0.55, and 0.35\,mm) bands with bolometers cooled to 0.1\,\hbox{K}.  Polarization is measured in all but the highest two bands \citep{Leahy2010, Rosset2010}.  A combination of radiative cooling and three mechanical coolers produces the temperatures needed for the detectors and optics \citep{planck2011-1.3}.  Two data processing centres (DPCs) check and calibrate the data and make maps of the sky \citep{planck2011-1.7, planck2011-1.6}.  \Planck's sensitivity, angular resolution, and frequency coverage make it a powerful instrument for Galactic and extragalactic astrophysics as well as cosmology.  Early astrophysics results are given in Planck Collaboration VIII--XXVI 2011, based on data taken between 13~August 2009 and 7~June 2010.  Intermediate astrophysics results are presented in a series of papers between the major data releases.

\begin{figure}[tbp]
\begin{center}
\includegraphics[scale=1.05]{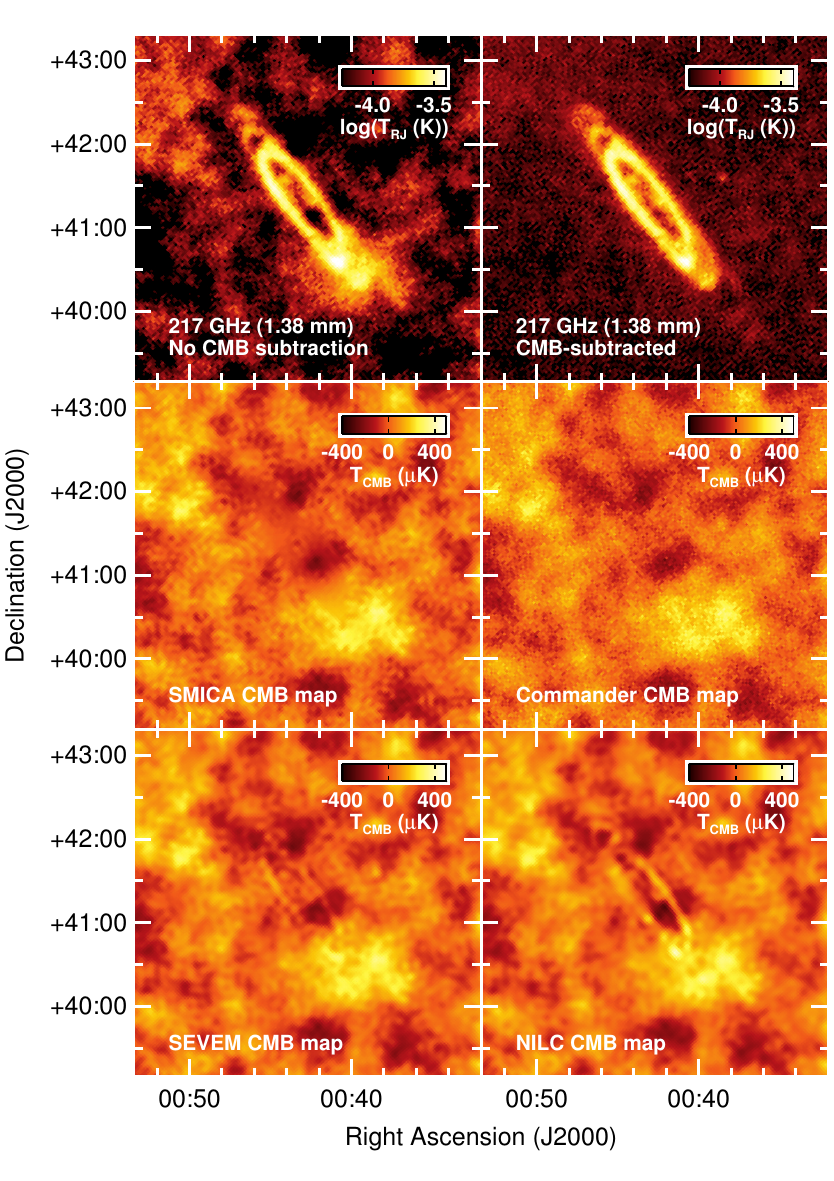}
\caption{{\it Top left}: M31 at 217\GHz\ (1.38\,mm), with no CMB subtraction. {\it Top right}: the {\tt SMICA} CMB-subtracted 217\GHz\ (1.38\,mm) map. {\it Middle left}: the {\tt SMICA} map of the CMB in the same region. {\it Middle right}: {\tt Commander} CMB map. {\it Bottom left}: {\tt SEVEM} CMB map. {\it Bottom right}: {\tt NILC} CMB map.}
\label{fig:m31cmb}
\end{center}
\end{figure}

\begin{figure*}[tbp]
\begin{center}
\includegraphics[scale=1.1]{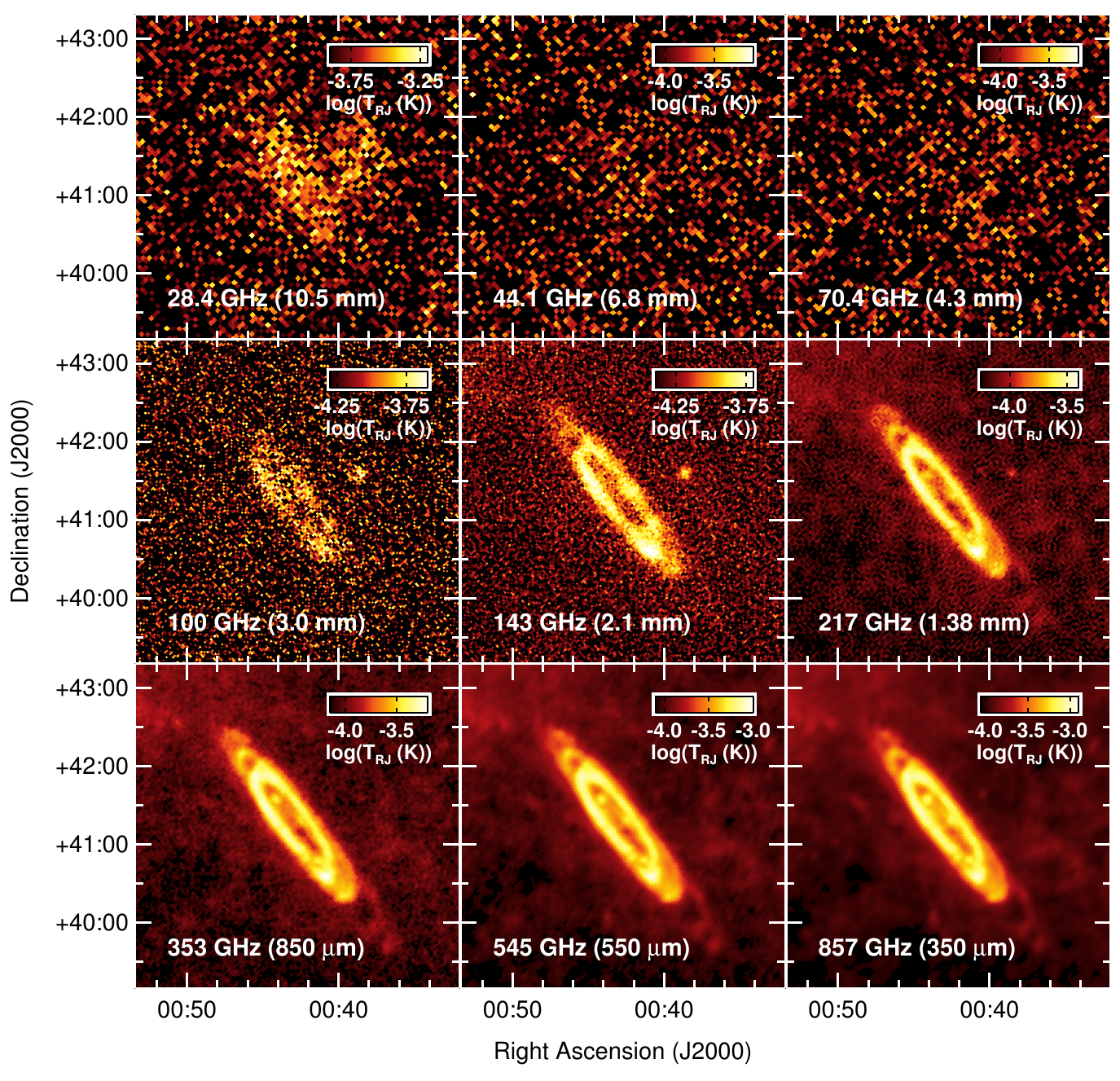}
\caption{Maps of M31 in total intensity from \Planck\ (after CMB subtraction). {\it Top to bottom, left to right}: \Planck\ 28.4, 44.1, and 70.4\GHz; \Planck\ 100, 143, and 217\ghz; \Planck\ 353, 545, and 857\ghz. All plots have units of Kelvin ($T_\mathrm{RJ}$), have a 1\deg\ equatorial graticule overlaid, are $250'\times250'$ with $0.5'\times0.5'$ pixels and are centred on RA 10\pdeg68, Dec 41\pdeg27, with North up and East to the left.}
\label{fig:m31planck}
\end{center}
\end{figure*}

In this paper we use \Planck\ data from the 2013 distribution of released products \citep{planck2013-p01}, which can be accessed via the \Planck\ Legacy Archive interface,\footnote{\url{http://pla.esac.esa.int/pla/}\\} based on the data acquired by \Planck\ during its ``nominal'' operations period from 13~August 2009 to 27 November 2010. In order to study M31 in the \Planck\ maps, the CMB needs to be subtracted. CMB maps in the M31 region are shown in Fig. \ref{fig:m31cmb}. The CMB-subtracted \Planck\ maps are presented in Fig. \ref{fig:m31planck} at their native resolution and their properties are summarized in Table \ref{tab:ancillarydata}. We use various combinations of the \Planck\ maps throughout this paper, and all maps are used to qualitatively study the emission at all \Planck\ frequencies in Sect. \ref{sec:morphology}. Maps at 217\GHz\ and higher frequencies (1.38\,mm and shorter wavelengths) are used to quantitatively investigate the emission in Sects. \ref{sec:colourratios} and \ref{sec:seds}, and all frequencies are used to measure the total emission in Sect. \ref{sec:lowressed}.

CMB subtraction is particularly important for M31, given the similarities in angular size between M31 and the anisotropies in the CMB. Additionally, there is an unfortunately large (\mbox{approximately 290\,$\upmu$K}) positive CMB fluctuation at the southern end of M31, which can be clearly seen in the CMB map panels of Fig. \ref{fig:m31cmb}. As part of the \Planck\ 2013 delivery, CMB maps from four component separation techniques were released, namely maps from the {\tt Commander}, {\tt NILC}, {\tt SEVEM} and {\tt SMICA} component separation methods \citep{planck2013-p06}. We specifically use the {\tt SMICA} CMB map to subtract the CMB from the \Planck\ data, as from a visual inspection this appears to be the cleanest map of the CMB in this region from the four methods (see Fig. \ref{fig:m31cmb}). The {\tt NILC} and {\tt SEVEM} maps are particularly contaminated by emission from M31; we also use the {\tt NILC} map to test the impact of residual foreground emission being subtracted out of the maps along with the CMB. Figure \ref{fig:m31cmb} shows the 217\GHz\ (1.38\,mm) map pre- and post-CMB subtraction, along with the four CMB maps of the M31 region.

The maps are converted from CMB temperature ($T_\mathrm{CMB}$) to Rayleigh-Jeans brightness temperature ($T_\mathrm{RJ}$) using the standard coefficients as described in \citet{planck2013-p28}; we also use the nominal frequencies for the bands, and (when fitting a spectral model to the data) colour corrections, depending on the spectra of the emission. The 100 and 217\ghz\ (3.0 and 1.38\,mm) \Planck\ bands include CO emission. The CO emission from M31 has been mapped with ground-based telescopes (e.g., \mbox{\citealp{Koper1991};} \citealp{Dame1993,Nieten2006}), but these do not include the full extent of M31 that is considered here. The CO emission is too weak to be reliably detected in the full-sky \planck\ CO maps \citep{planck2013-p03a}.\footnote{There is no detection in the \Planck\ Type 1 CO map \citep{planck2013-p03a}; the morphology is not consistent with the known structure in the Type 2 map, and although there is a detection in the Type 3 map and the ring morphology is visible, the detection does not have a high signal-to-noise ratio and may be contaminated by dust emission.} We do not correct for the CO emission in the maps; instead we omit the 217\GHz\ (1.38\,mm) channel from the SED fitting in Sect. \ref{sec:seds}, and we compare the level of CO emission expected from the ground-based CO map of \citet{Nieten2006} (described in Sect. \ref{sec:highres}) with the emission attributable to CO in the integrated \Planck\ measurements of M31 in Sect. \ref{sec:lowressed}.

To assess the uncertainty in the \Planck\ maps, we estimate the instrumental noise and cirrus contamination by measuring the scatter of flux densities in an adjacent background region of the \Planck\ maps (see Sect. \ref{sec:seds} for details). We conservatively assume calibration uncertainties of 10\,\% for 857 and 545\ghz\ (350 and 550\um) and 3\,\% for all other \Planck\ frequencies \citep[see][]{planck2013-p28}.

For the integrated spectrum analysis in Sect. \ref{sec:lowressed}, the \Planck\ full-sky maps are used directly in \healpix\footnote{\url{http://healpix.jpl.nasa.gov}} format \citep{Gorski2005}. For the higher resolution analyses, however, we use ``postage stamp'' 2-D projected maps centred on M31. To conserve the photometry of the data whilst repixelizing, we use the {\tt Gnomdrizz} package \citep{Paradis2012} to create the postage stamp maps from the \healpix\ data; since the \healpix\ Gnomview function uses nearest-neighbour interpolation, it does not necessarily conserve the photometry, although we tested that there is no significant difference in this case. The resulting postage stamp maps in equatorial coordinates are of size $250'\times250'$ with $0.5'\times0.5'$ pixels, centred on RA 10\pdeg68, Dec 41\pdeg27 ($l=121\pdeg2, b=-21\pdeg6$).

When quantitatively analysing the data, we first smooth to a common resolution of either 5\arcm\ (at 217\ghz\ and above / 1.38\,mm or lower, where the data have a native resolution of $4\parcm39$--$4\parcm87$) or 1\deg\ (at all frequencies) by convolving the map with a circular Gaussian beam with a full-width at half-maximum (FWHM) of $\theta_\mathrm{conv} = \left(\theta_\mathrm{new}^2-\theta_\mathrm{old}^2 \right)^{1/2}$, where $\theta_\mathrm{new}$ is the desired FWHM and $\theta_\mathrm{old}$ is the current FWHM of the maps. For some of the later analysis, we also repixelize the 5\arcm\ resolution data into 5\arcm\ pixels (see Sect. \ref{sec:colourratios}), while the 1\deg\ data are analysed at their native \healpix\ $N_\mathrm{side}$ resolution. The \Planck\ beams are not symmetric at the roughly 20\,\% level (e.g., see \citealp{planck2011-1.6,planck2011-1.7}), and the ancillary data sets used (see Sect.~\ref{sec:ancillary}) will also have non-Gaussian beams; smoothing the data to a common resolution reduces the effect of the asymmetry. However, a residual low-level effect will still be present in this analysis, for example in terms of introducing some correlation between adjacent 5\arcm\ pixels.

\section{Ancillary data} \label{sec:ancillary}
\begin{table*}
\begingroup
\newdimen\tblskip \tblskip=5pt
\caption{Sources of the data sets used in this paper, as well as their frequency, wavelength, resolution, calibration uncertainty, and rms on 5\arcm\ scales (for data with 5\arcm\ resolution or better only; see later). The Analysis column indicates whether the data set has been used in the 5\arcm\ high-resolution (H) and/or 1\deg\ low-resolution (L) analysis. The first part of the table describes the continuum data sets, and the second part describes the spectral line data sets.}
\label{tab:ancillarydata}
\nointerlineskip
\vskip -3mm
\footnotesize
\setbox\tablebox=\vbox{
    \newdimen\digitwidth 
    \setbox0=\hbox{\rm 0} 
    \digitwidth=\wd0 
    \catcode`*=\active 
    \def*{\kern\digitwidth}
    \newdimen\signwidth 
    \setbox0=\hbox{+} 
    \signwidth=\wd0 
    \catcode`!=\active 
    \def!{\kern\signwidth}
    \newdimen\pointwidth
    \setbox0=\hbox{{.}}
    \pointwidth=\wd0
    \catcode`?=\active
    \def?{\kern\pointwidth}
    \newdimen\notewidth
    \setbox0=\hbox{$^\mathrm{a}$}
    \notewidth=\wd0
    \catcode`@=\active
    \def@{\kern\notewidth}
    \halign{#\hfil\tabskip 10pt&
        #\hfil\tabskip 2pt&
        \hfil#\hfil\tabskip 2pt&
        \hfil#\hfil\tabskip 2pt&
        \hfil#\hfil\tabskip 2pt&
        \hfil#\hfil\tabskip 2pt&
        \hfil#\hfil\tabskip 2pt&
        \hfil#\hfil\tabskip 0pt\cr
        \noalign{\doubleline}
    \omit\hfil Source\hfil&$\nu$ [GHz]&$\lambda$ [mm]&Res.&Unc.& $\sigma$ (5\arcm, Jy) & Analysis & Reference\cr
\noalign{\vskip 4pt\hrule\vskip 6pt}
Haslam          & \phantom{\,}****0.408 & 734?**** & 60\parcm** & 10\,\% & \dots & L\phantom{H} & \citet{Haslam1982}\cr
Dwingeloo       & \phantom{\,}****0.820 & 365?**** & 72\parcm** & 10\,\% & \dots & L\phantom{H} & \citet{Berkhuijsen1972}\cr
Reich           & \phantom{\,}****1.4** & 214?**** & 35\parcm** & 10\,\% & \dots & L\phantom{H} & \citet{Reich2001}\cr
\WMAP\ 9-year   & \phantom{\,}***22.8** & *13?**** & 49\parcm** & *3\,\% & \dots & L\phantom{H} & \citet{bennett2012}\cr
\Planck         & \phantom{\,}***28.4** & *10.5*** & 32\parcm24 & *3\,\% & \dots & L\phantom{H} & \citet{planck2013-p02}\cr
\WMAP\ 9-year   & \phantom{\,}***33.0** & **9.0*** & 40\parcm** & *3\,\% & \dots & L\phantom{H} & \citet{bennett2012}\cr
\WMAP\ 9-year   & \phantom{\,}***40.7** & **7.4*** & 31\parcm** & *3\,\% & \dots & L\phantom{H} & \citet{bennett2012}\cr
\Planck         & \phantom{\,}***44.1** & **6.8*** & 27\parcm01 & *3\,\% & \dots & L\phantom{H} & \citet{planck2013-p02}\cr
\WMAP\ 9-year   & \phantom{\,}***60.7** & **4.9*** & 21\parcm** & *3\,\% & \dots & L\phantom{H} & \citet{bennett2012}\cr
\Planck         & \phantom{\,}***70.4** & **4.3*** & 13\parcm25 & *3\,\% & \dots & L\phantom{H} & \citet{planck2013-p02}\cr
\WMAP\ 9-year   & \phantom{\,}***93.5** & **3.2*** & 13\parcm** & *3\,\% & \dots & L\phantom{H} & \citet{bennett2012}\cr
\Planck         & \phantom{\,}**100?*** & **3.0*** & *9\parcm65 & *3\,\% & \dots & L\phantom{H} & \citet{planck2013-p03}\cr
\Planck         & \phantom{\,}**143?*** & **2.1*** & *7\parcm25 & *3\,\% & \dots & L\phantom{H} & \citet{planck2013-p03}\cr
\Planck         & \phantom{\,}**217?*** & **1.38** & *4\parcm99 & *3\,\% & 0.015 & LH & \citet{planck2013-p03}\cr
\Planck         & \phantom{\,}**353?*** & **0.85** & *4\parcm82 & *3\,\% & 0.05* & LH & \citet{planck2013-p03}\cr
\Planck         & \phantom{\,}**545?*** & **0.55** & *4\parcm68 & *7\,\% & 0.13* & LH & \citet{planck2013-p03}\cr
\Herschel\ SPIRE PLW & \phantom{\,}**600?*** & **0.50** &  *0\parcm59 & *4\,\% & 0.09* &\phantom{L}H & \citet{Fritz2012}\cr
\Planck         & \phantom{\,}**857?*** & **0.35** & *4\parcm33 & *7\,\% & 0.31* & LH & \citet{planck2013-p03}\cr
\Herschel\ SPIRE PMW & \phantom{\,}**857?*** & **0.35** &  *0\parcm40 &  *4\,\% & 0.23*  &\phantom{L}H & \citet{Fritz2012}\cr
\Herschel\ SPIRE PSW & \phantom{\,}*1200?*** & **0.25** &  *0\parcm29 &  *4\,\% & 0.29*  &\phantom{L}H & \citet{Fritz2012}\cr
\DIRBE          & \phantom{\,}*1249?*** & **0.240* & 40\parcm** & 13\,\% & \dots &L\phantom{H} & \citet{Hauser1998}\cr
\ISO            & \phantom{\,}*1763?*** & **0.175* & *1\parcm3* & 10\,\% & 0.8** &\phantom{L}H & \citet{Haas1998}\cr
\Spitzer\ MIPS B3 & \phantom{\,}*1870?*** & **0.16* & *0\parcm63 & 12\,\% & 0.53* &\phantom{L}H & \citet{Gordon2006,Bendo2012b}\cr
\Herschel\ PACS & \phantom{\,}*1874?*** & **0.16** &  *0\parcm22 &  *4\,\% & 2.4**  &\phantom{L}H & \citet{Fritz2012}\cr
\DIRBE          & \phantom{\,}*2141?*** & **0.14** & 40\parcm** & 13\,\% & \dots &L\phantom{H} & \citet{Hauser1998}\cr
\DIRBE          & \phantom{\,}*2997?*** & **0.10** & 40\parcm** & 13\,\% & \dots &L\phantom{H} & \citet{Hauser1998}\cr
\Herschel\ PACS & \phantom{\,}*2998?*** & **0.10** &  *0\parcm21 &  *4\,\% & 2.5**  &\phantom{L}H & \citet{Fritz2012}\cr
\IRAS\ (IRIS) B4 & \phantom{\,}*3000?*** & **0.10** & *4\parcm3* & 13\,\% & 0.37* &LH & \citet{Miville2005}\cr
\Spitzer\ MIPS B2 & \phantom{\,}*4280?*** & **0.07** & *0\parcm3* & 10\,\% & 0.12* &\phantom{L}H & \citet{Gordon2006,Bendo2012b}\cr
\DIRBE & \phantom{\,}*5000?*** & **0.06** & 40\parcm** & 13\,\% & \dots &L\phantom{H} & \citet{Hauser1998}\cr
\IRAS\ (IRIS) B3 & \phantom{\,}*5000?*** & **0.06** & *4\parcm0* & 13\,\% & 0.12*  &LH & \citet{Miville2005}\cr
\IRAS\ (IRIS) B2 & 12\,000?*** & **0.025* & *3\parcm8* & 13\,\% & 0.06* &LH & \citet{Miville2005}\cr
\Spitzer\ MIPS B1 & 12\,490?*** & **0.024* & *0\parcm1* & *4\,\% & 0.012 &\phantom{L}H & \citet{Gordon2006,Bendo2012b}\cr
\IRAS\ (IRIS) B1 & 25\,000?*** & **0.012* & *3\parcm8* & 13\,\% & 0.04*  &LH & \citet{Miville2005}\cr
\Spitzer\  IRAC & 83\,000?*** & **0.0036 & *1\parcs7* & *3\,\% & \dots  & \phantom{L}H & \citet{Barmby2006,IRAC2012}\cr
\noalign{\vskip 4pt\hrule\vskip 6pt}
DRAO \hi\ & \phantom{\,}****1.4** & \dots & *0\parcm37 & *5\,\% & \dots & \phantom{L}H & \citet{Chemin2009}\cr
IRAM CO & \phantom{\,}**115.*** & \dots & *0\parcm38 & 15\,\% & \dots & \phantom{L}H & \citet{Nieten2006}\cr
\noalign{\vskip 3pt\hrule\vskip 4pt}
}}
\endPlancktablewide
\endgroup
\end{table*}

The ancillary data that we use in this paper fall under two categories. For the higher resolution spatial analysis, we need observations with resolution equal to or greater than the \Planck\ high frequency resolution of 5\arcm; these are described in Sect. \ref{sec:highres}. For the integrated SED, we can make use of large-scale survey data with a resolution of 1\deg\ or higher; we describe these data sets in Sect. \ref{sec:lowres}. All of the data sets are summarized in Table \ref{tab:ancillarydata}.

We also make use of ancillary information about the distance and inclination of M31. M31 is at a distance of ($785\pm25$)\,kpc \citep{McConnachie2005}, with an optical major isophotal diameter of $190\parcm5\pm0\parcm1$ and a major-to-minor ratio of $3.09\pm0.14$ \citep{deVaucouleurs1991}. Following \citet{Xu1996}, we assume that it has an inclination angle $i=79$\deg\ and a position angle of 37\deg\ with respect to North (both in equatorial coordinates).

\subsection{The 5\arcm\ data set} \label{sec:highres}
We use six data sets in addition to the \Planck\ data in our 5\arcm\ resolution study. At high frequencies, we use \IRAS, \Spitzer, \Herschel, and \ISO\ data to trace the shorter wavelength dust emission. At lower frequencies, we use ground-based \hi\ and CO data sets to trace the gas within the galaxy. To match the pixelization and resolution of these other data to the \Planck\ data, we regrid the data onto 0.25\arcm\ pixels in the same coordinate system and sky region as the \Planck\ data and then smooth them to a common resolution of 5\arcm. Further repixelization to 5\arcm\ pixels is then carried out for the analysis in later sections.

To trace dust emission in the infrared, we use 24, 70, and 160\um\ data originally acquired by \citet{Gordon2006} using the Multiband Imaging Photometer for \Spitzer\ (MIPS; \citealp{Rieke2004}) on the \Spitzer\ {\it Space Telescope} \citep{Werner2004}. The 24, 70, and 160\um\ data have point spread functions with FWHM of 6\arcs, 18\arcs, and 38\arcs, respectively.  The data were processed using the MIPS Data Analysis Tools \citep{Gordon2005} as well as additional data processing steps described by \citet{Bendo2012b}. The data processing includes the removal of the effect of cosmic ray hits and drift in the background signal.  The background is initially subtracted from the individual data frames by characterizing the variations in the background signal as a function of time using data that fall outside of the galaxy.  This removes large-scale background/foreground structure outside the galaxy, including the cosmic infrared background (CIB) and zodiacal light, but residual foreground cirrus structures and compact sources remain in the data.  Any residual background emission in the final maps is measured outside the optical disc of the galaxy and subtracted from the data.  For additional details on the data reduction process, see the description of the data reduction by \citet{Bendo2012b}. PSF characteristics and calibration uncertainties are described by \citet{Engelbracht2007}, \citet{Gordon2007}, and \citet{Stansberry2007}. We use the data outside of the mask to assess the rms uncertainty in the \Spitzer\ data, including thermal noise, residual cirrus and other large-scale features that contribute to the uncertainty on 5\arcm\ scales.

We additionally utilise the 3.6\um\ data produced by \citet{Barmby2006} using the \Spitzer\ Infrared Array Camera (IRAC; \citealp{Fazio2004}) to trace the older stellar population in the bulge of M31; these data have a resolution of 1\parcs7.

We also include \herschel\ data from the \Herschel\ Exploitation of Local Galaxy Andromeda (HELGA) survey \citep{Fritz2012}. These data are at 500, 350, 250, 160 and 100\um\ (600, 857, 1200, 1874 and 2998\,GHz) and have resolutions between 0\parcm21 and 0\parcm59 \citep{Herschel2012,Herschel2014}. We re-align the data to match the \Spitzer\ 24\um\ data using three compact sources that are seen at all \Herschel\ frequencies. The \Herschel\ data have a flux calibration uncertainty of 4\,\% \citep{PACS,Bendo2013}. For the integrated SED, we smooth the \Herschel\ data to $1\deg$, and then regrid these data onto an oversampled $N_\mathrm{side}=$16\,384 \healpix\ map, which was then resampled down to \mbox{$N_\mathrm{side}=256$}.

\changed{We also make use of several data sets for comparison purposes, namely: the IRAS data at 12, 25, 60, and 100\um\ from \citet{Miville2005}; the \ISO\ 175\um\ data from \citet[][private communication]{Haas1998}; the \hi\ emission map from \citet{Chemin2009}; and the $^{12}$CO $J$=1$\rightarrow$0 map of M31 from \citet{Nieten2006}. These are described in Appendix \ref{sec:data}.}

\subsection{The 1\deg\ data set} \label{sec:lowres}

When looking at the integrated spectrum of M31, we make use of a number of large-area, low-resolution surveys. These are publicly available in \healpix\ format. We convolve the data sets to a resolution of 1\deg\ to match the resolution of the low-frequency data sets and perform the analysis directly on the \healpix\ maps.

At low frequencies, we use the low-resolution radio maps of the sky at 408\,MHz \citep[73.4\,cm; ][]{Haslam1981,Haslam1982}, 820\,MHz, \citep[36.5\,cm; ][]{Berkhuijsen1972} and 1.4\ghz\ \citep[21.4\,cm; ][]{Reich1982,Reich1986,Reich2001}. We assume that there is a 10\,\% uncertainty in these maps, which includes both the uncertainty in the flux density calibration (between 5 and 10\,\% depending on the survey) and also uncertainties arising from the morphology of the surrounding structure interacting with the aperture photometry technique we use. This choice of uncertainty has been shown to be reasonable for Galactic anomalous microwave emission (AME) clouds \citep{planck2013-XV}. For the 408\MHz\ (73.4\,cm) map, we also add 3.8\,Jy to the uncertainty to take into account the baseline striations in the map, which are at the level of $\pm3$\,K (\citealp{Haslam1982}; \citealp{planck2013-XV}). All of the maps have been calibrated on angular scales of around 5\deg, and consequently the difference between the main and full beams (the latter including sidelobes) needs to be taken into account when measuring the flux densities of more compact sources. This is significant for the 1.4\ghz\ (21.4\,cm) map, where the factor for objects on 1\deg\ scales is approximately 1.55. As M31 is on an intermediate scale, we adopt an intermediate correction factor of $1.3\pm0.1$, and also include the uncertainty in this correction factor in the flux density uncertainty. We assume that the value and uncertainty on the ratio for the 408\MHz\ (73.4\,cm) and 820\MHz\ (36.5\,cm) maps is small, and hence will be well within the existing calibration uncertainties, as per e.g., \citet{planck2013-XV}. We also make use of the integrated flux densities from higher-resolution surveys collated by \citet{Berkhuijsen2003} for comparison to those extracted from the maps.

\begin{figure*}[tbp]
\begin{center}
\includegraphics[width=120mm]{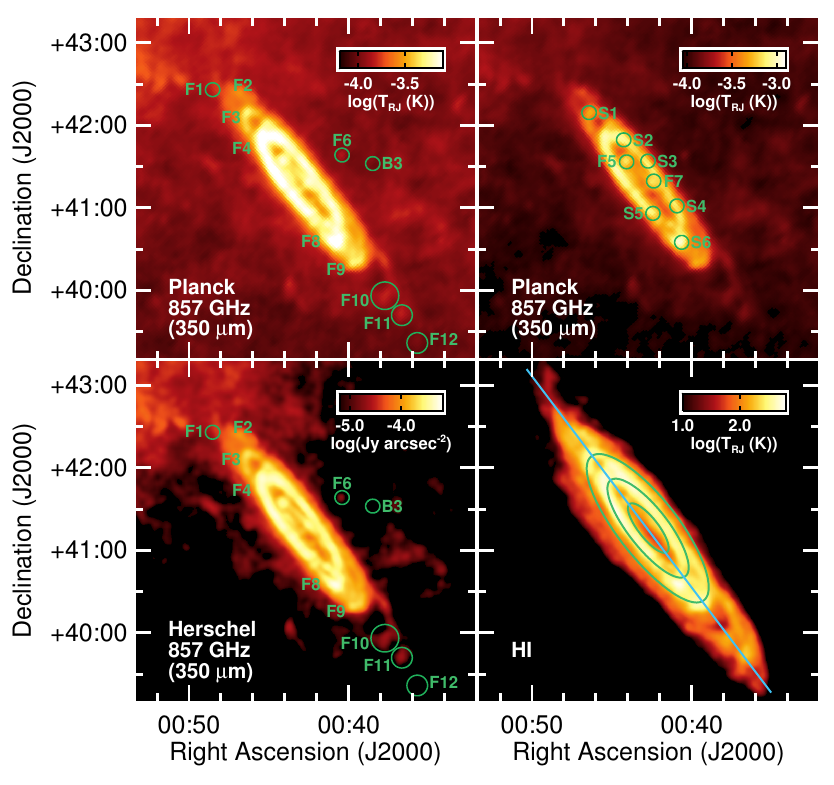}
\caption{{\it \changed{Top-left}}: The 857\ghz\ (350\um) \changed{\Planck} map of M31, highlighting the extension of the dust out to large distances. Fields of interest are labelled with ``F,'' and circled in some cases. The position of the radio source B3 0035+413 is also shown (labeled B3). {\it \changed{Top-right}}: The 857\ghz\ (350\um) map, with a scale chosen to highlight the brighter emission. Circles labelled with ``S'' indicate the sources included in the \Planck\ ERCSC (see Appendix \ref{sec:sources}); two fields in the centre region are also labelled with ``F''. \changed{{\it Bottom-left}: the 857\GHz\ (350\um) \Herschel\ map convolved to the same resolution as the \Planck\ data, with the same fields of interest labelled as in the \Planck\ map above.} {\it \changed{Bottom-right}}: A map of the \hi\ emission from M31 \citep{Chemin2009} convolved to the same resolution as the \Planck\ 857\ghz\ (350\um) map, with 5, 10, and 15\,kpc rings overlaid in green, and the cut used in Fig. \ref{fig:line} in blue.}
\label{fig:m31planck857}
\end{center}
\end{figure*}
\begin{figure}[tbp]
\begin{center}
\includegraphics[width=90mm]{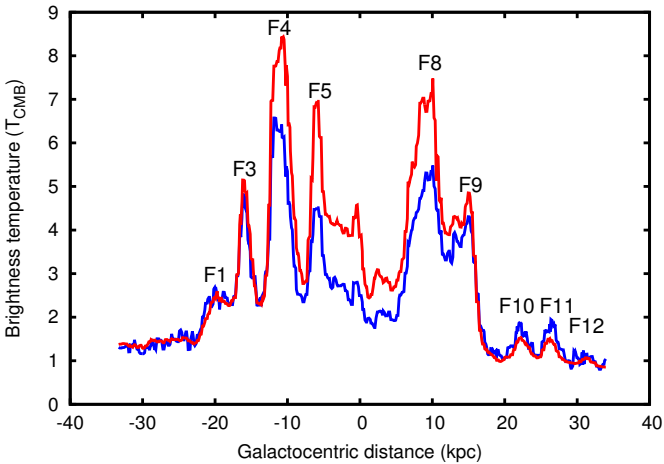}
\caption{Cut along the major axis of M31. Negative values are on the higher declination side of the galaxy. The red line is 857\GHz\ (350\um) and the blue line is 353\GHz\ (850\um) rescaled by a factor of 1000. Various spiral arm structures, including the outermost rings at 20--30\kpc, are clearly visible at both frequencies. Fields are numbered as per Fig. \ref{fig:m31planck857}.}
\label{fig:line}
\end{center}
\end{figure}

At intermediate frequencies, in addition to \Planck-LFI data, we use the deconvolved and symmetrized 1\deg-smoothed \WMAP\ 9-year data at 22.8, 33.0, 40.7, 60.7, and 93.5\ghz\ \citep[13, 9.0, 7.4, 4.9, and 3.2\,mm; ][]{bennett2012}\footnote{\url{http://lambda.gsfc.nasa.gov/product/map/dr4/maps_band_smth_r9_i_9yr_get.cfm}}. When fitting a model to these data, we apply colour corrections following the recipe in \citet{bennett2012}, and we conservatively assume a calibration uncertainty of 3\,\%.

At higher frequencies, we include the low-resolution \DIRBE\ data at 1249, 2141, and 2997\GHz\ (240, 140, and 100\um;  \citealp{Hauser1998}), in addition to the \changed{\IRAS\ data described in Appendix \ref{sec:data}.} We assume that these data have an uncertainty of 13\,\%. We also use the \Herschel\ data as described above.

\section{Infrared morphology} \label{sec:morphology}

The \Planck\ maps of M31 are shown in Fig. \ref{fig:m31planck}. At 100\GHz\ and all higher \Planck\ frequencies (3.0\,mm and shorter wavelengths), the prominent 10\,kpc dust ring of Andromeda can clearly be seen -- as expected based on previous infrared observations of M31 -- as well as a number of other extended features.

At high frequencies, where we have the highest signal-to-noise and highest spatial resolution, we can see features much further out than the 10\,kpc ring. The \changed{top} panels of Fig. \ref{fig:m31planck857} show the \Planck\ 857\,GHz (350\um) map labelled to show the locations of the key features in the dust emission; \changed{the bottom-left panel shows the \Herschel\ 857\GHz\ (350\um) data, and the bottom-right} panel of Fig. \ref{fig:m31planck857} shows the \hi\ map from \citet{Chemin2009} for comparison. A cut through the \changed{\Planck} 857 and 353\GHz\ (350 and 850\um) maps is shown in Fig. \ref{fig:line}. To the south, a total of four spiral arm or ring structures can clearly be seen -- the 10\,kpc ring (``F8''), as well as a second structure just outside of the ring (``F9'') and the more distant 21\,kpc (``F10'') and 26\,kpc (``F11'') arms. These features have also been identified by \mbox{\citet{Fritz2012}}\changed{, and can also be seen in the panel displaying the \Herschel\ 857\GHz\ data}. We see a hint of the emission at 31\,kpc (``F12''), which is also seen in \hi\ emission, but we are unable to confirm it as being part of M31, because of the large amount of surrounding cirrus emission from our Galaxy. To the north, we see three sets of spiral arm structures (``F4'', ``F3'', and ``F2'') and a wisp of emission at the northernmost end of the galaxy (``F1'', confirmed by cross-checking against \hi\ emission, as shown in the \changed{top-right} panel) before running into confusion from Galactic emission. These features can be seen most clearly at 857 and 545\ghz\ (350 and 550\um), but are also clearly visible in the 353, 217 and 143\ghz\ (0.85, 1.38 and 2.1\,mm) maps after CMB subtraction. The 10\kpc\ ring can still be seen clearly at 100\ghz, but at that frequency the more extended structure is not visible. We will use the terminology of ``ring'' for the 10\,kpc ring, since it has been clearly demonstrated to be a near-complete ring (e.g., see \citealp{Haas1998}), and ``arm'' for all other structures that may not completely circle the galaxy.

An area of particularly strong long-wavelength emission can be clearly seen at the southern end of the 10\,kpc ring, down to frequencies of 100\ghz\ (3.0\,mm; see Fig. \ref{fig:m31planck}; in Fig. \ref{fig:m31planck857} it is just to the right of ``F8'' and is also marked as ``S6''). This has the highest contrast with the rest of M31 at frequencies of 143 and 217\ghz\ (2.1 and 1.38\,mm); there are also hints of it down to 70.4\ghz\ (4.3\,mm). This is probably what is seen at the highest frequencies of the \WMAP\ data and in \Planck\ data by \citet{DePaolis2011} \changed{and} \citet{DePaolis2014}; i.e., the asymmetrical emission between the north and south parts of M31 that they detect is caused by either varying dust properties across the galaxy or CMB fluctuations, rather than being caused by galactic rotation.

Within the 10\,kpc ring, several features are also visible. A bright spot inside the Galactic ring to the north is marked as ``F5'' in the \changed{top-right} panel of Fig. \ref{fig:m31planck857}. This region corresponds to a limb of an asymmetric spiral structure within the 10\,kpc ring that has also been seen in the higher-resolution \Spitzer\ data \citep{Gordon2006}. \Planck\ does not detect the emission from the central nucleus, despite the prominence of this region in maps of M31 at higher frequencies. This implies that the majority of the dust in this region is at a higher dust temperature than average -- an implication which will be explored in later sections. \Planck\ does, however, detect a compact object to the right of the nucleus (marked as ``F7'') that has also been seen with \IRAS\ \citep{Rice1993}, \ISO\ \citep{Haas1998}, and \Spitzer\ \citep{Gordon2006}; the latter suggests that this emission is located where a bar and the spiral arm structure meet.

Outside of the ring, we see some cirrus dust emission from our own Galaxy, particularly at the higher \Planck\ frequencies. This is largely present on the north side of M31, towards the Galactic plane, but it can also be seen elsewhere, for example there is an arc of cirrus emission above the outermost rings at the south end of M31. Depending on the temperature and spectral properties of this emission, it may present problems for studies of the Sunyaev-Zeldovich effect in the halo of M31 (e.g., \mbox{\citealp{Taylor2003}}) unless the emission is properly taken into account or suitably masked.

At \Planck's lowest frequencies, M31 is clearly detected but is not resolved, because of the low resolution at these frequencies. It is clearest at 28.4\GHz\ (10.5\,mm), but can also be seen at 44.1 and 70.4\ghz\ (6.8 and 4.3\,mm) -- although at these frequencies CMB subtraction uncertainty starts to become important. The emission mechanism powering the source detections at these frequencies will be discussed in the context of the integrated SED in Sect. \ref{sec:lowressed}.

There are also several sources in the field near to M31 that are present in the maps shown here, including the dwarf galaxy M110/NGC205 (``F6'') and the blazar B3\,0035+413 (``B3''), additionally a number of components of M31 are included in the \Planck\ Early Release Compact Source Catalogue \citep[ERCSC; ][]{planck2011-1.10} and the \Planck\ Catalogue of Compact Sources \mbox{\citep[PCCS; ][]{planck2013-p05}}. We discuss these in Appendix \ref{sec:sources}.

\begin{figure*}[tbp]
\begin{center}
\includegraphics{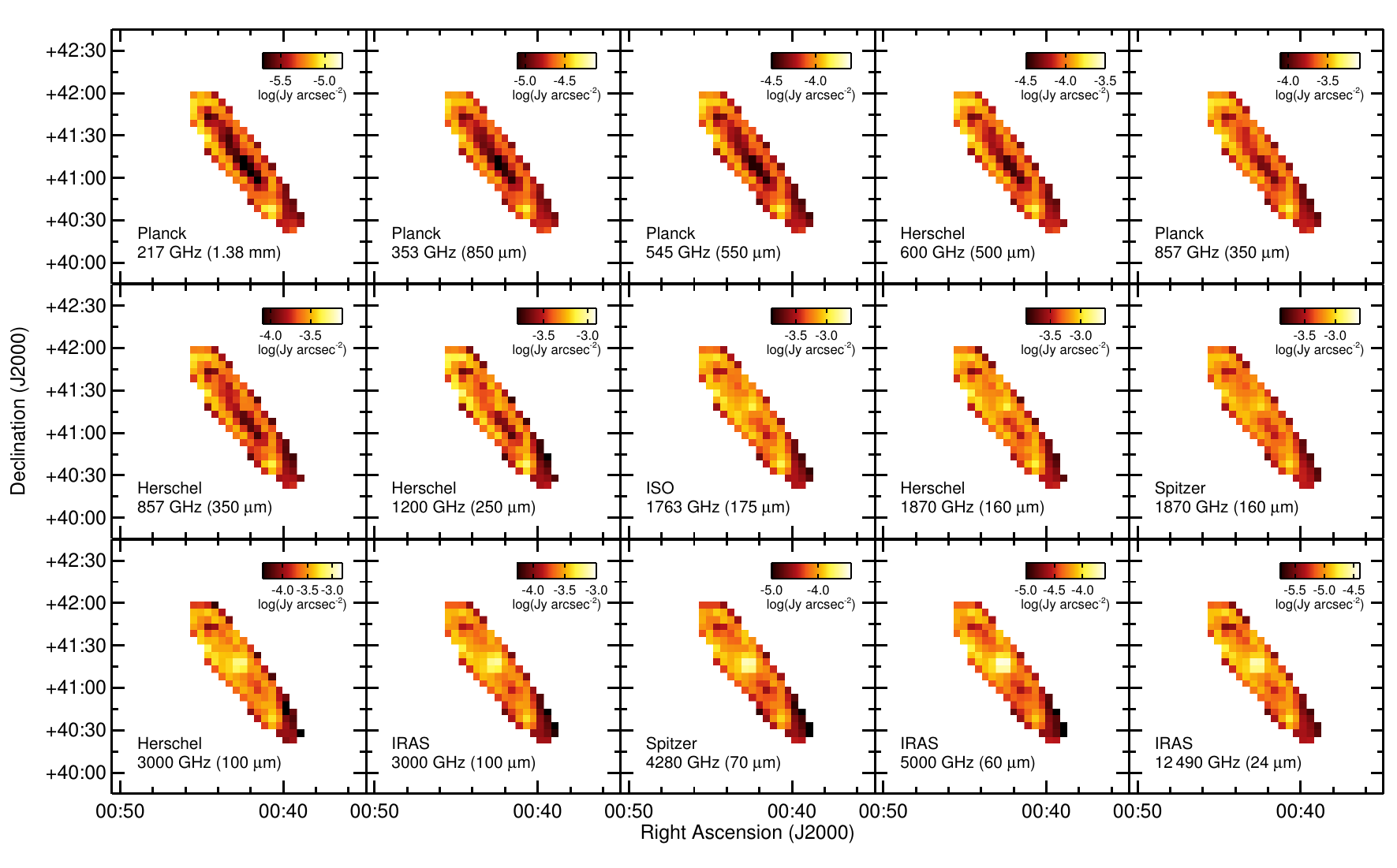}
\caption{Masked and pixelized 5\arcm\ data set ordered in increasing frequency (decreasing wavelength). The maps are $3\deg\times3\deg$ in size. There is a clear change of morphology as frequency increases, with the rings being most prominent at the lower frequencies, and the nucleus being highlighted at higher frequencies.}
\label{fig:m31_5arcmin}
\end{center}
\end{figure*}

\section{Colour ratios and implications for dust heating} \label{sec:colourratios}
\subsection{Colour ratios}
\begin{figure}[tbp]
\begin{center}
\includegraphics[scale=0.85]{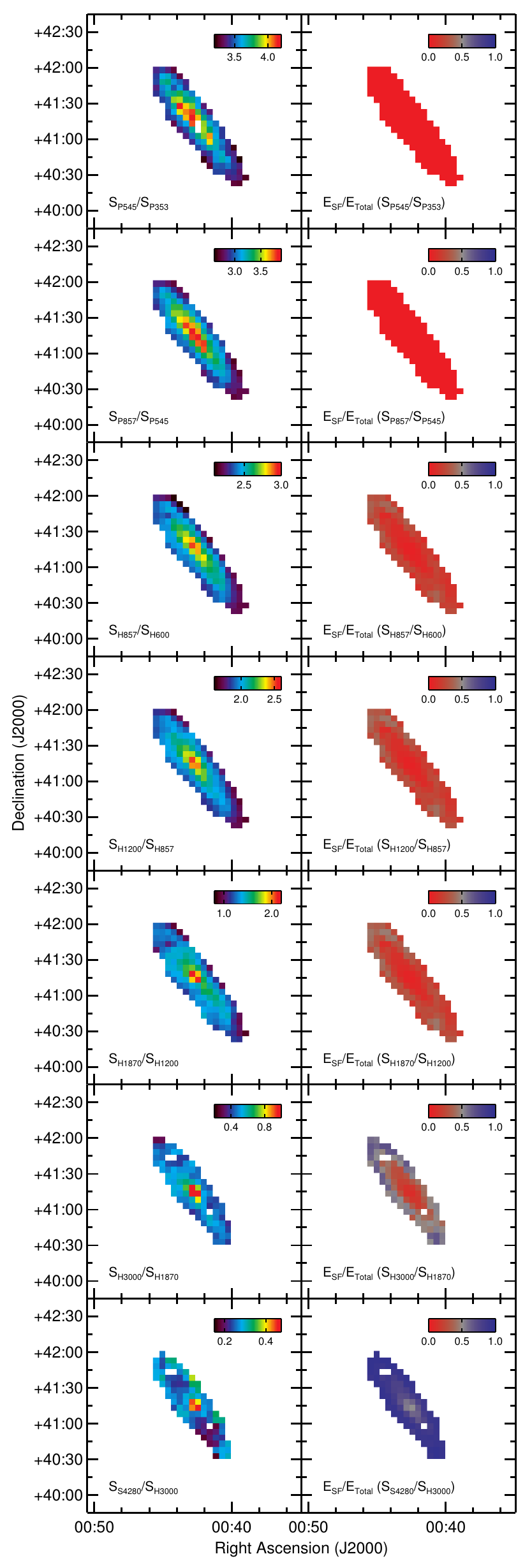}
\caption{{\it Left}: colour ratio plots with 5\arcm\ pixels. The \mbox{5\arcm\ pixels} are slightly bigger than the beam size, such that the signal they contain is largely independent. The colour ratios for the lower frequency plots do not trace the local features; rather, they depend on the galactocentric distance. The higher frequency plots -- in particular the highest frequency one -- trace the local structure much more. {\it Right}: images of the relative contribution of star formation to heating the dust traced by the specified pair of frequencies.  These maps were derived using Eq.~(\ref{eq:efrac}) and the parameters in Table~\ref{tab:efunc}. {\it Top to bottom}: $S_\mathrm{P545}/S_\mathrm{P353}$, $S_\mathrm{P857}/S_\mathrm{P545}$, $S_\mathrm{H857}/S_\mathrm{H600}$, $S_\mathrm{H1200}/S_\mathrm{H857}$, $S_\mathrm{H1874}/S_\mathrm{H1200}$, $S_\mathrm{H3000}/S_\mathrm{H1874}$, and $S_\mathrm{S4280}/S_\mathrm{H3000}$.}
\label{fig:m31_colourratio}
\end{center}
\end{figure}

A number of recent analyses based on \Herschel\ data have used FIR surface brightness ratios to examine dust heating sources within nearby galaxies.  \citet{Bendo2010} performed the first such analysis on M81; subsequent analyses were performed on M33 by \citet{Boquien2011}, on M83 and NGC\,2403 by \citet{Bendo2012}, and on a wider sample of galaxies by \citet{Bendo2014}.  These techniques have been shown to be more robust at identifying dust heating sources than comparing FIR surface brightnesses directly to star\changed{-}formation tracers, since brightness ratios normalize out the dust column density.

We studied the dust heating sources in M31 using surface brightness ratios measured in all \Planck\ data at frequencies of 217\ghz\ and above (1.38\,mm and shorter wavelengths), as well as in \Herschel, \ISO, \Spitzer, and \IRAS\ data. We convolved all of the high-resolution data (marked with ``H'' in Table \ref{tab:ancillarydata}) to 5\arcm\ resolution using a Gaussian kernel.  We then rebinned the data to 5\arcm\ pixels; we selected this size because it is the same size as the beam, and so the signal in each pixel should be largely independent of the others. It is important to do this repixelization in order to avoid the appearance of artificial correlations in the data. We masked out data from outside the optical disc of the galaxy by requiring the pixel centre to be less than 22\kpc\ away from the centre of M31 (which is equivalent to a 1\pdeg42 radius along the major axis).  To avoid pixels strongly affected by foreground or background noise, we only used data from pixels that had been detected in the 353\ghz\ (850\um) image at 10 times the rms uncertainty in the data (measured in a $10\times10$ pixel region of sky to the far bottom-left of the data set used here). We also removed pixels at the top-left corner of the images, where the data have been contaminated by bright cirrus structures in the foreground. Finally, we removed pixels at the edge of the optical disc, which primarily sample a combination of background emission and emission from the wings of the beams for sources in the dust ring. This leaves 126 independent data points. We look at colour ratios in adjacent bands of the same telescope to minimize the effect of different telescope beams on the results; the exception to this is the highest frequency comparison, where we have no alternative but to compare \Spitzer\ with \Herschel\ data.

The rebinned data are shown in Fig. \ref{fig:m31_5arcmin}.  The 10\kpc\ ring is present at all frequencies, but emission from the centre of the galaxy becomes more prominent as the frequency increases. At 1800\ghz\ (170\um), both sources are approximately similar in surface brightness.  The bright feature at the southern end of M31 (source ``S6'' in Fig. \ref{fig:m31planck857}) is clearly present at all frequencies, although it is more notable at $\nu<1700$\ghz. There is a  change of morphology apparent as frequency increases, with the rings being most prominent at the lower frequencies, and the nucleus being highlighted at higher frequencies.

Figure \ref{fig:m31_colourratio} shows the surface brightness ratios between adjacent frequency bands in the combined \Planck\ and ancillary data set, where e.g., $S_{545}/S_{353}$ denotes the colour ratio between 545 and 353\GHz\ (0.55 and 0.85\,mm). The  ratios of lower frequency bands (i.e., the \Planck\ bands) appear to smoothly decrease with radius, although the data are somewhat noisy at $S_{545}/S_{353}$.  The 10\,kpc ring is hardly noticable at all in the lower frequency ratio maps. In the higher frequency ratio maps (e.g., the $S_{4280}/S_{3000}$ map), the ring is much more prominent, and the colours also appear enhanced in a compact region around the nucleus.  

These results imply that the different frequencies are detecting dust heated by different sources.  At higher frequencies, the enhanced surface brightness ratios in the ring and the nucleus indicate that the dust is being heated by the star-forming regions in these structures.  At lower frequencies, however, the ratios vary more smoothly with radius, and the ring does not appear to be as enhanced as in the data for the higher frequencies.  This is consistent with dust heating being dominated by the total stellar population, including stars in the bulge of the galaxy, which should vary smoothly with radius (as has been suggested for other galaxies by \citealp{Bendo2010,Bendo2012}).

\subsection{Determination of dust heating mechanism}
To link the surface brightness ratios to heating sources, we compared the ratios to tracers of the total stellar population and star formation.  As a tracer of the total stellar population, we used the \Spitzer\ 3.6\um\ image, which is generally expected to be dominated by the Rayleigh-Jeans tail of the photospheric emission from the total stellar population \citep{Lu2003}.  While the 3.6\um\ band may also include roughly $1000$\,K dust emission from star-forming regions \citep{Mentuch2009, Mentuch2010}, this effect is usually only a major issue in late-type galaxies with relatively strong star formation compared to the total stellar surface density.  A high infrared-to-visible ratio or a dominant active galactic nuclei (AGN) could also contaminate the 3.6\um\ emission; however, neither of these issues are present in M31. We used the \Spitzer\ 24\um\ image as a star-formation tracer, as this has been shown to generally originate from SFR dust \citep{Calzetti2005, Calzetti2007, Prescott2007, Zhu2008, Kennicutt2009}, although we caution that it is possible for some 24\um\ emission to originate from dust heated by the diffuse interstellar radiation field from older stars \citep[e.g.,][]{Li2001,Kennicutt2009}.  The 24\um\ band also includes a small amount of stellar emission; this was removed by multiplying the 3.6\um\ image by 0.032 and then subtracting it from the 24\um\ image, as described by \citet{Helou2004}. While we could use 24\um\ data combined with \ha\ or ultraviolet emission to trace both obscured and unobscured star formation (as suggested by, e.g., \citealp{Leroy2008} and \citealp{Kennicutt2009}), the publicly-available data have problems that make them difficult to include in our analysis\footnote{The publicly-available H$\alpha$ data, such as from the Virginia Tech Spectral-Line Survey, \citet{Dennison1998}, contain artefacts from incompletely-subtracted foreground stars, as well as incompletely-subtracted continuum emission from the bulge of M31, and are therefore unsuitable for our analysis. The publicly-available ultraviolet data contain multiple bright foreground stars, as well as emission from older stellar populations within \changed{M31.}}. \citet{Bendo2014} have demonstrated that using the 24\um\ data as a star\changed{s}formation tracer for this analysis will yield results that are similar to using a composite \ha\ and 24\um\ star\changed{-}formation tracer \changed{(to within about 5\,\%)}, so given the issues with the ultraviolet and \ha\, we will use solely 24\um\ emission as \changed{our nominal} star\changed{-}formation tracer. \changed{We also use the combination of the 24\um\ and far-ultraviolet (FUV) data, using the foreground star-subtracted GALEX data from \citet{Ford2013} and \citet{Viaene2014}, to check whether this has a significant impact on our results. We return to the results of using this star-formation tracer later in this section.}

\begin{figure}
\begin{center}
\includegraphics[scale=0.17]{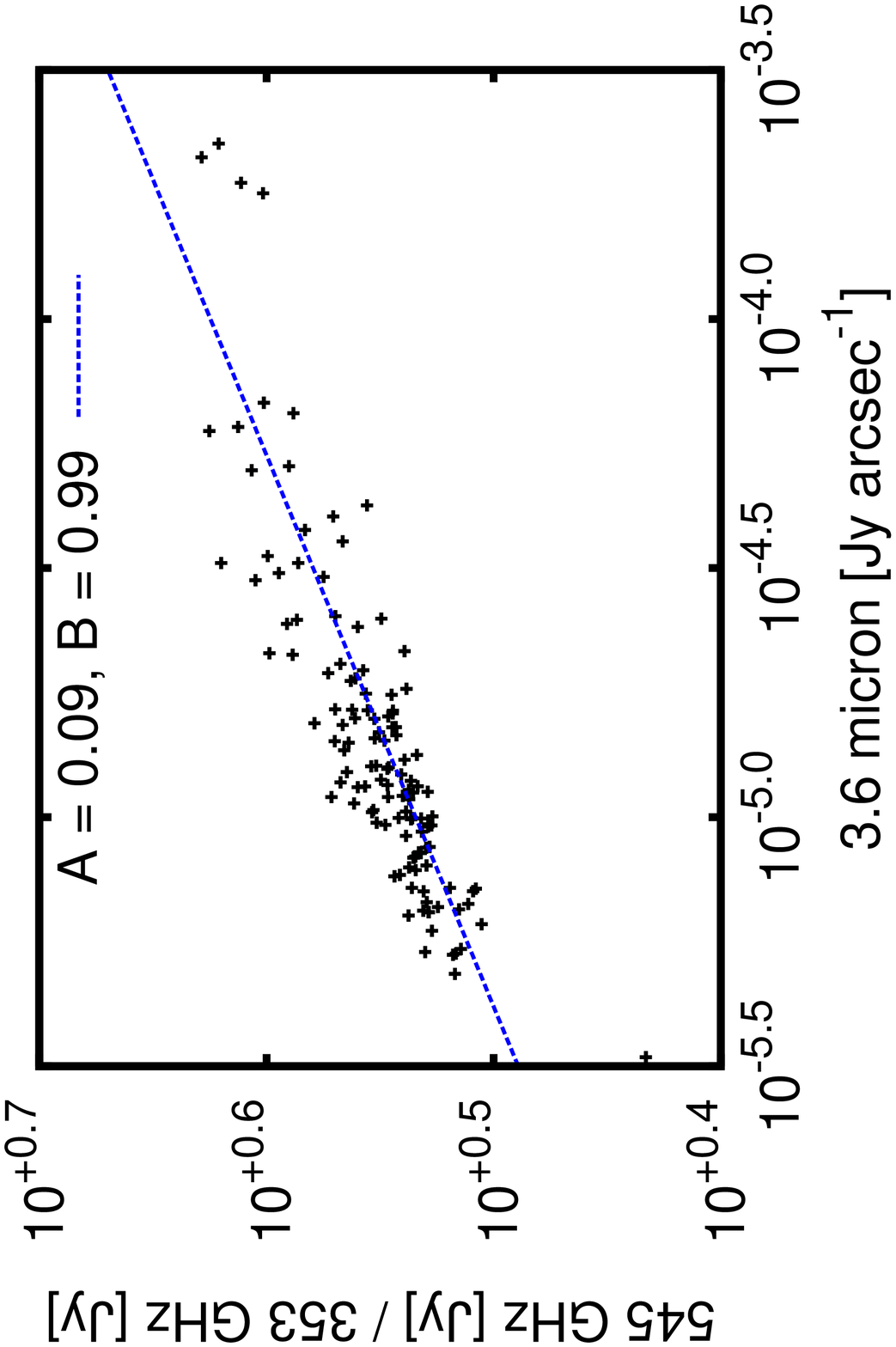}
\includegraphics[scale=0.17]{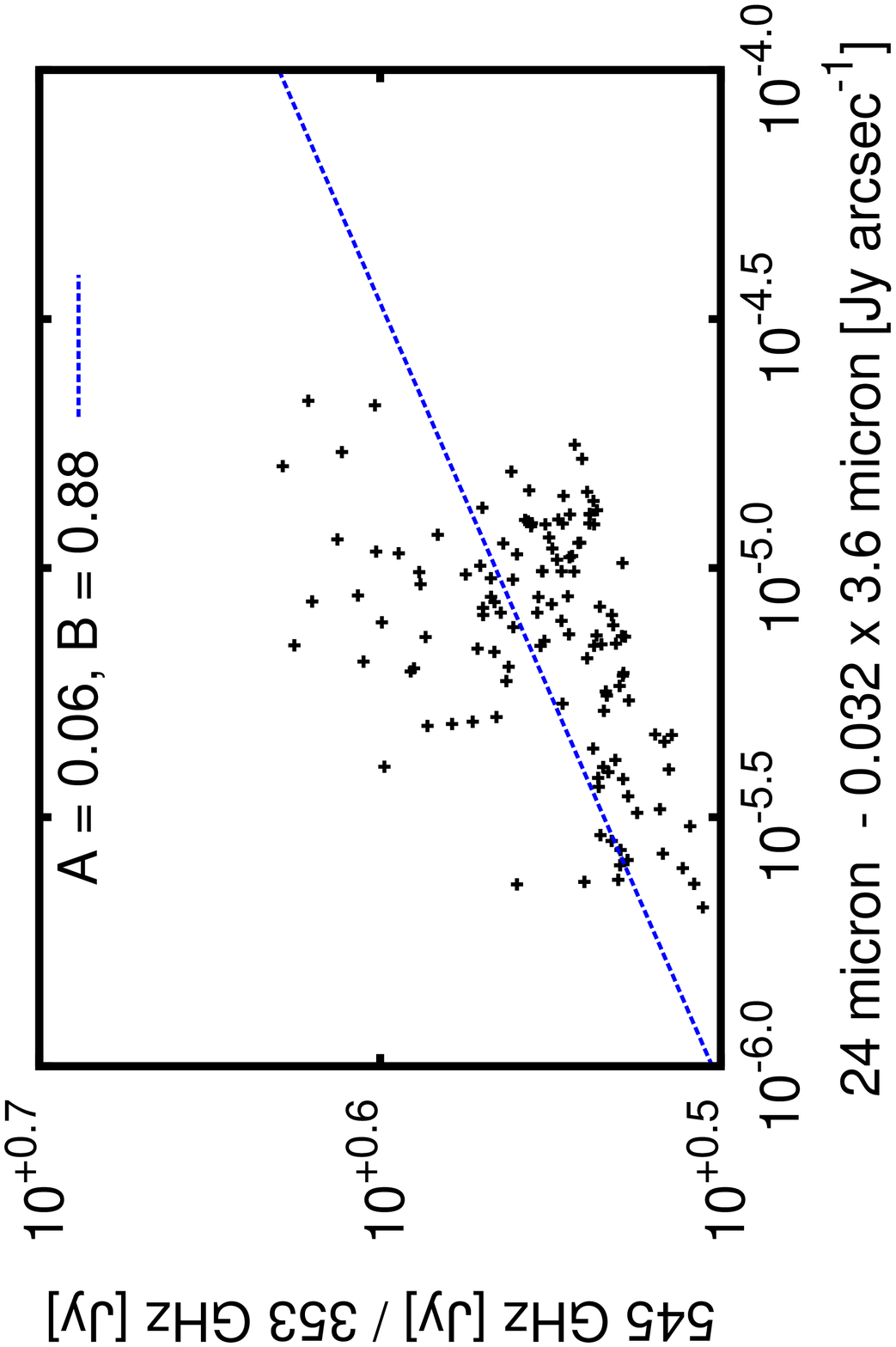}
\includegraphics[scale=0.17]{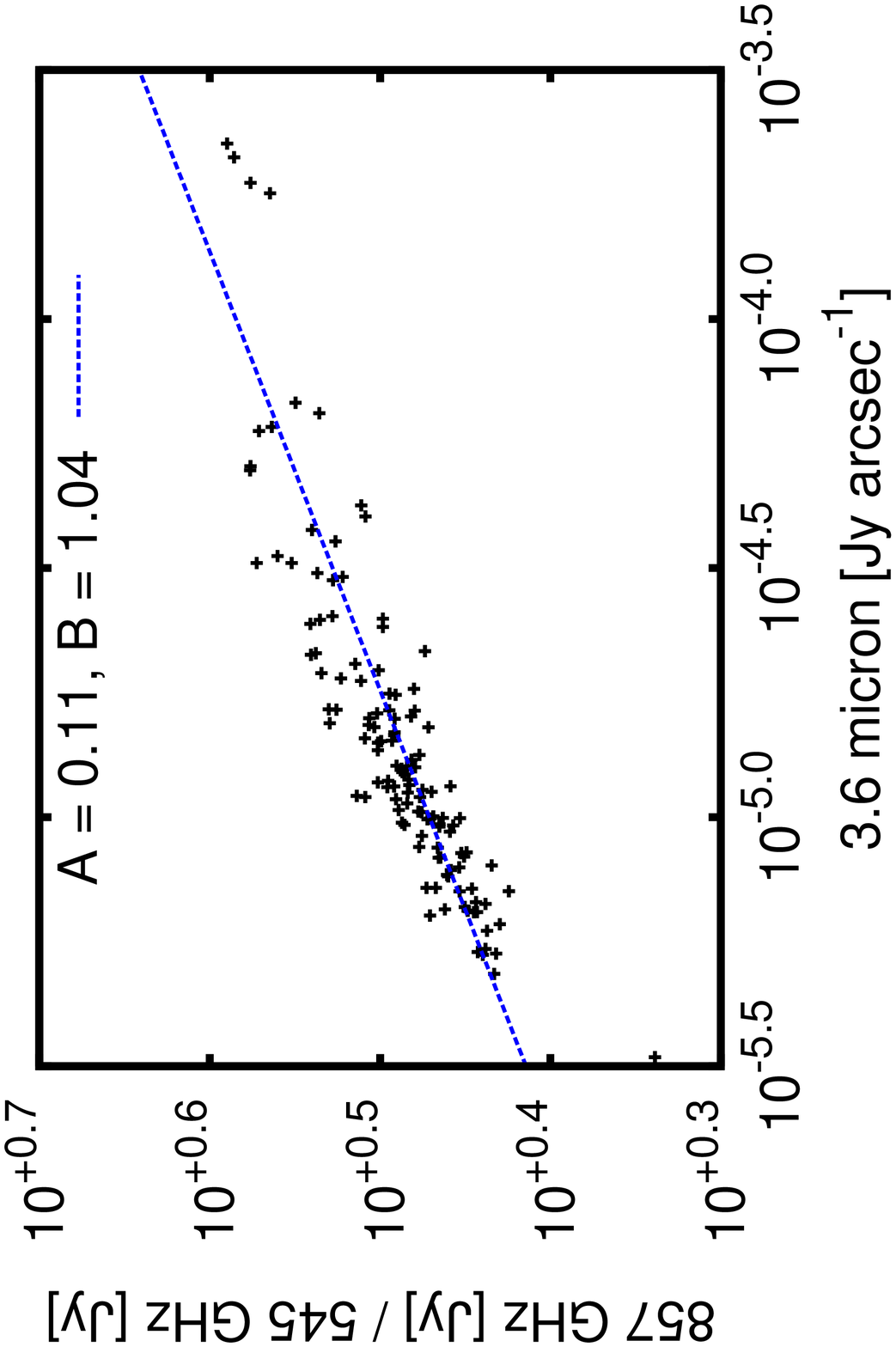}
\includegraphics[scale=0.17]{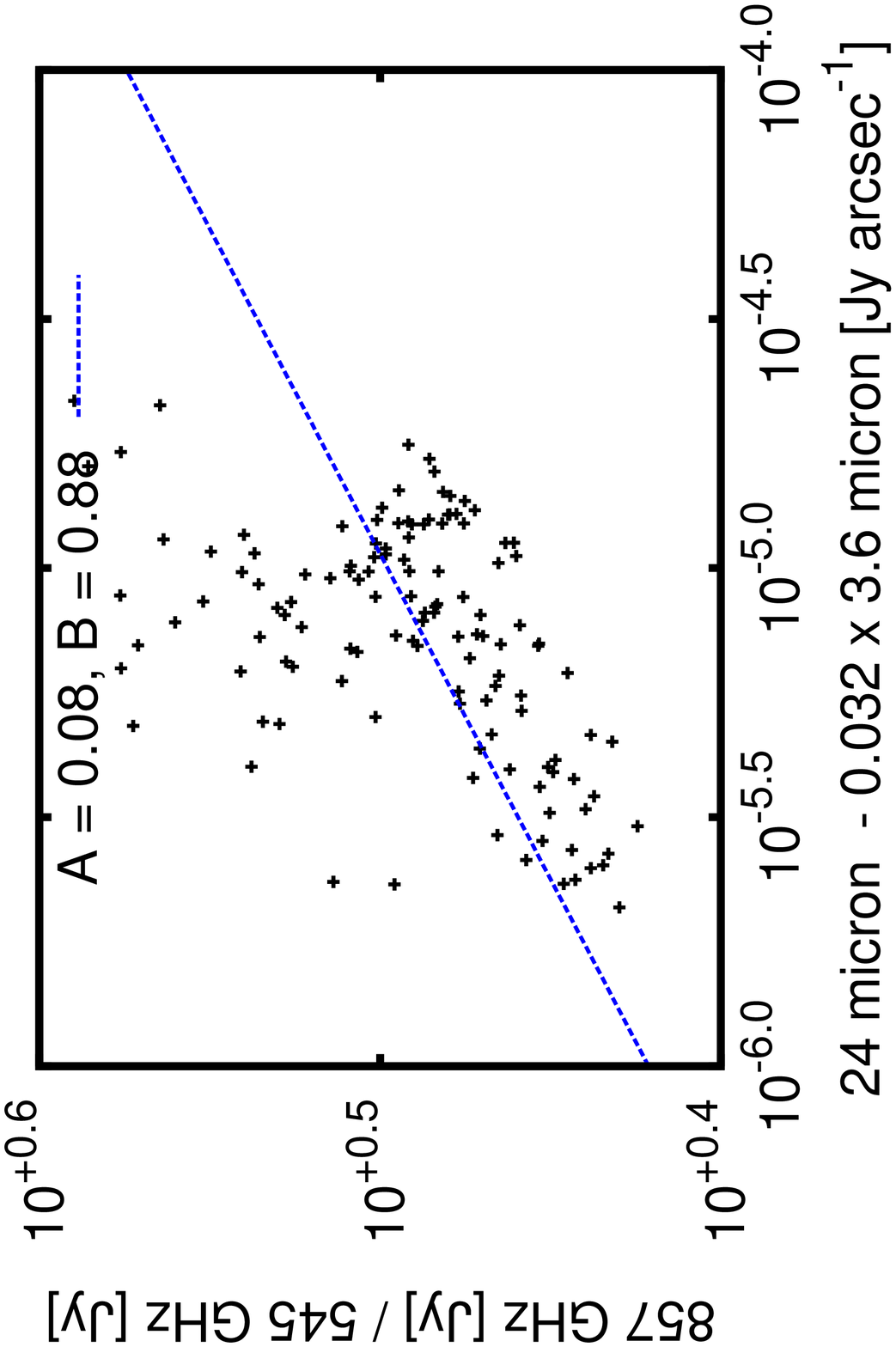}
\includegraphics[scale=0.17]{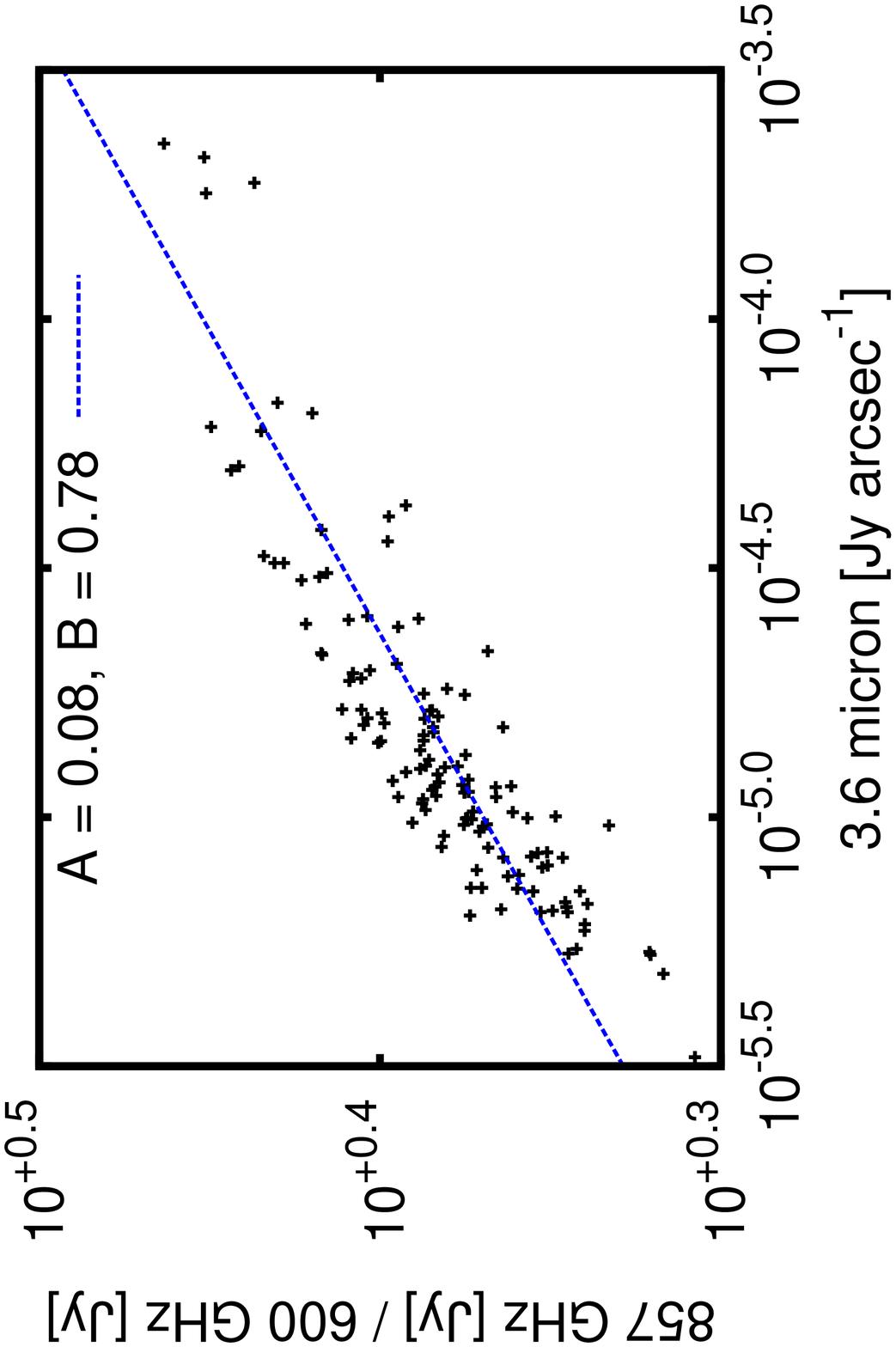}
\includegraphics[scale=0.17]{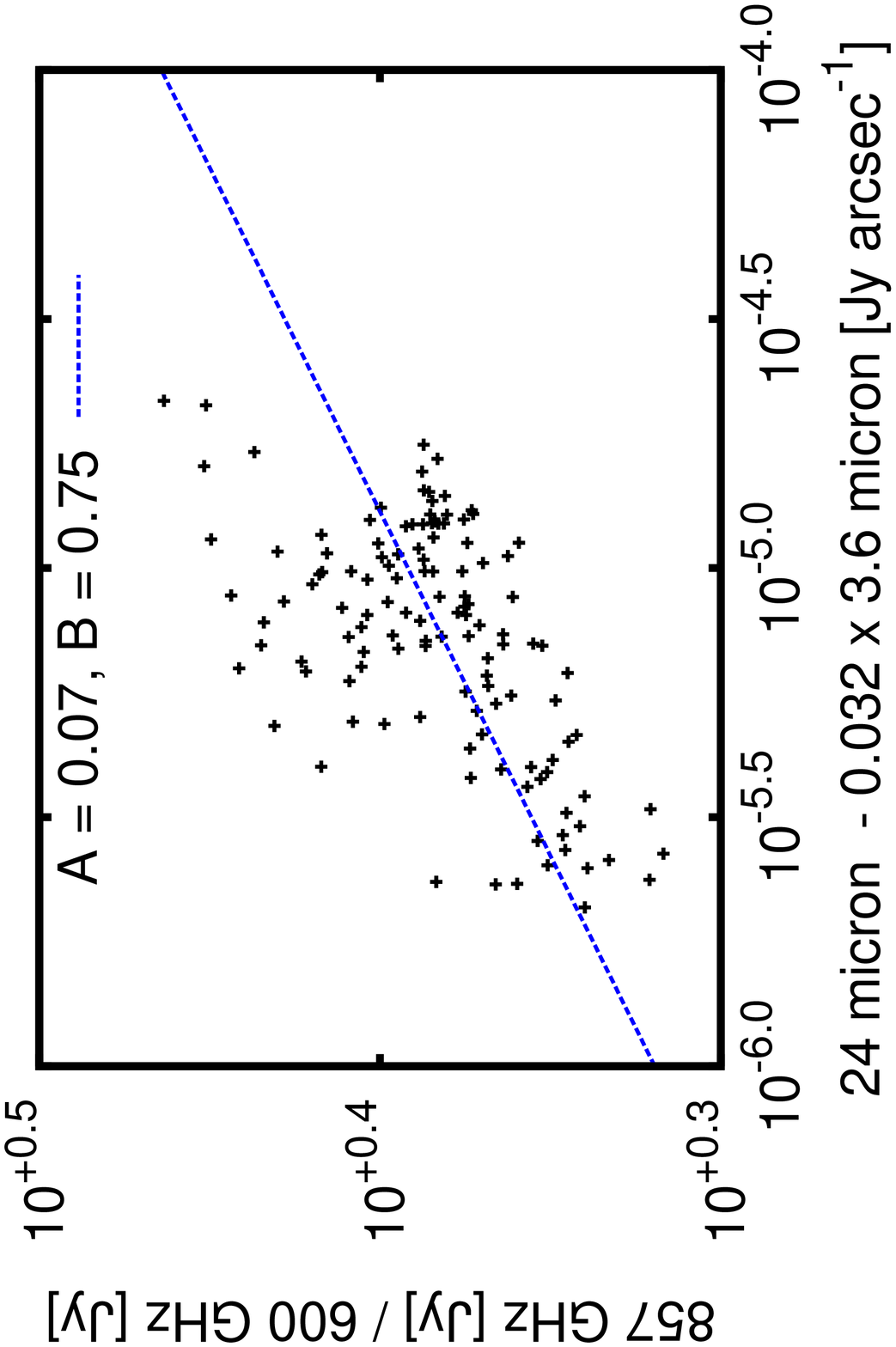}
\includegraphics[scale=0.17]{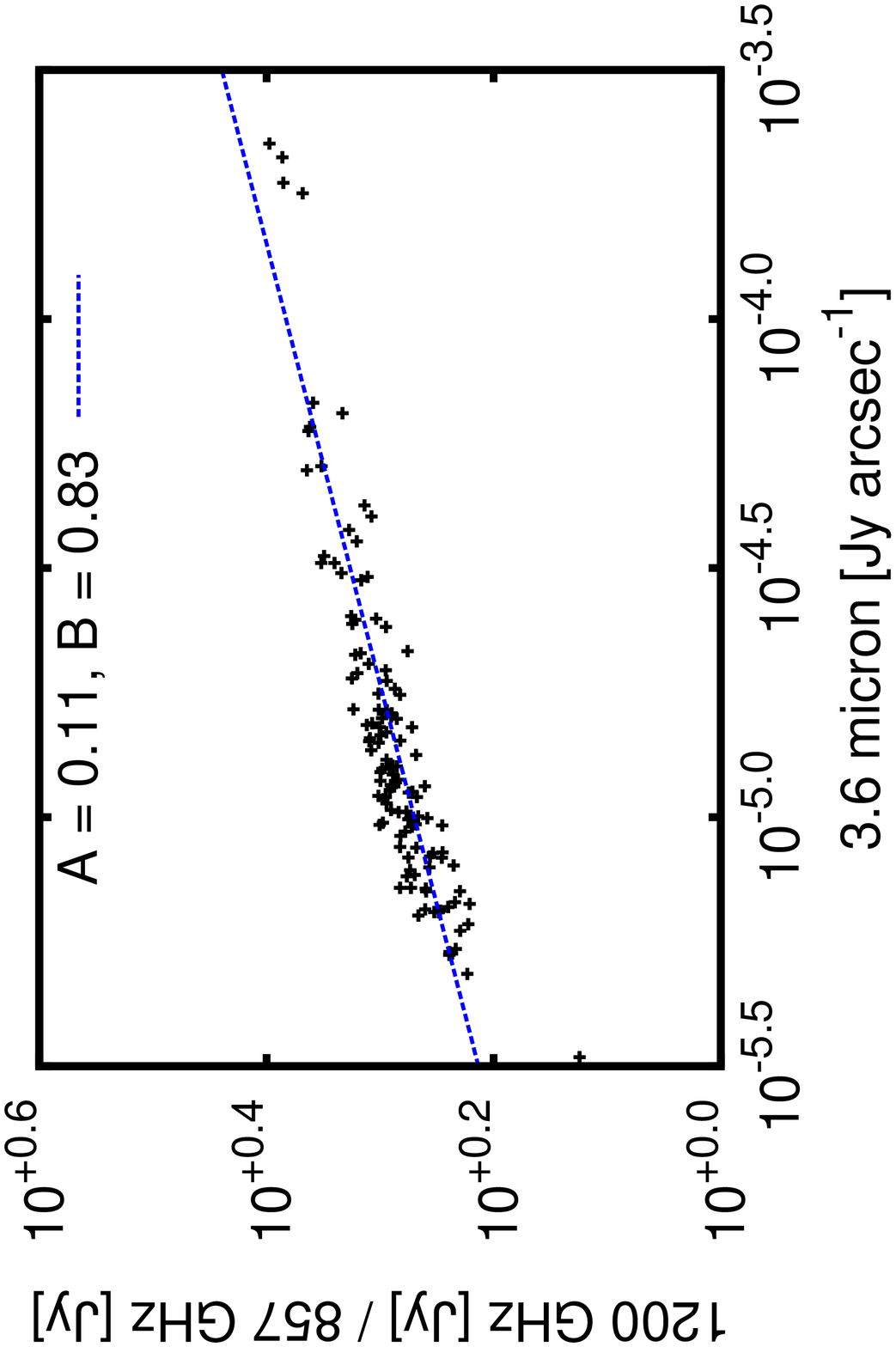}
\includegraphics[scale=0.17]{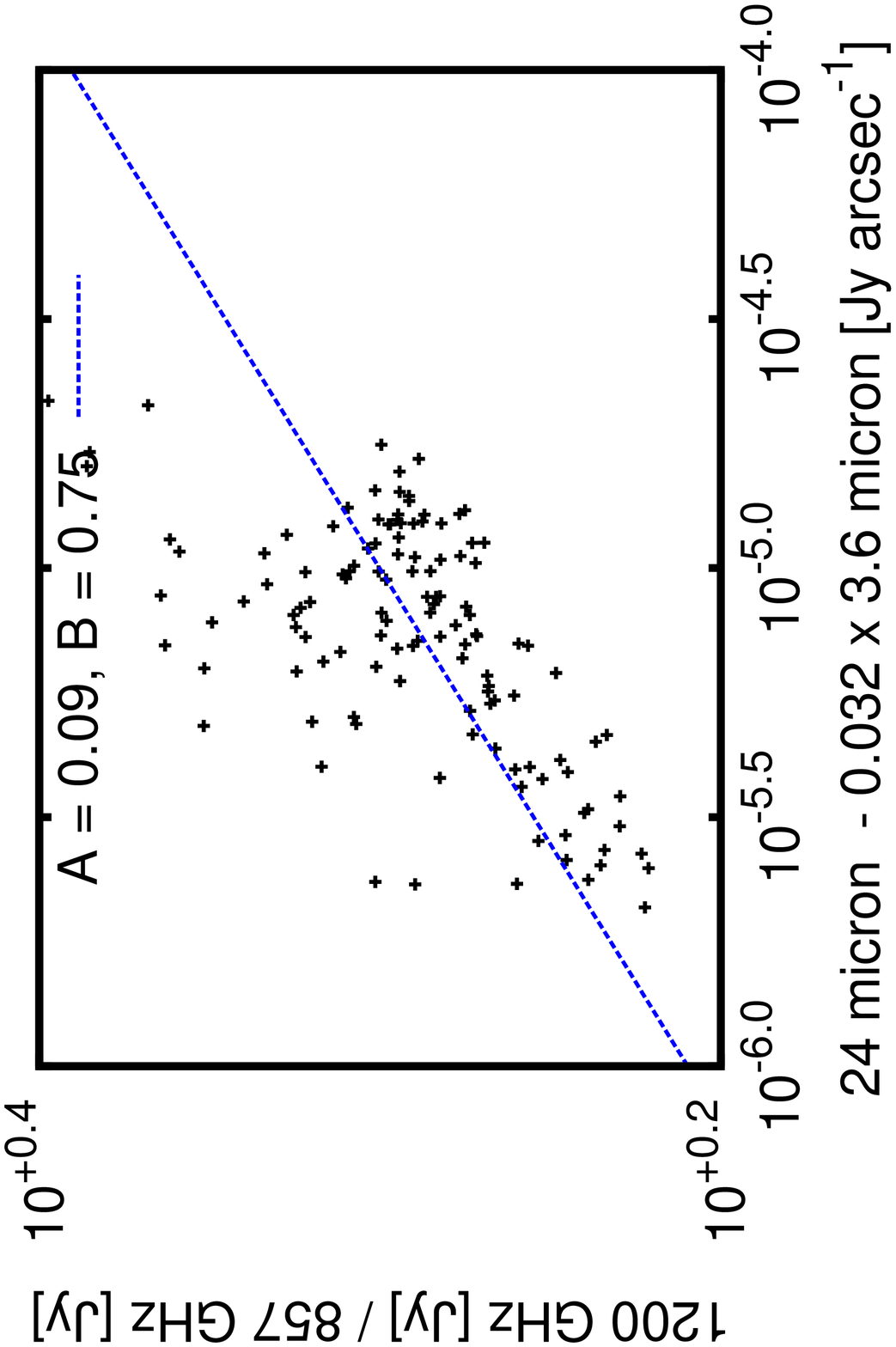}
\includegraphics[scale=0.17]{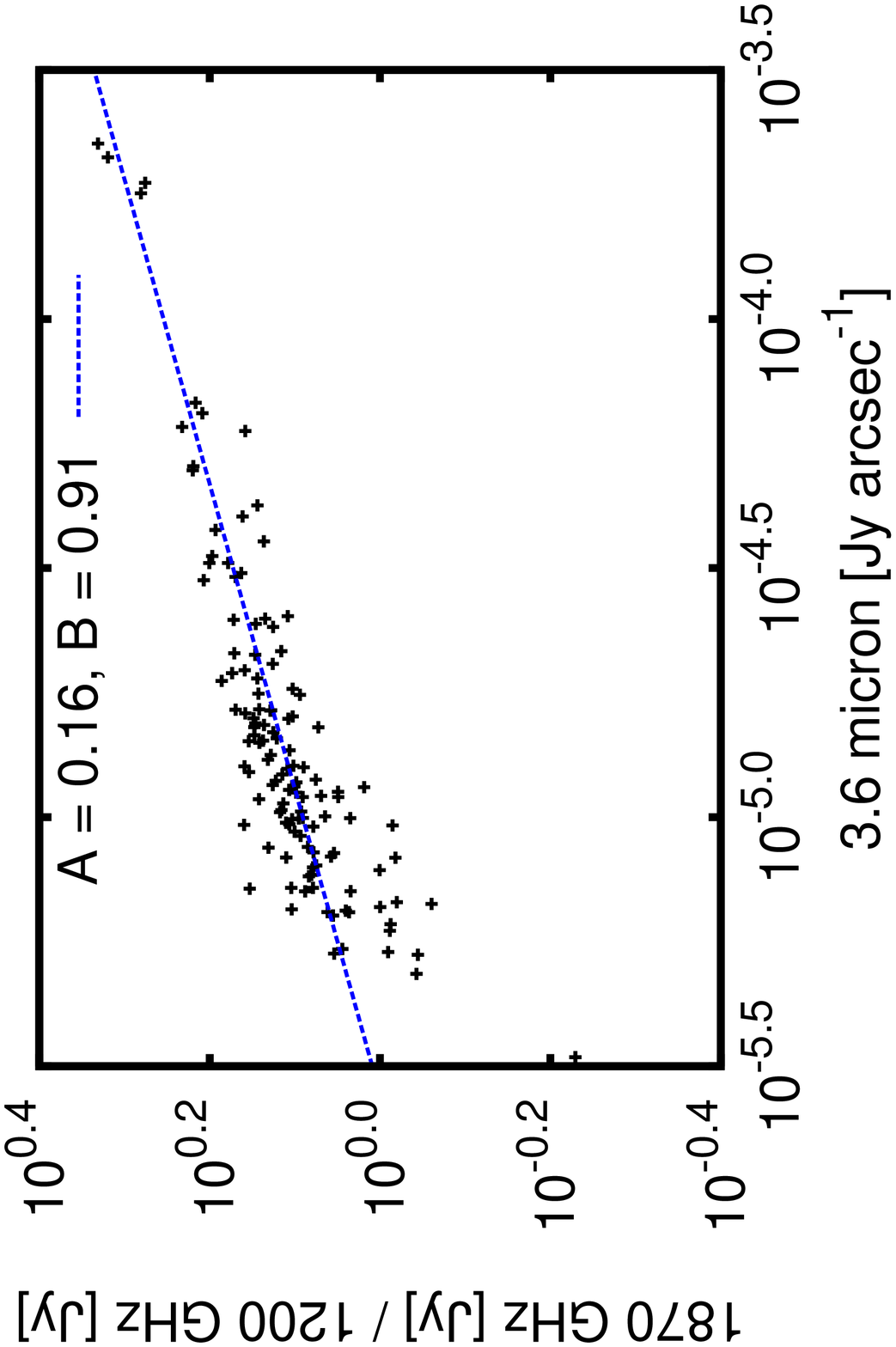}
\includegraphics[scale=0.17]{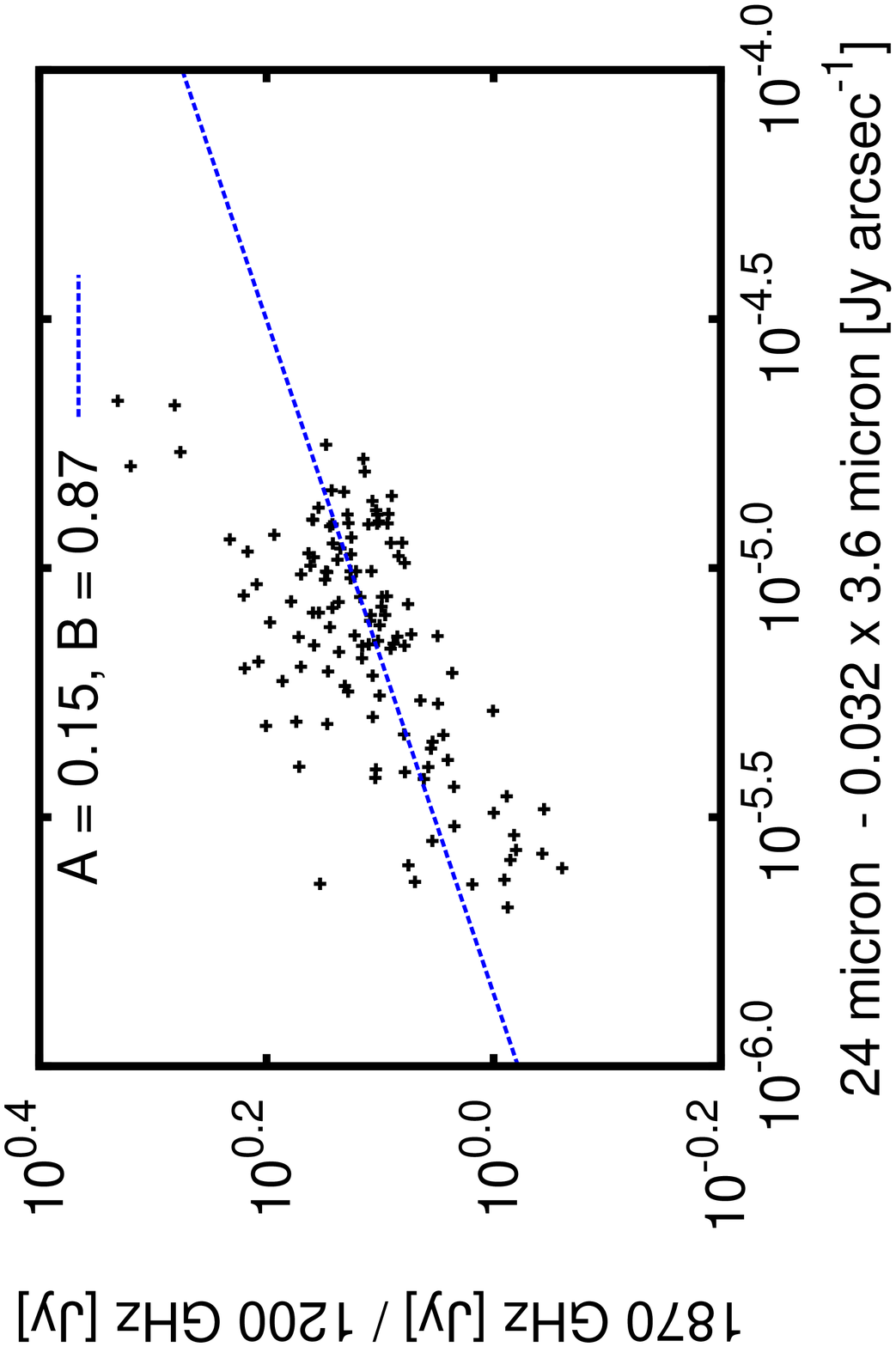}
\includegraphics[scale=0.17]{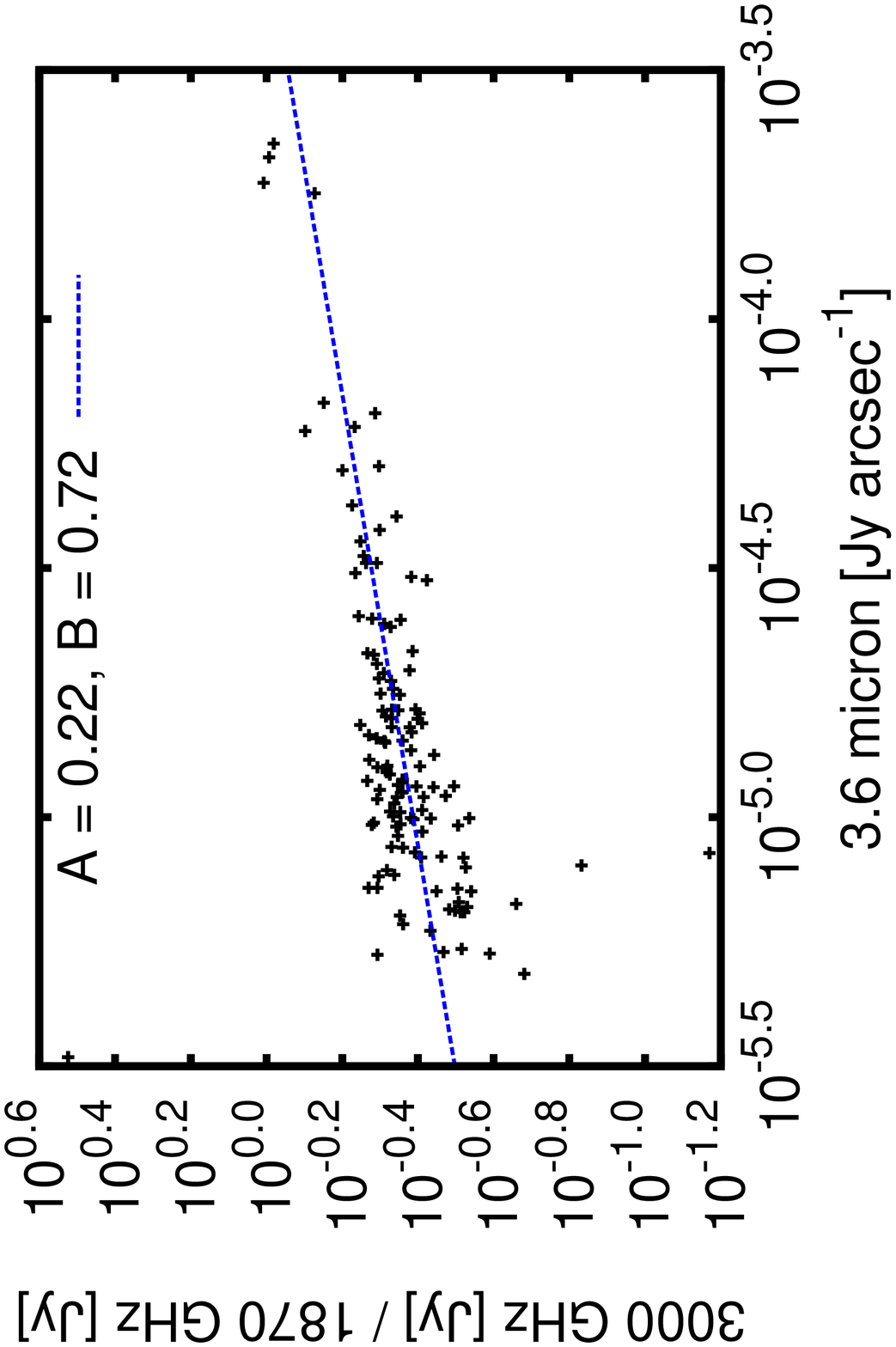}
\includegraphics[scale=0.17]{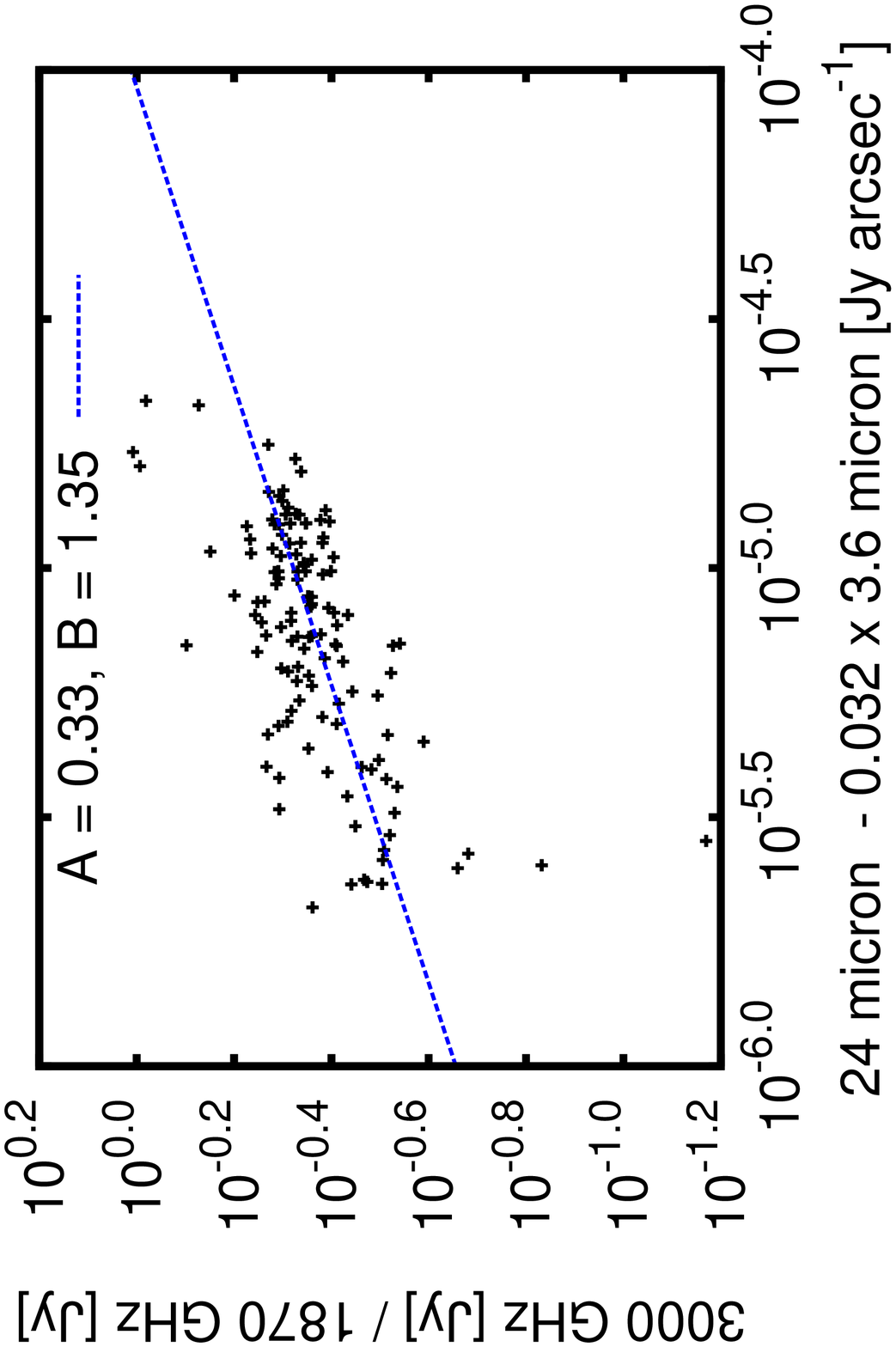}
\includegraphics[scale=0.17]{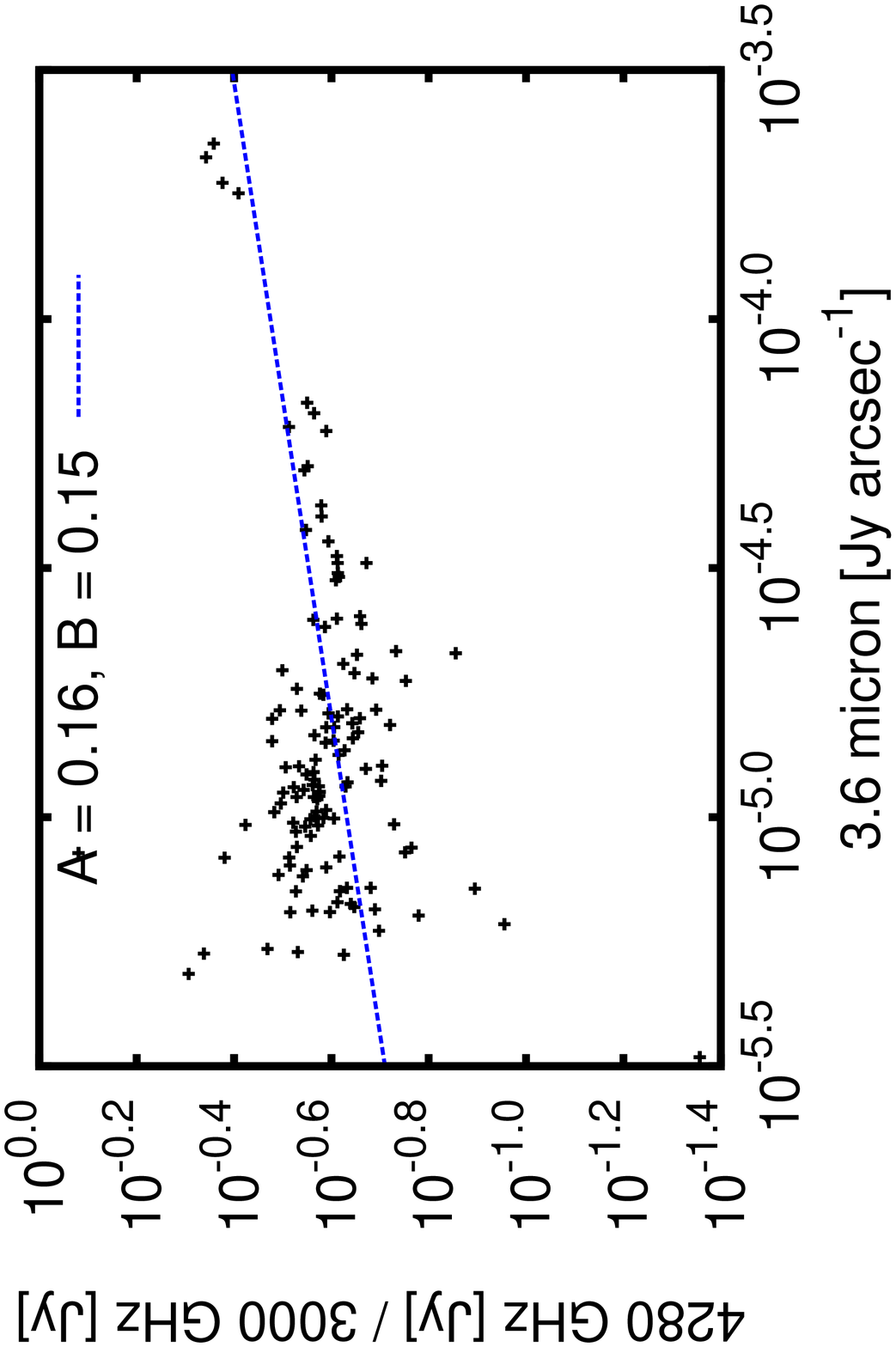}
\includegraphics[scale=0.17]{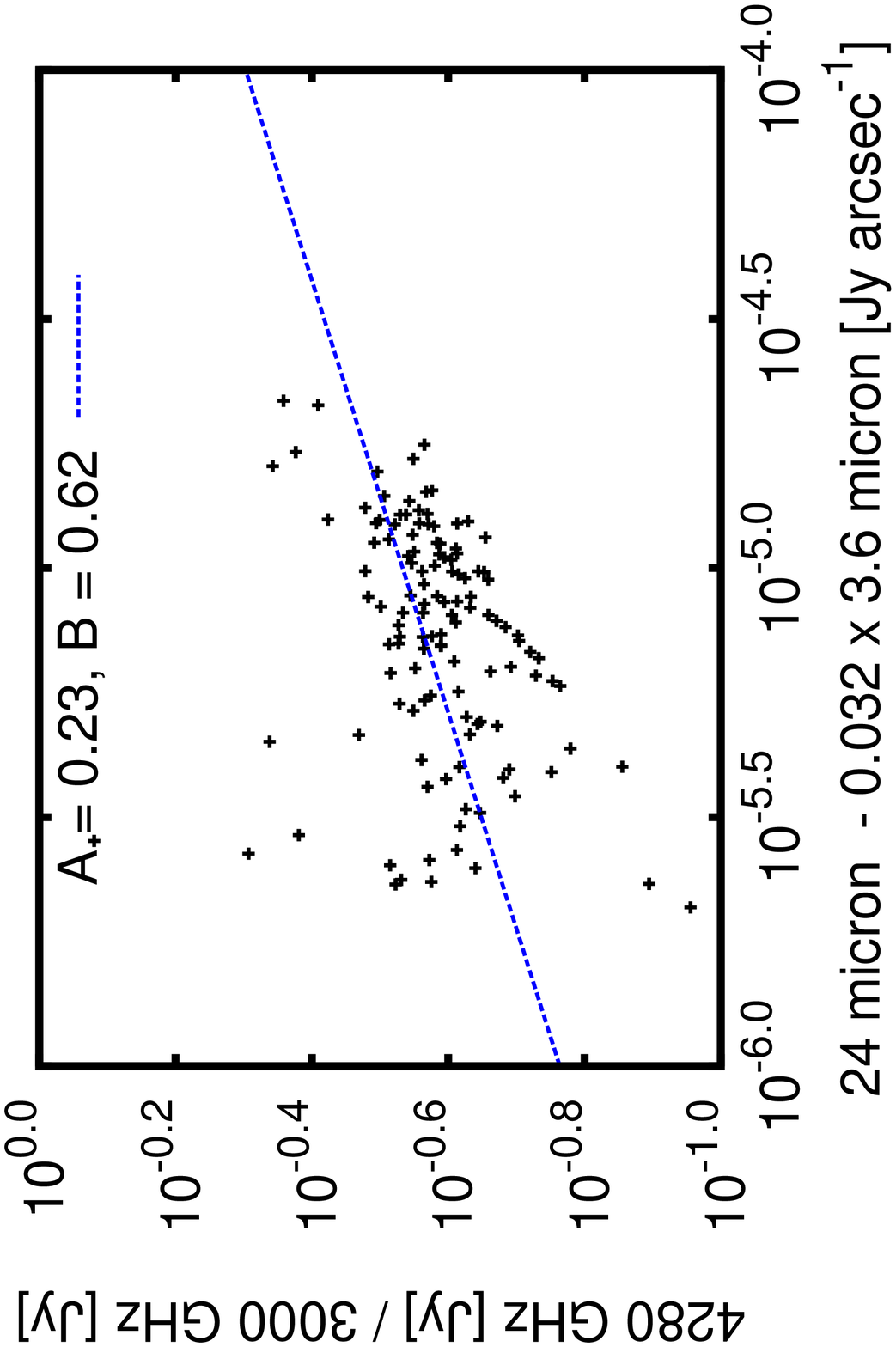}
\caption{Correlations between colour ratios and 3.6\um, tracing the stellar population (left), and 24\um, tracing star formation regions (right). The best fits of the form $\log_{10}(y) = A\log_{10}(x) + B$ are shown by the solid lines. The correlation with 3.6\um\ emission becomes less significant as the frequencies of the colour ratios increase, whilst the 24\um\ correlation becomes more significant. The correlation values are given in Table \ref{tab:efunc}.}
\label{fig:m31_correlations}
\end{center}
\end{figure}
\begin{table*}
\begingroup
\newdimen\tblskip \tblskip=5pt
\caption{Results from fitting Equation~\ref{eq:efunc} to the colour ratio data, including the covariance terms between the fitted parameters. Upper limits are 1\,$\sigma$. The last two columns are the correlation coefficients ($C$) from Fig. \ref{fig:m31_correlations}. Top: using {\tt SMICA} CMB subtraction. \changed{Middle}: using {\tt NILC} CMB subtraction. \changed{Bottom: using the combination of 24\um\ and FUV emission as the star\changed{-}formation tracer, with {\tt SMICA} CMB subtraction.} \Planck\ data are marked by ``P,'' \Herschel\ data by ``H,'' and \Spitzer\ data by ``S.''}
\label{tab:efunc}
\nointerlineskip
\vskip -3mm
\footnotesize
\setbox\tablebox=\vbox{
   \newdimen\digitwidth 
   \setbox0=\hbox{\rm 0} 
   \digitwidth=\wd0 
   \catcode`*=\active 
   \def*{\kern\digitwidth}
   \newdimen\signwidth 
   \setbox0=\hbox{+} 
   \signwidth=\wd0 
   \catcode`!=\active 
   \def!{\kern\signwidth}
    \halign{#\tabskip 1.0em&
            \hfil#\hfil\tabskip 2pt&
            \hfil#\hfil\tabskip 2pt&
            \hfil#\hfil\tabskip 2pt&
            \hfil#\hfil\tabskip 2pt&
            \hfil#\hfil\tabskip 2pt&
            \hfil#\hfil\tabskip 2pt&
            \hfil#\hfil\tabskip 2pt&
            \hfil#\hfil\tabskip 2pt&
            \hfil#\hfil\tabskip 2pt\cr
    \noalign{\doubleline\vskip 2pt}
    Ratio & $\alpha$ & $A_1$ & $A_2$ & $E_\mathrm{SF}/E_\mathrm{Total}$ & Cov($\alpha$, $A_1$) & Cov($\alpha$, $A_2$) & Cov($A_1$, $A_2$) & {\it C}(3.6) & {\it C}(24)\cr
\noalign{\vskip 4pt\hrule\vskip 6pt}
$S_{P545}/S_{P353}$ & $0.071\pm0.01*$ & \dots & $\phantom{-}1.25\pm0.01$ & 0 & \dots & $-4.7\times10^{-5}$ & \dots & 0.85 & 0.51\cr
$S_{P857}/S_{P545}$ & $0.103\pm0.018$ & \dots & $\phantom{-}1.10\pm0.01$ & $<0.13$ & \dots & $-9.3\times10^{-5}$ & \dots & 0.86 & 0.71\cr
$S_{H857}/S_{H600}$ & $0.085\pm0.009$ & **2.40* & $\phantom{-}0.77\pm0.04$ & $0.19\pm0.12$ & $-0.0069$* & $\phantom{-}5.7\times10^{-5}$ & $-0.096$** & 0.88 & 0.57\cr
$S_{H1200}/S_{H857}$ & $0.108\pm0.008$ & **2.04* & $\phantom{-}0.54\pm0.04$ & $0.22\pm0.09$ & $-0.0048$* & $\phantom{-}8.3\times10^{-5}$ & $-0.058$** & 0.90 & 0.61\cr
$S_{H1874}/S_{H1200}$ & $0.180\pm0.013$ & **1.82* & $\phantom{-}0.06\pm0.05$ & $0.23\pm0.08$ & $-0.0063$* &  0.00022 & $-0.047$** & 0.85 & 0.61\cr
$S_{H3000}/S_{H1870}$ & $0.30*\pm0.03*$ & **0.69* & $-0.95\pm0.04$ & $0.44\pm0.07$ & $-0.0039$* & 0.00036 & $-0.0076$* & 0.66 & 0.62\cr
$S_{S4280}/S_{H3000}$ & $0.37*\pm0.04*$ & **0.077 & $-1.35\pm0.02$ & $0.85\pm0.04$ & $-0.00049$ & 0.00015 & $-0.00052$ & 0.34 & 0.65\cr
\noalign{\vskip 4pt\hrule\vskip 6pt}
$S_{P545}/S_{P353}$ & $0.063\pm0.014$ & \dots & $\phantom{-}1.27\pm0.01$ & 0 & \dots & $-5.0\times10^{-5}$ & \dots & 0.85 & 0.52\cr
$S_{P857}/S_{P545}$ & $0.102\pm0.018$ & \dots & $\phantom{-}1.10\pm0.01$ & $<0.15$ & \dots & $-9.3\times10^{-5}$ & \dots & 0.86 & 0.71\cr
$S_{H857}/S_{H600}$ & $0.095\pm0.009$ & **2.40* & $\phantom{-}0.77\pm0.04$ & $0.19\pm0.12$ & $-0.0075$* & $\phantom{-}6.1\times10^{-5}$ & $-0.093$** & 0.88 & 0.57\cr
$S_{H1200}/S_{H857}$ & $0.108\pm0.008$ & **2.04* & $\phantom{-}0.54\pm0.04$ & $0.22\pm0.09$ & $-0.0048$* & $\phantom{-}8.3\times10^{-5}$ & $-0.058$** & 0.90 & 0.61\cr
$S_{H1874}/S_{H1200}$ & $0.180\pm0.013$ & **1.82* & $\phantom{-}0.06\pm0.05$ & $0.23\pm0.08$ & $-0.0063$* &  0.00022 & $-0.047$** & 0.85 & 0.61\cr
$S_{H3000}/S_{H1870}$ & $0.30*\pm0.03*$ & **0.69* & $-0.95\pm0.04$ & $0.44\pm0.07$ & $-0.0039$* & 0.00036 & $-0.0076$* & 0.66 & 0.62\cr
$S_{S4280}/S_{H3000}$ & $0.37*\pm0.04*$ & **0.077 & $-1.35\pm0.02$ & $0.85\pm0.04$ & $-0.00049$ & 0.00015 & $-0.00052$ & 0.34 & 0.65\cr
\noalign{\vskip 4pt\hrule\vskip 6pt}
$S_{P545}/S_{P353}$ & $0.071\pm0.014$ & \dots & $-0.28\pm0.18$ & 0 & \dots & $-4.9\times10^{-5}$ & \dots & 0.85 & 0.38\cr
$S_{P857}/S_{P545}$ & $0.101\pm0.017$ & \dots & $\phantom{-}0.36\pm0.69$ & $<0.13$ & \dots & $0.00421$ & \dots & 0.86 & 0.60\cr
$S_{H857}/S_{H600}$ & $0.085\pm0.009$ & **1.71* & $\phantom{-}0.80\pm0.05$ & $0.16\pm0.11$ & $-0.0056$* & $\phantom{-}1.4\times10^{-4}$ & $-0.074$** & 0.88 & 0.46\cr
$S_{H1200}/S_{H857}$ & $0.109\pm0.009$ & **1.27* & $\phantom{-}0.59\pm0.05$ & $0.21\pm0.09$ & $-0.0048$* & $\phantom{-}1.8\times10^{-4}$ & $-0.054$** & 0.90 & 0.50\cr
$S_{H1874}/S_{H1200}$ & $0.183\pm0.014$ & **1.05* & $\phantom{-}0.16\pm0.05$ & $0.24\pm0.08$ & $-0.0047$* &  0.00041 & $-0.027$** & 0.85 & 0.51\cr
$S_{H3000}/S_{H1870}$ & $0.31*\pm0.03*$ & **0.38* & $-0.77\pm0.05$ & $0.44\pm0.06$ & $-0.0021$* & 0.00076 & $-0.0049$* & 0.66 & 0.56\cr
$S_{S4280}/S_{H3000}$ & $0.39*\pm0.05*$ & **0.055 & $-1.34\pm0.04$ & $0.81\pm0.04$ & $-0.00030$ & 0.00013 & $-0.00046$ & 0.34 & 0.41\cr
\noalign{\vskip 3pt\hrule\vskip 4pt}
}}
\endPlancktable
\endgroup
\end{table*}

In Fig. \ref{fig:m31_correlations}, we show the surface brightness ratios for the binned data as a function of the 3.6 and 24\um\ data.  Statistics on the relations are given in Table \ref{tab:efunc}. Only the $S_{4280}/S_{3000}$ ratio shows a stronger correlation with the 24\um\ emission than the 3.6\um\ emission, which implies that only the dust dominating emission at $\geq4280$\ghz\ ($\leq$\,70\um) is heated by star-forming regions. All the other ratios are more strongly correlated with 3.6\um\ emission, which demonstrates that the total stellar populations, and not solely the star forming regions, are heating the dust observed below 3000\ghz\ (100\um).

As an additional assessment of the heating sources of the dust observed at these frequencies, we fit the $S_{545}/S_{353}$, $S_{857}/S_{545}$, $S_{1200}/S_{857}$, $S_{1874/1200}$, $S_{3000/1870}$, and $S_{4280}/S_{3000}$ ratio data to the 3.6 and 24\um\ data using the equation
\begin{equation}
\ln \left( \frac{S_{\nu_1}}{S_{\nu_2}} \right)    = \alpha \ln(I_\mathrm{SFR} + A_1 I_\mathrm{stars}) + A_2 .
\label{eq:efunc}
\end{equation}
This equation, derived from the Stefan-Boltzmann law by \citet{Bendo2012}, relates the surface brightness ratios of the dust with the dust heating sources.  The ratio $S_{\nu_1}/S_{\nu_2}$ on the left side of the equation is related to the dust temperature through a power law (assuming that the dust seen in any pair of frequencies has only one temperature or a narrow range of temperatures).  Therefore,  $S_{\nu_1}/S_{\nu_2}$ can ultimately be expressed as a function of the total energy emitted by the dust using a version of the Stefan-Boltzmann law modified for dust with an emissivity function that varies as $\nu^{\,\beta}$.  The $I_\mathrm{SFR}$ and $I_\mathrm{stars}$ terms represent the energy absorbed by the SFR dust (traced by the 24\um\ band) and the ISRF dust (traced by the 3.6\um\ band).  The slope $\alpha$ and the scaling terms $A_1$ and $A_2$ are free parameters in the fit.  The scaling terms will adjust the $I_\mathrm{SFR}$ and $I_\mathrm{stars}$ data to account for the fraction of light from those sources that is absorbed by the dust.  

After Eq.~(\ref{eq:efunc}) has been fit to the data, the relative magnitudes of the $I_\mathrm{SFR}$ and $A_1 I_\mathrm{stars}$ terms can be used to estimate the fraction of dust heating (traced by the variation in the surface brightness ratios for any pair of frequencies) that can be related to each heating source.  We calculated the fraction of dust heating from star-forming regions using
\begin{equation}
\frac{E_\mathrm{SF}}{E_\mathrm{Total}} = \frac{I_\mathrm{SFR}}{I_\mathrm{SFR} + A_1 I_\mathrm{stars}}.
\label{eq:efrac}
\end{equation}

The resulting parameters from fitting Eq.~(\ref{eq:efunc}) to the data are given in Table~\ref{tab:efunc}.  We also give the resulting $E_\mathrm{SF}/E_\mathrm{Total}$ values based on the flux density measurements integrated over the disc of the galaxy.  Uncertainties in the $\alpha$\changed{, $A_2$,} and $E_\mathrm{SF}/E_\mathrm{Total}$ values are estimated using a Monte Carlo approach, but we did not report uncertainties for \changed{the $A_1$ term because the $A_1$ and $A_2$ terms} become degenerate when $I_\mathrm{SFR} \ll A_1 I_\mathrm{stars}$ and because the $A_1$ term is poorly constrained where $I_\mathrm{SFR} \gg A_1 I_\mathrm{stars}$, resulting in high uncertainties for the scaling terms.  Figure~\ref{fig:m31_colourratio} shows the $E_\mathrm{SF}/E_\mathrm{Total}$ maps calculated using Eq.~(\ref{eq:efrac}) and the best fitting $A_1$ terms.  What all of these data demonstrate is that the relative fraction of SFR dust decreases when going from higher frequencies to lower frequencies.  The results for the $S_{4280}/S_{3000}$ ratio show that around 90\,\% of the colour variations can be related to SFR dust.  In the centre of the galaxy, however, more than 50\,\% of the emission at these frequencies originates from dust heated by the total stellar population, including stars in the bulge.  The $S_{857}/S_{545}$ and $S_{545}/S_{353}$ colours show that the dust emission at these frequencies primarily originates from ISRF dust. The contribution of SFR dust, even in the ring, is relatively weak.  In fact, when Eq.~(\ref{eq:efunc}) was fit to the $S_{545}/S_{353}$ data, the resulting parameters indicated that the contribution from the $I_\mathrm{SFR}$ component was negligible; the terms in Table~\ref{tab:efunc} were adjusted to normalize $A_1$ to 1.  The transition between SFR dust and ISRF dust is at around 3000\GHz.

We have also carried out this analysis using data where the {\tt NILC} map was used to subtract the CMB. We find that the results are not very sensitive to the contamination of the CMB map by M31. The results from fitting Eq.~(\ref{eq:efunc}) to the {\tt NILC}-subtracted data are shown in the \changed{middle section} of Table \ref{tab:efunc} and are well within the stated uncertainties. The apparently different values for $A_1$ and $A_2$ for the $S_{857}/S_{545}$ ratios are due to the degeneracy in these parameters, as described above.

\changed{We have also run this analysis using the combination of 24\um\ and far-ultraviolet data to trace both the obscured and unobscured star formation. We combine the two datasets to obtain the equivalent corrected 24\um\ emission (for ease of comparison with the results given above) using}
\begin{equation}
\changed{I_\mathrm{SFR}=\left(I_{24}-0.032\,I_{3.6}\right) + 25.3 \left(I_\mathrm{FUV}-8.0\times10^{-4}\,I_{3.6}\right),}
\end{equation}
\changed{where the corrections for the old stellar population emission are as per \citet{Ford2013}, and the rescaling of the FUV emission uses the combination of the 24\um\ and FUV coefficients given in the calculation of the combined star formation rate in \citet{Leroy2008}. The results from this are given in the bottom section of Table \ref{tab:efunc}. The addition of FUV data to the star\changed{-}formation tracer changes our results by less than 5\,\%, which is a similar level to the effect that \citet{Bendo2014} found when switching between uncorrected \ha, combined \ha\ and 24\um\ emission, and 24\um\ emission by itself as star\changed{-}formation tracers.}

\citet{Groves2012} and \citet{Smith2012} have also examined dust heating in M31 using \Herschel\ data that overlaps with the \Planck\ frequencies.  Both groups found that the dust emission in the centre of M31 appears to be heated by the total stellar population, including the bulge stars, which agrees with our results.  However, \citet{Smith2012} found that the dust temperatures in the ring may be weakly correlated with star formation activity and is not correlated with the total stellar emission, whereas we find that about 50\,\% of the dust in the ring seen below 3000\ghz\ is still heated by the total stellar population.  This is probably related to the different analyses and underlying assumptions used. \citet{Smith2012} assume that a single modified blackbody spectrum can accurately describe all data between 600 and 1870\ghz\ (500 and 160\um), whereas we treated each pair of frequency bands independently and found that the 545--1870\ghz\ (550--160\um) range contains dust from two different thermal components heated by different sources.  Hence, the \citet{Smith2012} fits may have been unduly influenced by relatively small amounts of hot dust that may have masked the presence of the colder dust within the ring and therefore may not have been able to detect the dust primarily emitting at $<870$\ghz\ ($>350$\um) that is heated by the diffuse interstellar radiation field.

The conclusion that the dust in M31 seen at \Planck\ frequencies is primarily heated by the total stellar population and not just the star forming regions is broadly consistent with the recent \Herschel\ results using similar techniques \citep{Bendo2010, Boquien2011, Bendo2012}, as well as some of the older \IRAS\ studies of dust heating in our Galaxy and other nearby galaxies (e.g., \mbox{\citealp{Lonsdale1987}}; \citealp{Walterbos1987,Sauvage1992,Walterbos1996}).  Some dust emission and radiative transfer models for galaxies \citep[e.g., ][]{Draine2007, Popescu2011} include two dust components: diffuse dust heated by a general interstellar radiation field; and dust heated locally by star forming regions.  The results here are broadly consistent with the characterisation of the dust emission by these models.  However, we observe the transition between SFR dust and ISRF dust at 70--100\um, whereas in the model fits shown by \citet{Draine2007}, the transition generally takes place at shorter wavelengths (below 70\um)\changed{, while \citet{DeLooze2014} finds the transition to be around 60\um, and the model of \citet{Natale2015} has the transition at around 30\um. These results are, of course, dependent on the implementation details of the modelling of the two components.}

\section{SEDs at 5\arcm\ resolution} \label{sec:seds}
We now move on to investigate and quantify the distribution of dust temperatures and spectral indices within M31 by fitting thermal dust spectra to the 5\arcm\ data set (as described in the previous section) comprised of \Planck, \Spitzer, \IRAS, and \ISO\ data. We use the dust heating fraction that was quantified in the previous section in order to focus only on the colder dust that is heated by the total stellar population.

\begin{figure}[tbp]
\begin{center}
\includegraphics[scale=1.7]{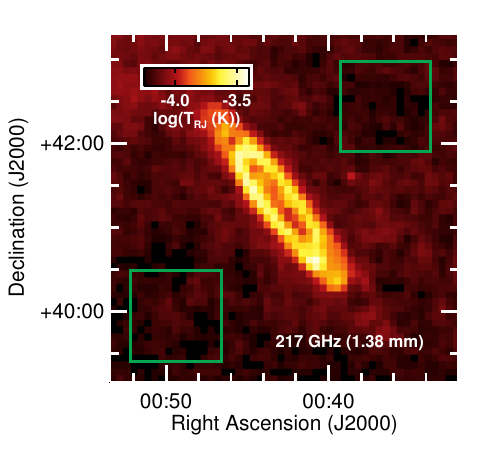}
\caption{\Planck\ 217\GHz\ map at 5\arcm\ resolution showing the two regions used to calculate the background and rms (top-right and bottom-left).}
\label{fig:backgroundregions}
\end{center}
\end{figure}

To remove the effect of the background, including offsets in the map, the average contribution from the CIB, and the average Galactic cirrus contribution, we background-subtract the \planck\ and \iras\ maps. To do this we use the mean flux density in two $13\times13$ pixel areas to the bottom-left and top-right of the 2D maps, which straddle the galaxy at locations that are well away from both the emission from M31 and our Galaxy (see Fig. \ref{fig:backgroundregions}). We also use the rms scatter within these regions to estimate the uncertainty in the data caused by noise and cirrus contamination. As the \Spitzer\ and \ISO\ data have already been background-subtracted, and as the maps do not extend to the background subtraction regions that we are using for the other data, we do not background-subtract these further. We assess the uncertainty on each pixel by combining the calibration uncertainties (as per Table \ref{tab:ancillarydata}) in quadrature with the rms of the map estimated in the same regions as the background subtraction or, for the \Spitzer\ data, the rms of the data outside the mask. We have estimated the background and rms using different regions and found consistent results. These uncertainties also take into account the contribution of the cirrus and CIB structure on smaller scales; as the rms is in the range 0.015--0.37\,Jy (depending on the frequency), this is generally small compared to the signal from M31, except in the outermost rings.

\begin{figure*}[tbp]
\begin{center}
\includegraphics[scale=0.23]{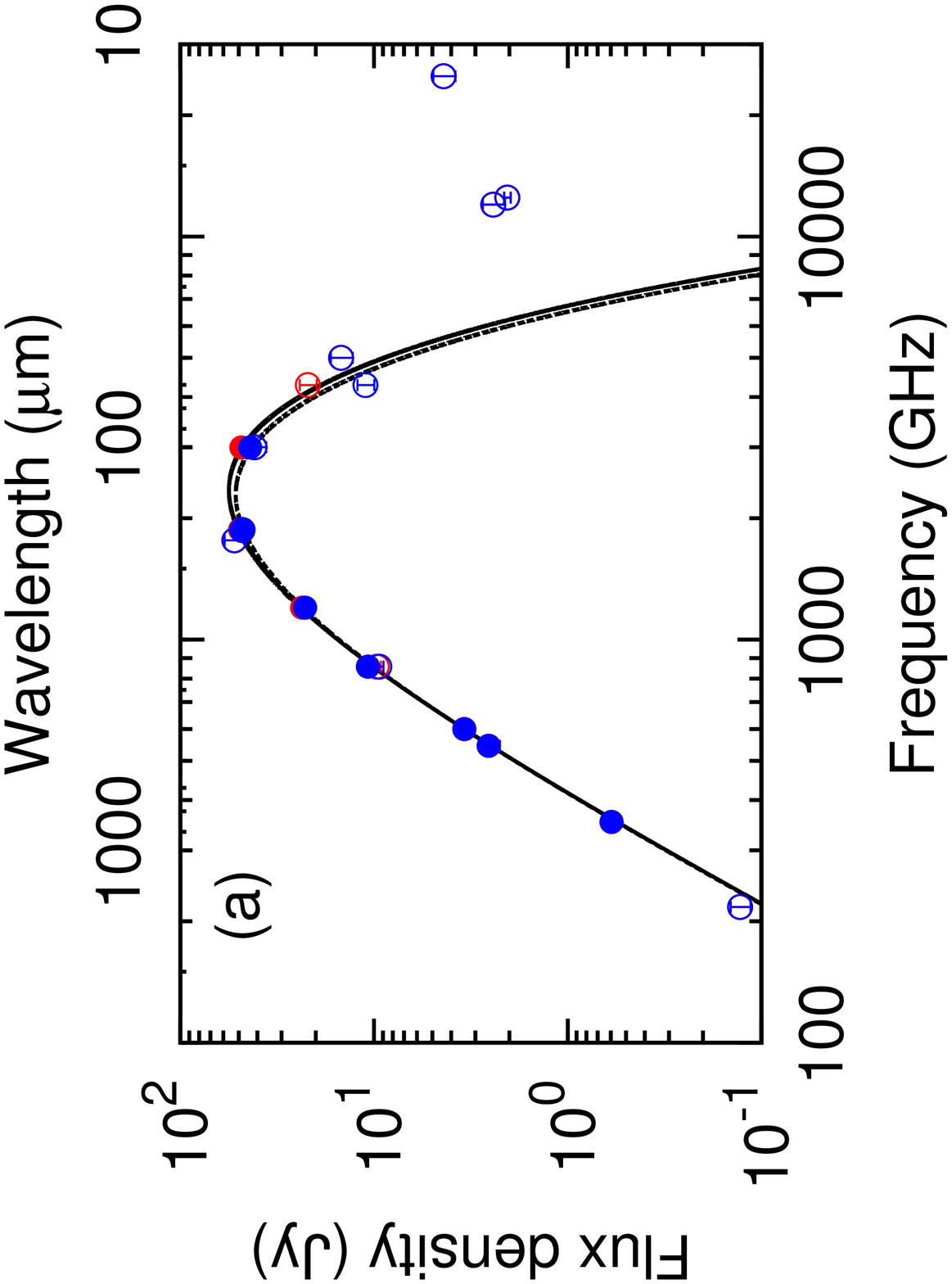}
\includegraphics[scale=0.23]{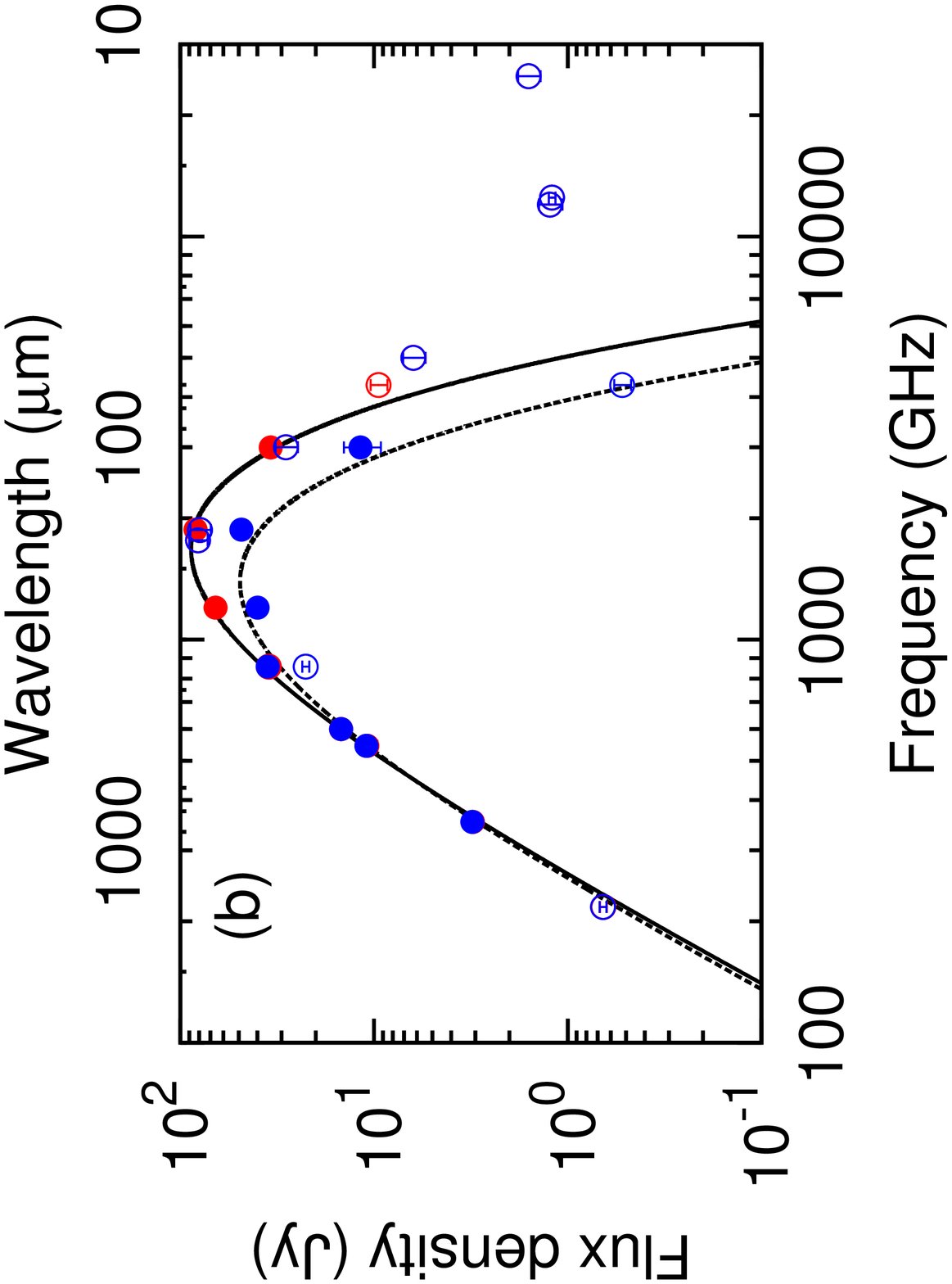}
\includegraphics[scale=0.23]{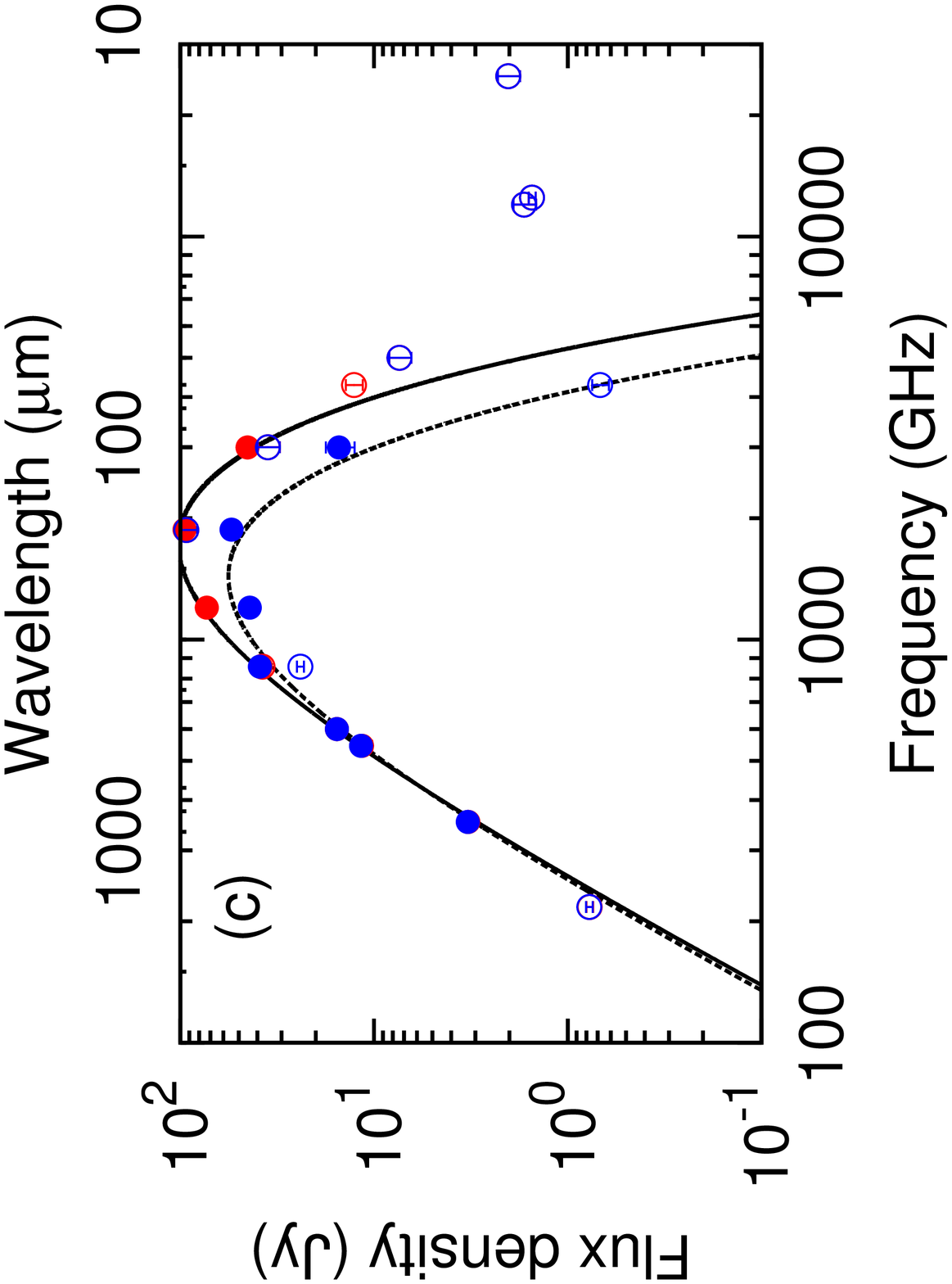}
\includegraphics[scale=0.23]{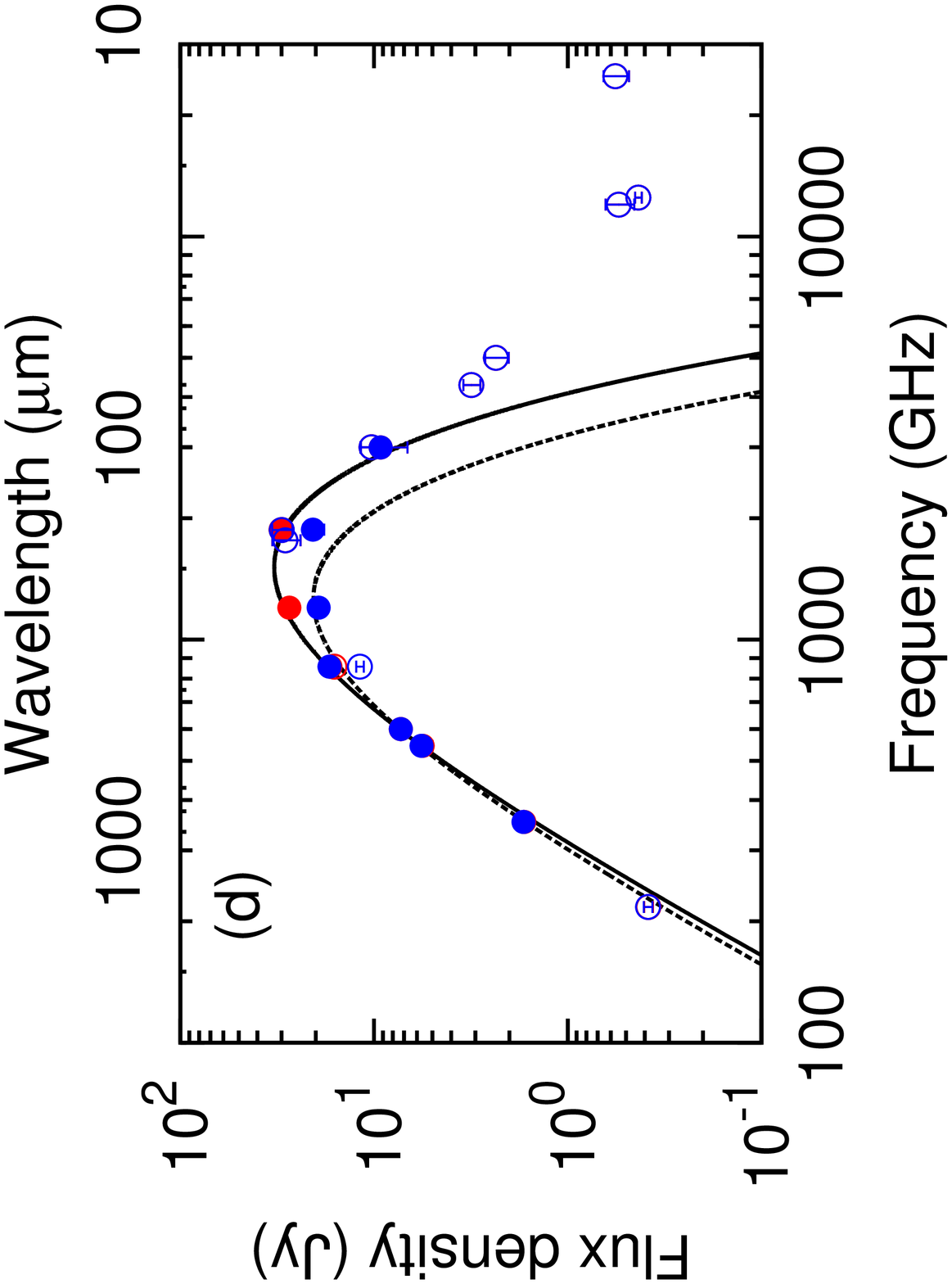}
\includegraphics[scale=0.23]{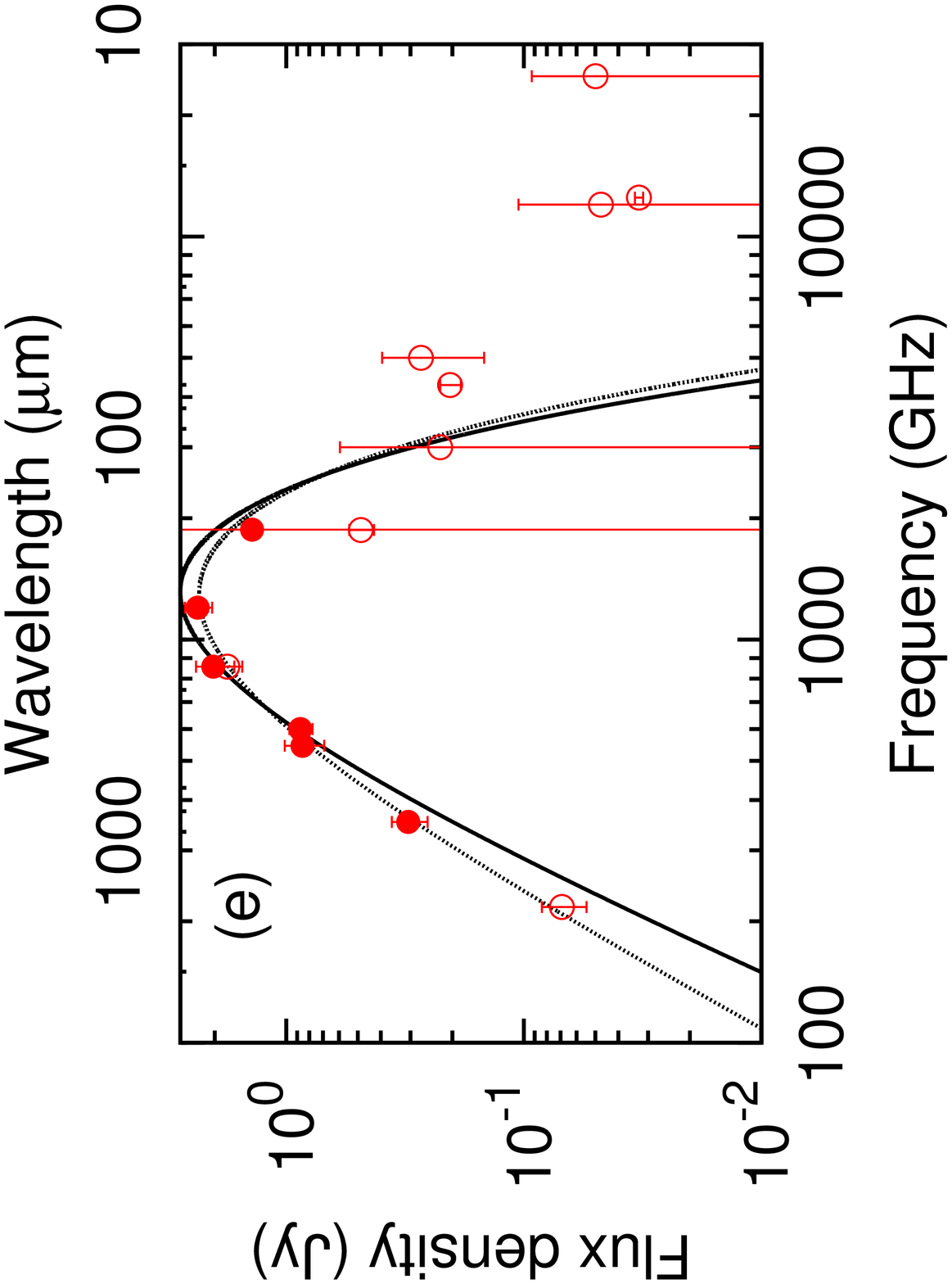}
\includegraphics[scale=0.23]{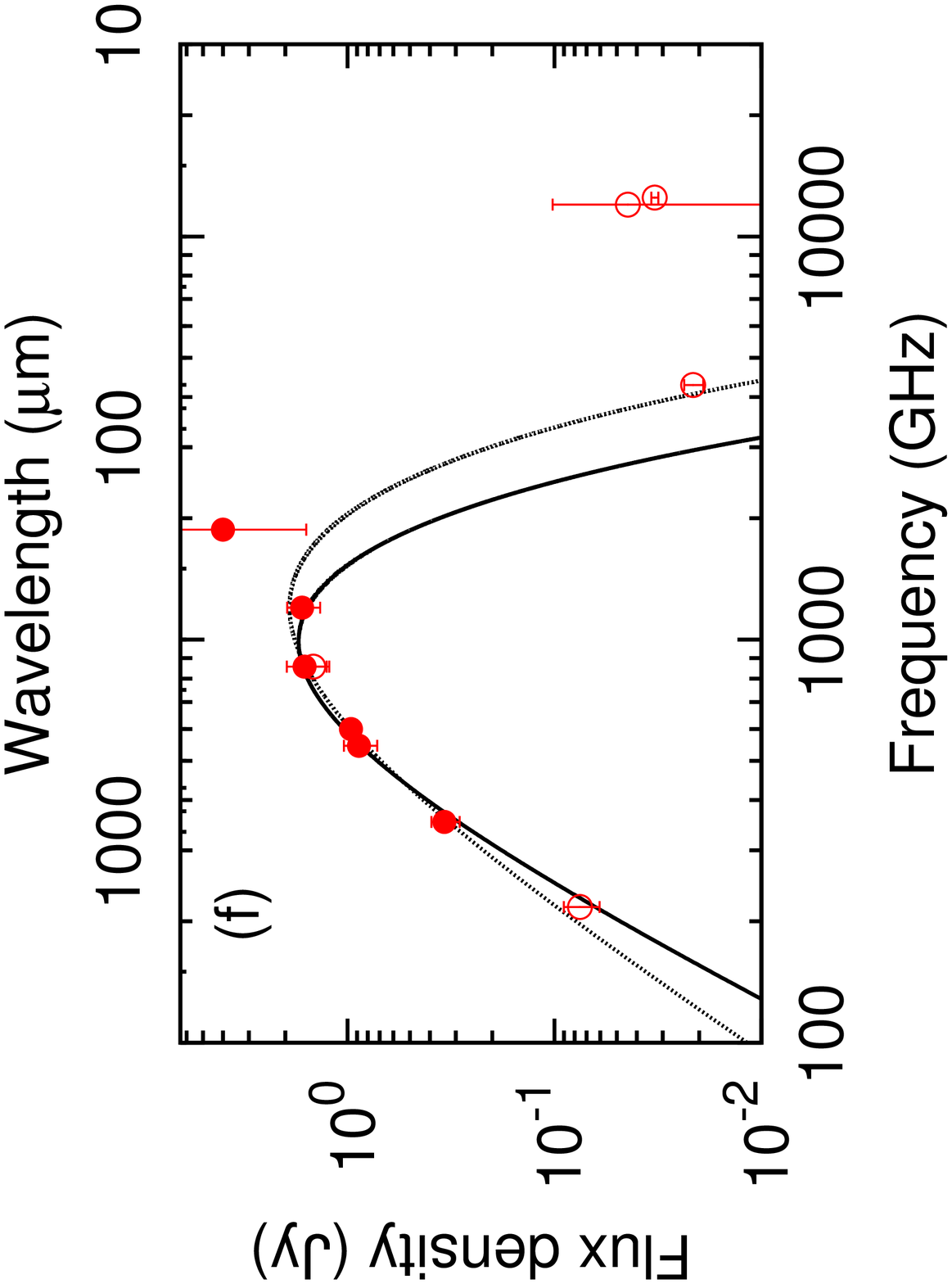}
\caption{Representative SEDs in various parts of M31. The solid line shows the fit of the modified blackbody with $\beta=2$ to the full flux densities (shown in red), and the dashed line shows the fit to the rescaled flux densities (shown in blue). Only the filled data points are used in the fits. The fit parameters are discussed in the text. {\it Top row}: a) The SED of a pixel in the nuclear region (00$^{\rm h}$42$^{\rm m}$30$^{\rm s}$, +41\deg09\arcm); b) the northern edge of the 10\,kpc ring (00$^{\rm h}$45$^{\rm m}$40$^{\rm s}$, +41\deg54\arcm); c) the bright pixel in the southern end of M31 (00$^{\rm h}$40$^{\rm m}$45$^{\rm s}$, +40\deg39\arcm). {\it Bottom row}, Outer spiral arms: d) 14.8\,kpc arm (00$^{\rm h}$39$^{\rm m}$53$^{\rm s}$, +40\deg23\arcm); e) 22\,kpc arm (00$^{\rm h}$37$^{\rm m}$43$^{\rm s}$, +39\deg58\arcm); and f) 26\,kpc arm (00$^{\rm h}$36$^{\rm m}$52$^{\rm s}$, +39\deg43\arcm). The last two SEDs also show a variable $\beta_\mathrm{dust}$ fit with a dotted line.}
\label{fig:pixelfits}
\end{center}
\end{figure*}

\subsection{Fitting method}
We focus on fitting the SEDs of the ISRF dust rather than the total SED. The ISRF dust will be the most massive component of the interstellar dust, as it is the coldest dust component in the galaxy and also the most dominant source at $\leq$3000\,GHz ($\geq$100\um, see e.g., \citealp{Draine2007}).

We use the maps of the ratio of the contribution of star formation and the old stellar population shown in Fig. \ref{fig:m31_colourratio} to rescale the flux densities of the 545, 857, 1200, 1874, 3000, and 4280\ghz\ (550, 350, 250, 160, 100, and 70\um) data points. We use the \Planck\ 353\GHz\ (850\um) and \Herschel\ 600\GHz\ (500\um) maps as zero points and rescale the higher frequency flux densities based on the ratio maps (e.g., the 857\ghz/350\um\ map is rescaled by $[1-R^{857}_{545}]$) where $R=E_\mathrm{SF}/E_\mathrm{Total}$ between the indicated frequencies within the pixel; we refer to these values as ``rescaled flux densities" in the following analysis. At the lowest frequencies, nearly all of the dust emission is attributed to the global stellar population component, while at higher frequencies the ratio will indicate how much energy in the higher frequency band originates from SFR dust relative to the lower frequency band.  If the relative fraction of SFR dust is very low, then the ratios can be used as an approximate way to rescale the flux densities for the higher frequency bands. We also fit the global SED to demonstrate the difference that the rescaling makes on the measured dust properties.

We carry out a least-squares fit using a single modified blackbody function with a normalization based on the 353\GHz\ (850\um) optical depth, i.e.
\begin{equation}
S_\mathrm{dust}(\nu) =  2 \, h \, \frac{\nu^3}{c^2} \frac{1}{e^{h\nu/kT_\mathrm{dust}}-1} \, \tau_{353} \, \left(\frac{\nu}{\mathrm{353\,GHz}}\right)^{\,\beta_\mathrm{dust}} \,\Omega.  \label{eq:sdust}
\end{equation}
Independent fits within each 5\arcm\ pixel (with solid angle $\Omega$) are used to derive the optical depth at 353\ghz\ (850\um), $\tau_{353}$, and the dust temperature $T_\mathrm{dust}$. We use $T_\mathrm{dust}=15$\,K as the starting value for the fit. We fix the dust spectral index at $\beta_\mathrm{dust}=2.0$, as used by \citet{Li2001} (see discussion in the next paragraph); we also look at the consequences of allowing this to vary in Sect. \ref{sec:dustparammaps}.  We exclude the \Planck\ 217\ghz\ (1.38\,mm) data from the fit because of CO contamination, and we do not use data above 3000\ghz\ (100\um), because we are only considering the large dust grain population rather than small dust grains or PAH emission. We also exclude data where we have \Planck\ or \Herschel\ data at the same frequency. In practice, this means that we fit to the \Planck\ 353, 545, and 857\GHz\ (850, 550, and 350\um) bands, and the \Herschel\ 600, 1200, 1874, and 3000\GHz\ (500, 250, 160, and 100\um) bands. We use the other data sets for comparison only. During the fitting process we iteratively colour correct the data based on the model.

We also calculate the dust mass per arcminute based on $\tau_\mathrm{353}$ (chosen to reduce the sensitivity of the calculation to the dust temperature) via
\begin{equation}
M_\mathrm{dust} = \frac{\tau_{353} \Omega D^2}{\kappa_{353}} \label{eq:mass}
\end{equation}
\citep{Hildebrand1983}, where we use the dust mass absorption coefficient at 353\GHz\ (850\um), $\kappa_{353} = 0.0431$\,m$^2$\,kg$^{-1}$, as being representative of the diffuse interstellar medium (\citealp{Li2001}; this value agrees well with that measured by \citealp{planck2013-XVII}, and is also close to the value measured in \citealp{planck2013-p06b}). The value of $\kappa_{353}$ depends on $\beta_\mathrm{dust}$ (see \citealp{planck2013-XVII}); however the value of $\kappa_{353}$ that we use here is from a specific model and should not be rescaled. As such we only use Eq.~(\ref{eq:mass}) to calculate the dust masses for fits where we have fixed $\beta_\mathrm{dust}=2.0$, which is approximately the spectral index for the carbon dust in the model of \citet{Li2001}. When we allow $\beta_\mathrm{dust}$ to vary we show the amplitudes of the best-fitting functions ($\tau_{353}$). For the fits where $\beta_\mathrm{dust}=2.0$, the dust mass per pixel can be converted into optical depth by dividing by $1.45\times10^{10}$\,\Msolar. We quote dust mass surface densities in units of \mkpc; along the major axis of M31 there is 228\,pc per arcmin, which after adjusting for inclination gives 0.273\kpc$^2$ per arcmin$^2$.

\subsection{Representative SEDs}

We show representative SEDs of individual pixels in Fig. \ref{fig:pixelfits}, looking at the nuclear region, positions on the north and south sides of the 10\,kpc ring, and pixels containing the outer spiral arm structure to the south of the galaxy. These plots include both SED fits to the entire spectrum, and to the emission that is related to heating by the stellar population (with both original and rescaled flux densities shown in the SED).

We find that the long wavelength data are generally well-fitted by a single modified blackbody with a fixed spectral index of 2.0, with $\chi^2$ values between 0.9 and 20.7 and an average value of $\bar{\chi}^2=7.6$ ($N_\mathrm{dof}=5$). In general we find a similar $\chi^2$ for the rescaled SEDs compared to the original SEDs. These values of $\chi^2$ indicate that the estimates of the uncertainties in the data are conservative; however, the data points are not independent, as the maps are correlated due to their common calibration schemes, an estimate of which has been incorporated in these uncertainties (see Sects. \ref{sec:planck} and \ref{sec:ancillary}).

In the nuclear region, the data fit the model well (see the left-hand panel of Fig. \ref{fig:pixelfits}), and there is a clear separation between the large grain contribution compared with the small grains, since the large grain contribution peaks at around 55\,Jy at approximately 140\um\ (2\,THz) compared with only 2\,Jy at 25\um\ (12\,THz). This pixel is dominated by ISRF dust up to high frequencies (as also seen by \citealp{Viaene2014}), such that there is not a large difference between the original and rescaled SED fits. The model residual at 217\GHz\ (1.38\,mm), which could be ascribed to CO emission, has the same amplitude as the uncertainty on the data point. For the original SED, we find a mass surface density of $(0.95\pm0.05)\times10^4$\,\mkpc\ and a temperature of ($22.7\pm0.3$)\,K with $\chi^2=1.31$ ($N_\mathrm{dof}=5$). We find very similar results for the rescaled SED; we find a mass surface density of $(1.03\pm0.06)\times10^4$\,\mkpc\ and a temperature of ($21.9\pm0.3$)\,K, with $\chi^2=1.25$ ($N_\mathrm{dof}=5$). This temperature agrees well with that of \citet{Fritz2012}, who found ($22.2\pm0.4$)\,K from \Herschel\ data. However, it is considerably less than the approximately 35\,K in the 14\arcs\ study by \citet{Groves2012}, which is due to: the lower resolution used here; that the core of M31 is spread out across four pixels in this analysis; and the steep temperature gradient in the core of M31. The nuclear region is the warmest in the entire galaxy, which is unsurprising, since warmer dust is present in spiral galaxies with large bulges \citep{Sauvage1992,Engelbracht2010}.

\begin{figure*}[tbp]
\begin{center}
\includegraphics[scale=1.5]{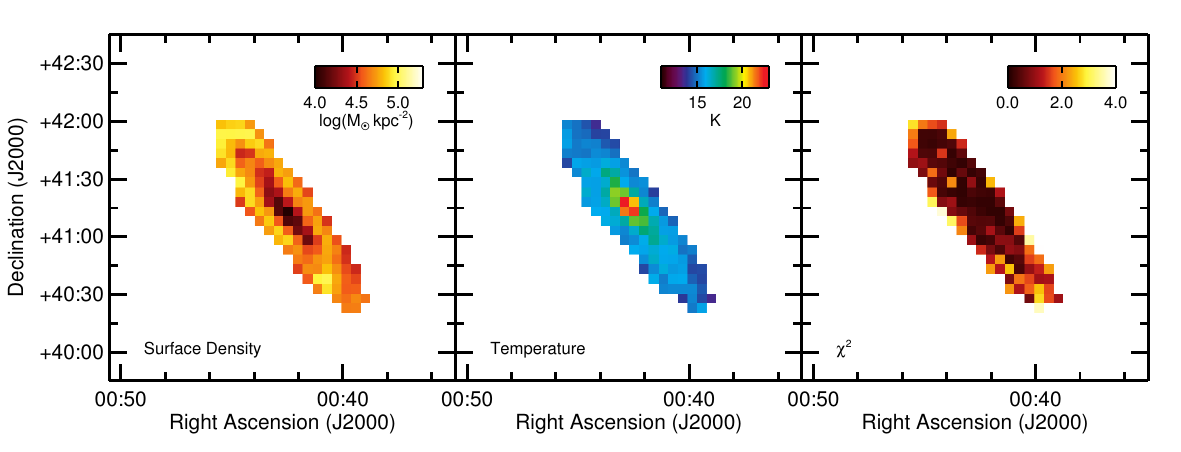}
\caption{Maps of the fitted dust mass surface density ({\it left}), temperature ({\it middle}) and reduced $\chi^2$ ({\it right}) for a one-component fit to the 5\arcm\ data set using the rescaled SEDs. The ring morphology can be clearly seen in the dust mass density, while the dust temperature decreases with galactocentric distance.}
\label{fig:1component}
\end{center}
\end{figure*}

Moving to the 10\,kpc ring, we find that this has higher optical depth/dust masses than the rest of M31, but that the dust heated by the global stellar population has lower temperatures than the dust heated by the global stellar population in the nucleus. However, there are multiple heating processes present, as demonstrated by the rescaled SEDs being much lower than the unscaled SEDs, and the SFR dust will be warmer than the dust emission studied here. We present an example on the northern edge of the ring in the top-centre panel of Fig. \ref{fig:pixelfits}. In this pixel, the rescaled [original] SEDs are fitted by a mass surface density of $(8.0\pm0.3)\times10^{4}$\,\mkpc\ [$(8.9\pm0.4)\times10^{4}$\,\mkpc], and a temperature of ($16.2\pm0.08$)\,K [($15.5\pm0.2$)\,K], with $\chi^2$ of 10.4 [9.8] ($N_\mathrm{dof}=5$). As with the nucleus, the large dust grain component is significantly brighter than the small dust grain emission. The residual at 217\GHz\ (1.38\,mm) is ($0.08\pm0.03$)\,Jy, implying that there is some CO emission in this pixel.
We also look at a pixel on the southern edge, which is particularly bright at low frequencies, shown in the right-hand panel of Fig. \ref{fig:pixelfits}. Once more, there is a good separation between the lower-frequency large dust grain and the higher-frequency small dust grain populations. The temperature and spectral indices are similar to the northern pixel, with the rescaled [original] SEDs fitted by a mass surface density of $(8.3\pm0.3)\times10^{4}$\,\mkpc\ [$(9.2\pm0.4)\times10^{4}$\,\mkpc] and a temperature of ($16.7\pm0.1$)\,K [($15.7\pm0.1$)\,K], and $\chi^2$ of 13.6 [9.6] ($N_\mathrm{dof}=5$). The residual at 217\GHz\ (1.38\,mm) in this pixel is ($0.16\pm0.03$)\,Jy [($0.14\pm0.03$)\,Jy], implying that there is some CO emission detected by \Planck\ in this (and neighbouring) pixels. This is 22\,\% [18\,\%] of the total emission at 217\GHz\ (1.38\,mm) in that pixel.

\begin{figure*}
\begin{center}
\includegraphics[scale=0.35]{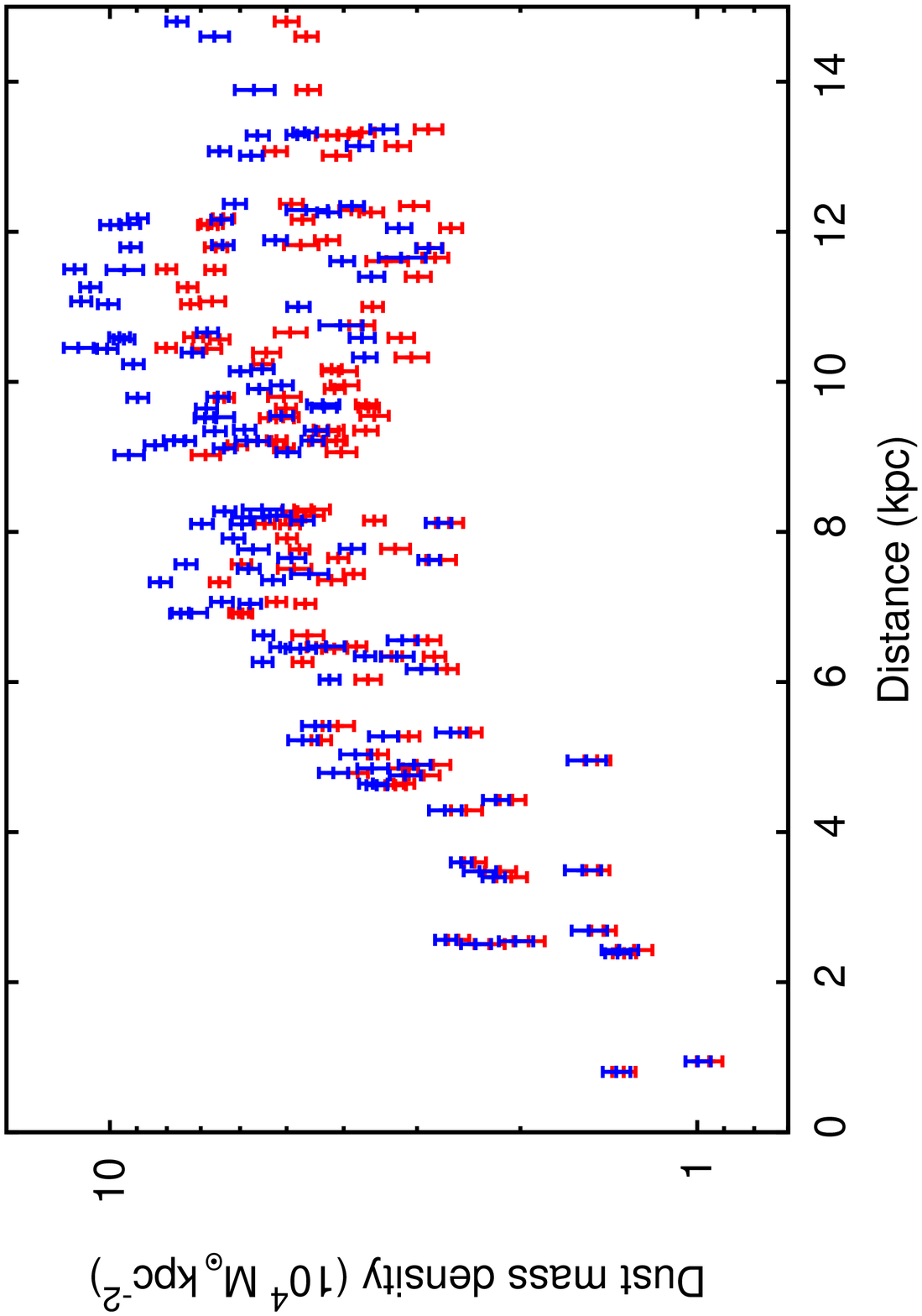}
\includegraphics[scale=0.35]{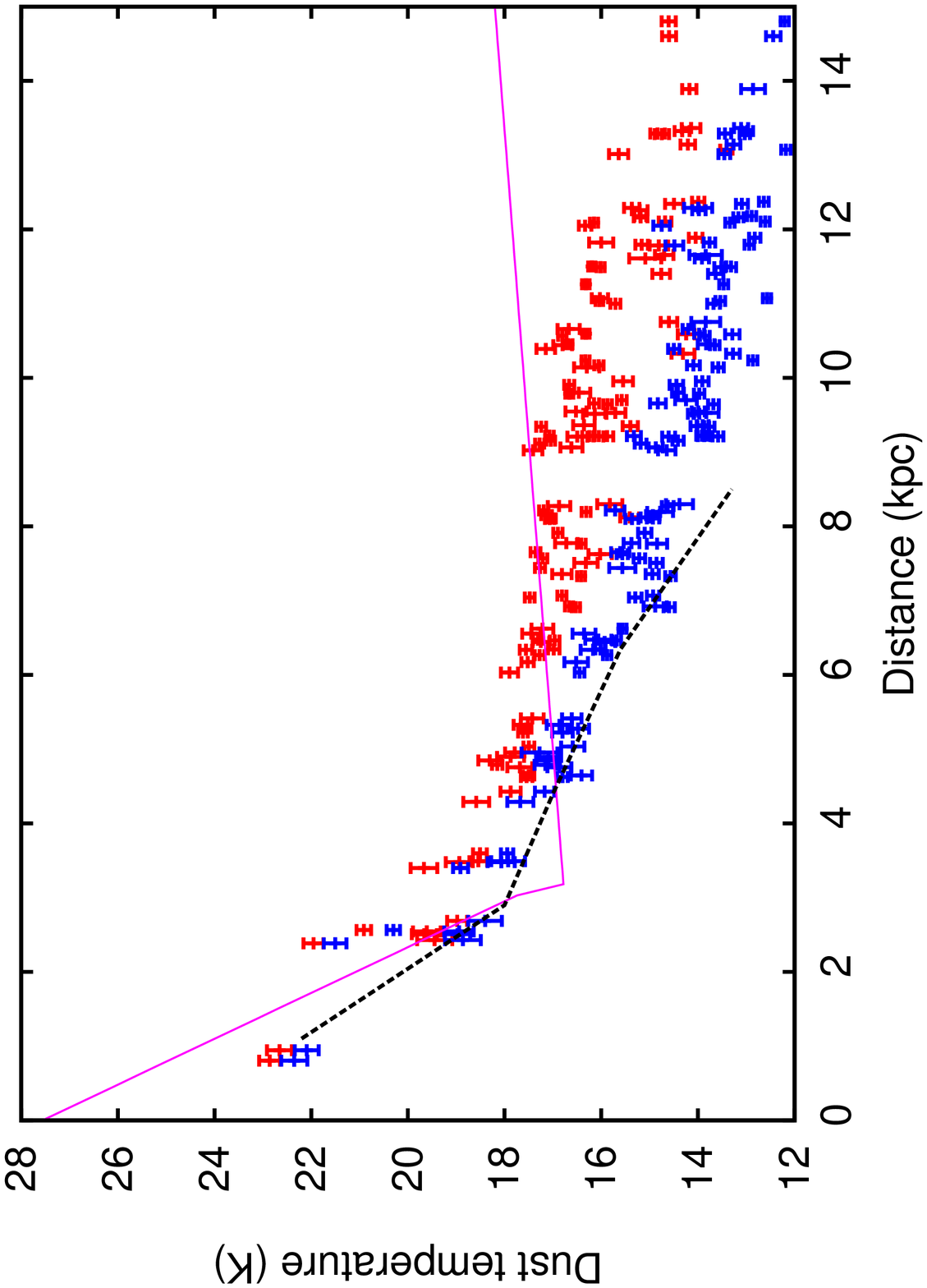}
\caption{The fitted dust mass surface density ({\it left}) and dust temperature ({\it right}) as a function of galactocentric distance. Red points are from fitting the entire SED, while blue points are fitting to the fraction of the flux density that can be attributed to the dust heated by the global stellar population. Temperatures from \citet{Fritz2012} and \citet{Smith2012} are shown by the dashed black line and the solid magenta line, respectively; these differ from the analysis here due to the different dust models used and whether the dust spectral index has been allowed to vary or not. There is a systematic shift towards lower dust temperatures and higher dust masses with the rescaled SEDs.}
\label{fig:1comp_distance}
\end{center}
\end{figure*}

We also look at the dust properties in the outermost spiral arms of M31. \Herschel\ observations of M31 \citep{Fritz2012} find dust distributed in three ring-shaped structures extending out to 21, 26, and 31\,kpc, with dust temperatures of 19.2, 18.2, and 17.1\,K (all with uncertainties of approximately $\pm0.2$\,K) respectively, derived using a single component $\beta_\mathrm{dust}=2$ dust model. We look at three pixels on the 14.8, 22 and 26\,kpc spiral arms, the SEDs for which are shown in the bottom row of Fig. \ref{fig:pixelfits}. We do not include \Spitzer\ or \ISO\ data in the fits for the outermost two spiral arm SEDs. The background subtraction method used for the \Spitzer\ data removed large-scale background structures in a way that was different than the background subtraction in other maps, and \ISO\ did not map out that far. In addition, the 22 and 26\,kpc arms were not included in the colour ratio analysis, so we consider the original rather than rescaled SED for those arms. The temperatures for these, using the \Planck\ and \Herschel\ data and $\beta_\mathrm{dust}=2.0$, are
($14.8\pm0.2$)\,K ($\chi^2=11.8$, , $N_\mathrm{dof}=5$) [rescaled, ($15.6\pm0.3$)\,K, $\chi^2=24.9$, $N_\mathrm{dof}=4$],
($12.7\pm0.4$)\,K ($\chi^2=7.9$, $N_\mathrm{dof}=4$) and
($9.6\pm0.2$)\,K ($\chi^2=3.9$,  $N_\mathrm{dof}=4$). The SEDs are also poorly fitted with a fixed $\beta_\mathrm{dust}=2.0$, so we also try a variable $\beta_\mathrm{dust}$ fit for the outermost arms. We find a much flatter spectral index for these,
with the 22\,kpc ring at ($15.0\pm1.0$)\,K and $\beta_\mathrm{dust}=1.2\pm0.3$ ($\chi^2=2.2$, $N_\mathrm{dof}=3$),
and the 26\,kpc ring at ($14.7\pm0.4$)\,K and $\beta_\mathrm{dust}=1.0\pm0.3$ ($\chi^2=2.4$, $N_\mathrm{dof}=3$), although these fits lie above the 217\GHz\ data point that is not fitted in the data. These results for $T_\mathrm{dust}$ and $\beta_\mathrm{dust}$ could be biased as there are only a few higher frequency data points that constrain the peak of the SED. The lower temperatures found here (compared with those by \citealp{Fritz2012}) are driven by the inclusion of \planck\ data in the analysis.

\subsection{Maps of the dust parameters}\label{sec:dustparammaps}
In Fig. \ref{fig:1component} we plot the best-fit parameter values for the dust heated by the old stellar population, as well as the reduced $\chi^2$ values for all pixels, with the dust mass density deprojected so that the pixels are in units of $\Msolar$\,kpc$^{-2}$. The fitted dust temperature traces the stellar population, with the dust decreasing in temperature with distance from the centre of M31. The dust mass traces the 10\,kpc ring structure, with the highest masses in the north part of the ring and in the bright region in the south part of the ring. We have compared our dust masses with those from \citet{planck2014-XXIX}; after correcting their map for the inclination of M31 we find that their dust mass \changed{estimates} are approximately 30\,\% higher than ours, which is due to the different dust modelling technique used as well as the different zero levels in the maps.

Figure \ref{fig:1comp_distance} shows the dust mass density and temperature per pixel as a function of galactocentric distance, for both the entire and rescaled SEDs. The fits to the rescaled SEDs have lower temperatures and higher dust masses than the fits to the full SEDs, and the difference increases in significance as the galactocentric distance increases. The dust mass clearly increases from the nucleus outwards, peaking at the position of the ring (around 10\,kpc), before decreasing again. The dust that is heated by the evolved stellar population is hottest in the centre of M31 at about 22\,K, and the temperature decreases down to around 14\,K at the distance of the ring. These trends are broadly comparable to those found by \citet{Fritz2012}, which are shown by the dashed black line in the plot, and have the opposite trend at larger distances than the results of \citet{Smith2012}, shown by the solid magenta line. The dust temperature behaviour as a function of distance most likely differs between these analyses due to the different dust models used, in particular whether the dust spectral index is fixed or allowed to vary; this will be discussed later in this section.

The temperatures from the rescaled and unscaled SEDs agree well with each other in the centre of the galaxy, but diverge from each other beyond about 5\,kpc. We expect the SFR to contribute more to dust heating at larger distances from the centre as the bulge stars peak in the centre of the galaxy and the SFR peaks in the ring.  At the centre of the galaxy, the old stellar population in the bulge dominates the ISRF, and the dust seen at wavelengths as short as 70~$\mu$m is heated by the ISRF, which means that the rescaled vs. unscaled SEDs should be essentially the same. The rescaled and unscaled SEDs may converge outside of the 10\,kpc ring, but we lack the sensitivity to accurately measure the SED at larger radii and therefore cannot see this reconvergence.

We also combine the values from all pixels and fit a fixed $\beta_\mathrm{dust}$ model to the overall SED from M31. We find that this has an average temperature for the rescaled [original] SED of (16.96$\pm$0.13)\,K [(16.89$\pm$0.12)\,K] and a combined dust mass of $(6.6\pm0.3)\times10^{7}$\,\Msolar\ [$(7.0\pm0.3)\times10^{7}$\,\Msolar]. The combined dust mass from the fits to the individual pixels is $(3.6\pm0.3)\times 10^{7}$\,\Msolar\ [$(4.1\pm0.3)\times 10^{7}$\,\Msolar], around a factor of two lower. This is the opposite of the results of \citet{Galliano2011}, who found 50\,\% higher dust masses when fitting data for the LMC at higher resolution compared to the integrated SED, and \citet{Aniano2012} who find up to 20\,\% higher dust masses from fitting the resolved data for NGC\,628 and NGC\,6946. However it is in agreement with \citet{Smith2012} who, like us, find a significantly higher dust mass from the integrated SED than the individual pixel SEDs in M31, and also \citet{Viaene2014} who find a dust mass 7\,\% higher using an integrated SED. The difference between the masses is caused by different analyses using varying data sets and assumptions, as well as different weightings resulting from the morphology and temperature distribution within M31.  In the case of the LMC and other dwarf galaxies, warm dust in star forming regions are embedded within a dust component of cool large grains heated by a relatively weak diffuse interstellar radiation field.  When integrating over large areas, the resulting temperatures are warmer and dust surface densities are lower than what would be found within individual subregions.  In M31 and other galaxies with large bulges \citep[e.g., M81;][]{Bendo2010,Bendo2012}, the temperature of large dust grains heated by the diffuse interstellar radiation field in the centres of the galaxies may be higher than the temperature of the large dust grains heated locally by star-forming regions in the outer disc.  When integrating over an area that includes both the centre of the galaxy and the star forming ring, the inferred dust temperatures may appear lower and the dust surface densities may appear higher than what would be found in individual subregions within the galaxies. The combined dust mass from the individual pixels will be more physically accurate, and is in agreement with the overall dust mass for M31 found by \citet{Draine2013}.

However, \citet{Smith2012} find about $2.8\times10^7$\,\Msolar\ with their higher resolution analysis, but this is likely to be related to the different approaches with SED fitting.  Smith et al. effectively assumed that all emission above $100$\,$\mu$m originates from dust heated by a single source, whereas we show that some of the emission above 1000\GHz\ originates from dust heated by star formation.  The \citet{Smith2012} dust temperatures will therefore be higher, which will bias the resulting dust masses lower.

\begin{figure*}[tbp]
\begin{center}
\includegraphics[scale=1.0]{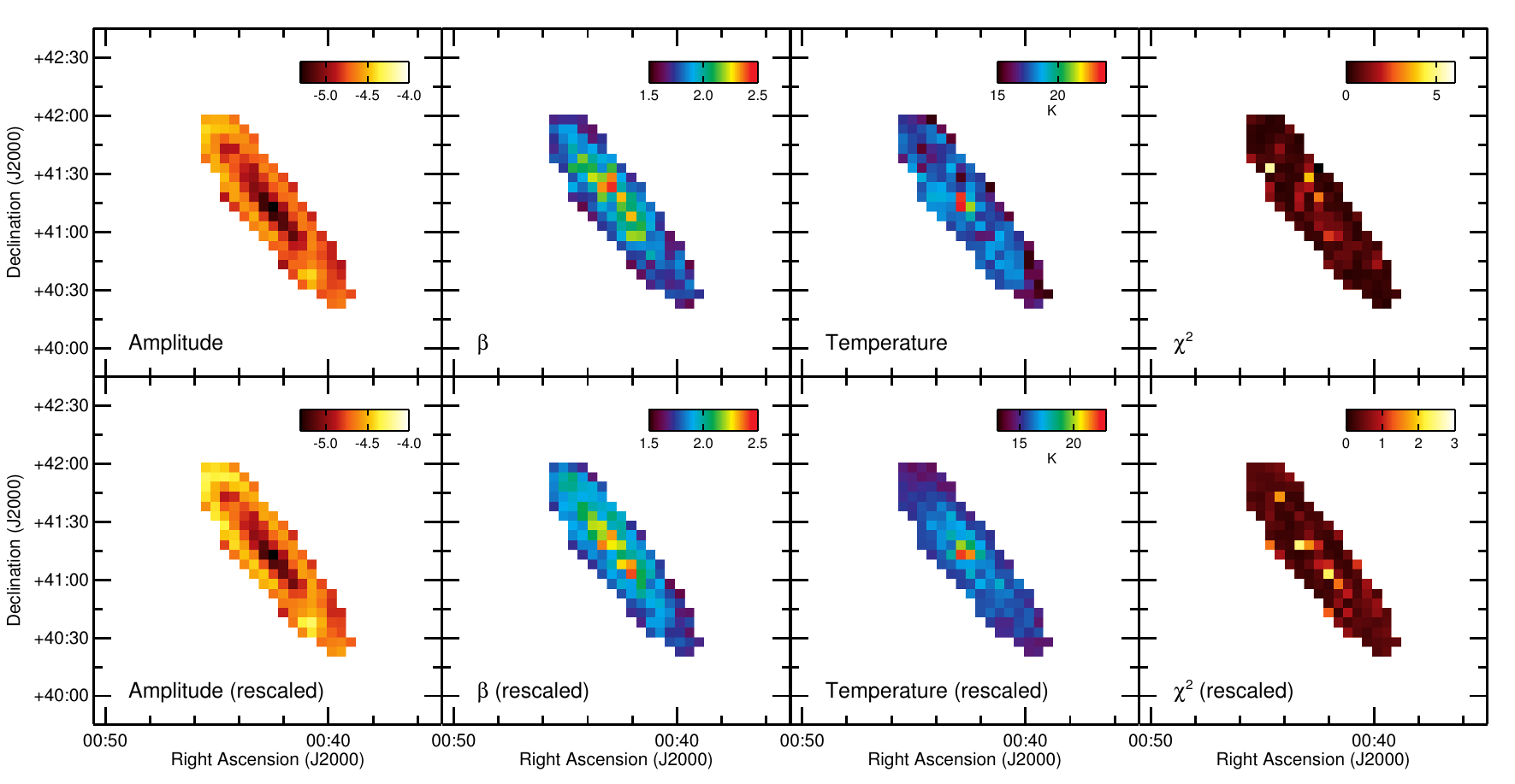}
\caption{Results from allowing $\beta_\mathrm{dust}$ to vary. {\it Top}: fitting for the entire SED. {\it Bottom}: Fitting to the rescaled SED (i.e., only the dust component that is heated by the stellar population). {\it Left to right}: the fitted dust optical depth, spectral index, temperature, and reduced $\chi^2$. Note the difference in morphology for the dust temperature and indices.}
\label{fig:varbeta}
\end{center}
\end{figure*}

\begin{figure}[tbp]
\begin{center}
\includegraphics[scale=0.35]{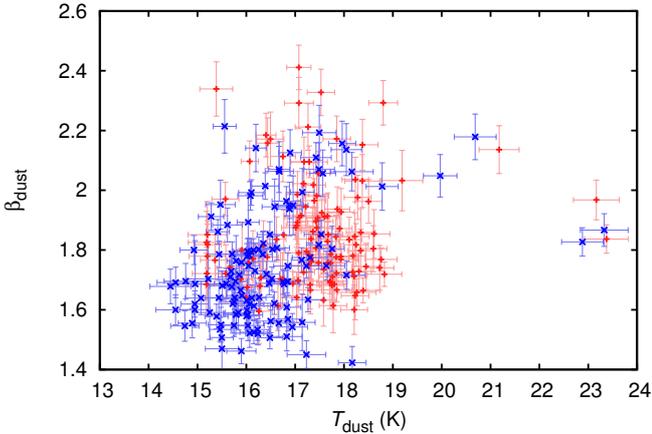}
\caption{$T_\mathrm{dust}$ vs. $\beta_\mathrm{dust}$ for the entire SED (red plus symbols) and the rescaled SED (blue crosses). There is a systematic shift to lower dust temperatures with the rescaled SEDs, while the distribution of $\beta_\mathrm{dust}$ does not \changed{change as much}.}
\label{fig:tempbeta}
\end{center}
\end{figure}

We have also fitted the data using a variable spectral index. The results of these fits for both the full and rescaled SEDs are shown in Fig. \ref{fig:varbeta}. The behaviour of the temperature and spectral index are somewhat different in these two situations. When fitting the entire SED, the resulting temperature distribution is much more uniform than for the fixed $\beta_\mathrm{dust}$ case, whilst the spectral index decreases with galactocentric distance. The same behaviour is still evident when the rescaled SEDs are fitted; however, it is less significant. This helps explain some of the differences between our analysis and that using the \Herschel\ data described in \citet{Smith2012}, where the entire SED has been fitted without taking into account the heating mechanism. The behaviour of flatter spectral indices further out from the galactic centre was noted by \citet{Smith2012}, and has also been seen in our Galaxy \citep{Paradis2012a,planck2013-p06b}. The fitted spectral indices are steep, with a range of 1.8--2.5 for M31 both here and in \citet{Smith2012}; in comparison, our Galaxy has much flatter spectral indices of 1.4--1.8 \citep{planck2013-p06b}.

We plot the temperatures vs. $\beta_\mathrm{dust}$ for both the entire SED and the rescaled SED in Fig. \ref{fig:tempbeta}. There is a systematic shift towards colder temperatures of about 9\,\% when fitting the rescaled SED compared with the entire SED, while the $\beta_\mathrm{dust}$ distribution steepens slightly from an average of \changed{1.80} for the entire SED to \changed{1.70} for the rescaled SED. It is well known that there is a significant degeneracy present when both the temperature and spectral index are allowed to vary (e.g., see \citealp{Shetty2009,Galametz2012,Juvela2012}), but we do not see evidence for this here.

We have also carried out this analysis using the {\tt NILC} CMB subtraction and find very similar results, with differences primarily in the outermost pixels where the signal is weaker compared to the CMB. As seen in Sect. \ref{sec:colourratios}, the colour ratio analysis is negligibly affected by the different CMB subtraction; the SED fitting is more sensitive to differences in the 353\GHz\ (850\um) flux density. For the fixed $\beta$ case, the dust mass changes by an average of 0.37\,$\sigma$, and a maximum of 1.8\,$\sigma$, and the temperature changes by an average of 0.4\,$\sigma$ and a maximum of 2.6\,$\sigma$, with the largest changes being at high galactocentric distances. \changed{Repeating this analysis using the combination of 24\um\ and FUV data to trace the star formation also does not have a significant effect, with dust temperature estimates from the rescaled SEDs changing by a maximum of 1.6\,K.}

\section{Integrated SED} \label{sec:lowressed}

\begin{figure}[tbp]
\begin{center}
\includegraphics[scale=1.1]{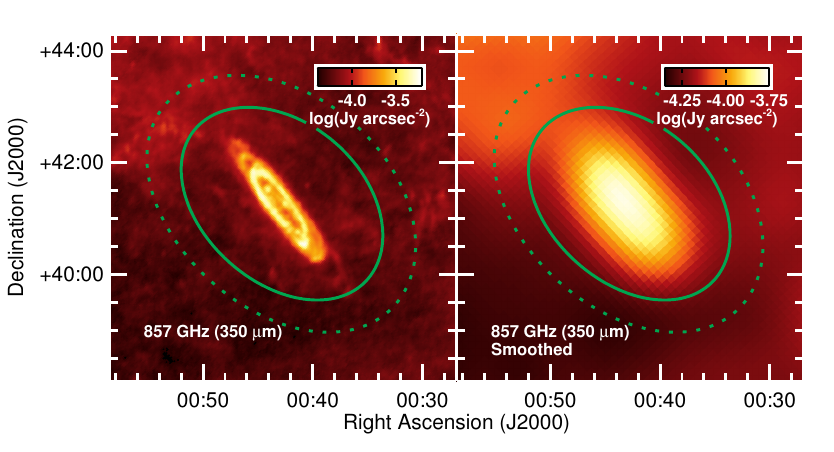}
\caption{Aperture used to extract the total emission from M31, superimposed on the raw ({\it left}) and 1\degr-convolved ({\it right}) 857\ghz\ map.}
\label{fig:aperture}
\end{center}
\end{figure}

We use aperture photometry to construct an integrated spectrum of M31 as a whole, using the same code developed for \citet{planck2011-7.2} and \mbox{\citet{planck2013-XV}}, with modifications to enable us to have an elliptical aperture. The aperture is shown in Fig. \ref{fig:aperture}. We use an inner radius of 100\arcm, and a background annulus between 100\arcm\ and 140\arcm, with a ratio of the major to minor axes of 0.7 and an angle of 45\deg. We convolve the data set (in \healpix\ format) to 1\deg\ resolution prior to extracting the flux densities to have a matched resolution data set across all frequencies, which allows us to accurately compile the measured flux densities across the spectrum. The measured flux densities for M31 at the various frequencies (prior to colour correction) are listed in Table \ref{tab:fitflux}.

\begin{table}
\begingroup
\newdimen\tblskip \tblskip=5pt
\caption{Flux densities for the integrated spectrum from aperture photometry. Note that the flux densities have not been colour-corrected. Values marked with $\dagger$ were not included in the fit.}
\label{tab:fitflux}
\nointerlineskip
\vskip -3mm
\footnotesize
\setbox\tablebox=\vbox{
   \newdimen\digitwidth 
   \setbox0=\hbox{\rm 0} 
   \digitwidth=\wd0 
   \catcode`*=\active 
   \def*{\kern\digitwidth}
   \newdimen\signwidth 
   \setbox0=\hbox{+} 
   \signwidth=\wd0 
   \catcode`!=\active 
   \def!{\kern\signwidth}
    \halign{\hbox to 0.9in{#\leaderfil}\tabskip 0.5em&
            \hfil#\hfil\tabskip 3pt &
            \hfil#\hfil\tabskip 3pt &
            \hfil#\hfil\tabskip 0pt\cr
    \noalign{\doubleline\vskip 2pt}
    \omit\hfil Survey \hfil& $\nu$ [GHz] \hfil & $\lambda$ [mm]& Flux density [Jy]\cr
\noalign{\vskip 4pt\hrule\vskip 6pt}
Haslam & ****0.408 & 734\phantom{.000} & $***23.1*\pm***4.5*$\cr
Dwingeloo & ****0.82* & 365\phantom{.000} & $***13.0*\pm***2.4*$\cr
Reich & ****1.42* & 214\phantom{.000} & $****7.5*\pm***0.8*$\cr
\WMAP& ***22.8** & *13\phantom{.000} & $****2.09\pm***0.10$\cr
\Planck& ***28.4** & *10.5** & $****2.05\pm***0.11$\cr
\WMAP& ***33.0** & **9.0** & $****1.88\pm***0.14$\cr
\WMAP& ***40.7** & **7.4** & $****1.73\pm***0.18$\cr
\Planck& ***44.1** & **6.8** & $****1.31\pm***0.21$\cr
\WMAP& ***60.7** & **4.9** & $****2.90\pm***0.38$\cr
\Planck& ***70.4** & **4.3** & $****3.27\pm***0.51$\cr
\WMAP& ***93.5** & **3.2** & $****4.32\pm***0.78$\cr
\Planck& **100\phantom{.}*** & **3.0** & $****7.3*\pm***1.2*$\cr
\Planck& **143\phantom{.}*** & **2.1** & $***18.2*\pm***1.5*$\cr
\Planck& **217\phantom{.}*** & **1.38* & $***76.0*\pm***8.2*$\cr
\Planck& **353\phantom{.}*** & **0.85* & $**298\phantom{.}**\pm**13\phantom{.}**$\cr
\Planck& **545\phantom{.}*** & **0.55* & $*1018\phantom{.}**\pm*104\phantom{.}**$\cr
\Herschel& **600\phantom{.}*** & **0.50* & $*1290\phantom{.}**\pm*130\phantom{.}**$\cr
\Planck& **857\phantom{.}*** & **0.35* & $*3050\phantom{.}**\pm*310\phantom{.}**$\cr
\Herschel$\dagger$& **857\phantom{.}*** & **0.35* & $*3040\phantom{.}**\pm*310\phantom{.}**$\cr
\Herschel& *1200\phantom{.}*** & **0.25* & $*5840\phantom{.}**\pm*770\phantom{.}**$\cr
\DIRBE$\dagger$& *1249\phantom{.}*** & **0.24* & $*5700\phantom{.}**\pm*770\phantom{.}**$\cr
\Herschel& *1874\phantom{.}*** & **0.16* & $*6900\phantom{.}**\pm*900\phantom{.}**$\cr
\DIRBE& *2141\phantom{.}*** & **0.14* & $*7300\phantom{.}**\pm1000\phantom{.}**$\cr
\DIRBE$\dagger$& *2997\phantom{.}*** & **0.10* & $*3600\phantom{.}**\pm*500\phantom{.}**$\cr
\IRAS$\dagger$& *2997\phantom{.}*** & **0.10* & $*3330\phantom{.}**\pm*440\phantom{.}**$\cr
\Herschel& *2998\phantom{.}*** & **0.10* & $*2960\phantom{.}**\pm*390\phantom{.}**$\cr
\DIRBE$\dagger$& *4995\phantom{.}*** & **0.06* & $**710\phantom{.}**\pm*100\phantom{.}**$\cr
\IRAS$\dagger$& *4995\phantom{.}*** & **0.06* & $**750\phantom{.}**\pm*100\phantom{.}**$\cr
\IRAS$\dagger$& 12000\phantom{.}*** & **0.025 & $**134\phantom{.}**\pm**17\phantom{.}**$\cr
\IRAS$\dagger$& 25000\phantom{.}*** & **0.012 & $**184\phantom{.}**\pm**24\phantom{.}**$\cr
\noalign{\vskip 3pt\hrule\vskip 4pt}
}}
\endPlancktable
\endgroup
\end{table}

We use a Bayesian Markov Chain Monte Carlo (MCMC) technique, as described by \citet{planck2011-6.4a}, to fit the multiple components of the spectrum between 0.408 and 3,000\GHz\ (73.4\,cm and 100\um) via
\begin{equation}
S(\nu) = A_\mathrm{sync} \nu^\alpha + S_{\rm ff}(\nu) + S_{\rm AME}(\nu) + S_\mathrm{dust}(\nu) + S_{\rm cmb},
\end{equation}
where the synchrotron emission is characterized by a power law with amplitude $A_\mathrm{sync}$ and index $\alpha$. $S_\mathrm{dust}$ is given by a single modified blackbody spectrum as
\begin{equation}
S_\mathrm{dust}(\nu) =  2 \, h \, \frac{\nu^3}{c^2} \frac{1}{e^{h\nu/kT_\mathrm{dust}}-1} \, \tau_\mathrm{250} \, \left(\frac{\nu}{\mathrm{250 \mu m}}\right)^{\,\beta_\mathrm{dust}} \,\Omega,
\end{equation}
with free parameters $\tau_\mathrm{250}$, $T_\mathrm{dust}$ and $\beta_\mathrm{dust}$. Note that this is fitted to the measured SED, rather than a rescaled SED as in previous sections.

The free-free flux density, $S_{\rm ff}$, is calculated from the brightness temperature, $T_{\rm ff}$, based on the optical depth, $\tau_{\rm ff}$, using
\begin{equation}
S_{\rm ff} = \frac{2  k T_{\rm ff}  \Omega  \nu^2}{c^2},
\end{equation} 
where $\Omega$ is the solid angle, $\nu$ is the frequency,
\begin{equation} \label{eq:freefree}
T_{\rm ff} = T_{\rm e}(1-e^{-\tau_{\rm ff}}),
\end{equation}
and the optical depth, $\tau_{\rm ff}$, is given by
\begin{equation}
\tau_{\rm ff} = 5.468 \times 10^{-2} \, T_\mathrm{e}^{-1.5} \, \nu^{-2} \,  {\rm EM} \, g_{\rm ff}~,
\end{equation}
in which the Gaunt factor is approximated by \citep{Drain2011b}
\begin{equation}
g_{\rm ff} = {\rm ln} \left( {\rm exp} \left[5.960 - \frac{\sqrt{3}}{\pi} {\rm ln} (Z_i \,\nu_9 \,T_4^{-3/2}) \right] + e \right)~,
\end{equation}
where $e=2.71828$. We assume a fixed electron temperature of $T_\mathrm{e}=8000$\,K, which is a typical value from our Galaxy, and fit for the emission measure (EM).

We also fit for anomalous microwave emission using
\begin{equation}
S_{\rm AME} = A_{\rm AME} \, j(\nu) \,\Omega,
\end{equation}
where we use a warm ionized medium (WIM) model for $j (\nu)$ calculated from {\tt SPDUST} \citep{Ali-Hamoud2009,Silsbee2011}. The amplitude $A_{\rm AME}$ has units of aperture area times the column density (Sr\,cm$^{-2}$).

The CMB is fitted using the differential of a blackbody  at $T_{\rm CMB}=2.7255$\,K \citep{fixsen2009}:
\begin{equation}
S_{\rm cmb} = \left(\frac{2  k \Omega \nu^2}{c^2}\right) \Delta T_{\rm CMB},
\end{equation}
where $\Delta T_{\rm CMB}$ is the CMB fluctuation temperature in thermodynamic units. We do not attempt to remove CMB from the maps before carrying out the aperture photometry and analysis of the SED. The CMB amplitude can be either negative or positive.

\begin{figure*}[tbp]
\begin{center}
\includegraphics[scale=0.6]{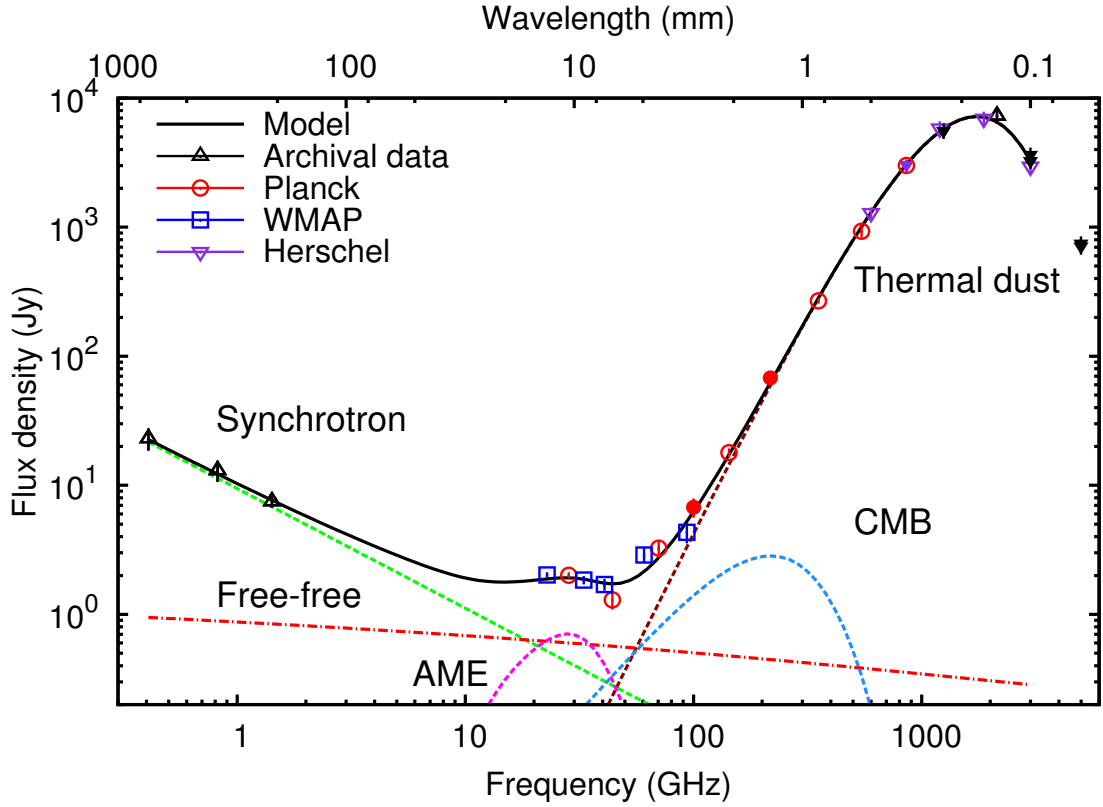}
\caption{The integrated SED from radio to far-infrared, with the best-fit model from Bayesian analysis (see Sect. \ref{sec:lowressed}). Data points are from \WMAP\ (blue), \Planck\ (red), \Herschel\ (purple) and other archival data (black). Filled points represent data that were not included in the fit. The best fit is shown in black; the green line shows the synchrotron fit, the red line shows the free-free emission, the magenta curve shown the AME, the dark red line shows the thermal dust, and the light blue line shows the CMB.}
\label{fig:integratedspectrumbayesian}
\end{center}
\end{figure*}

\begin{figure*}[tbp]
\begin{center}
\includegraphics[scale=0.45]{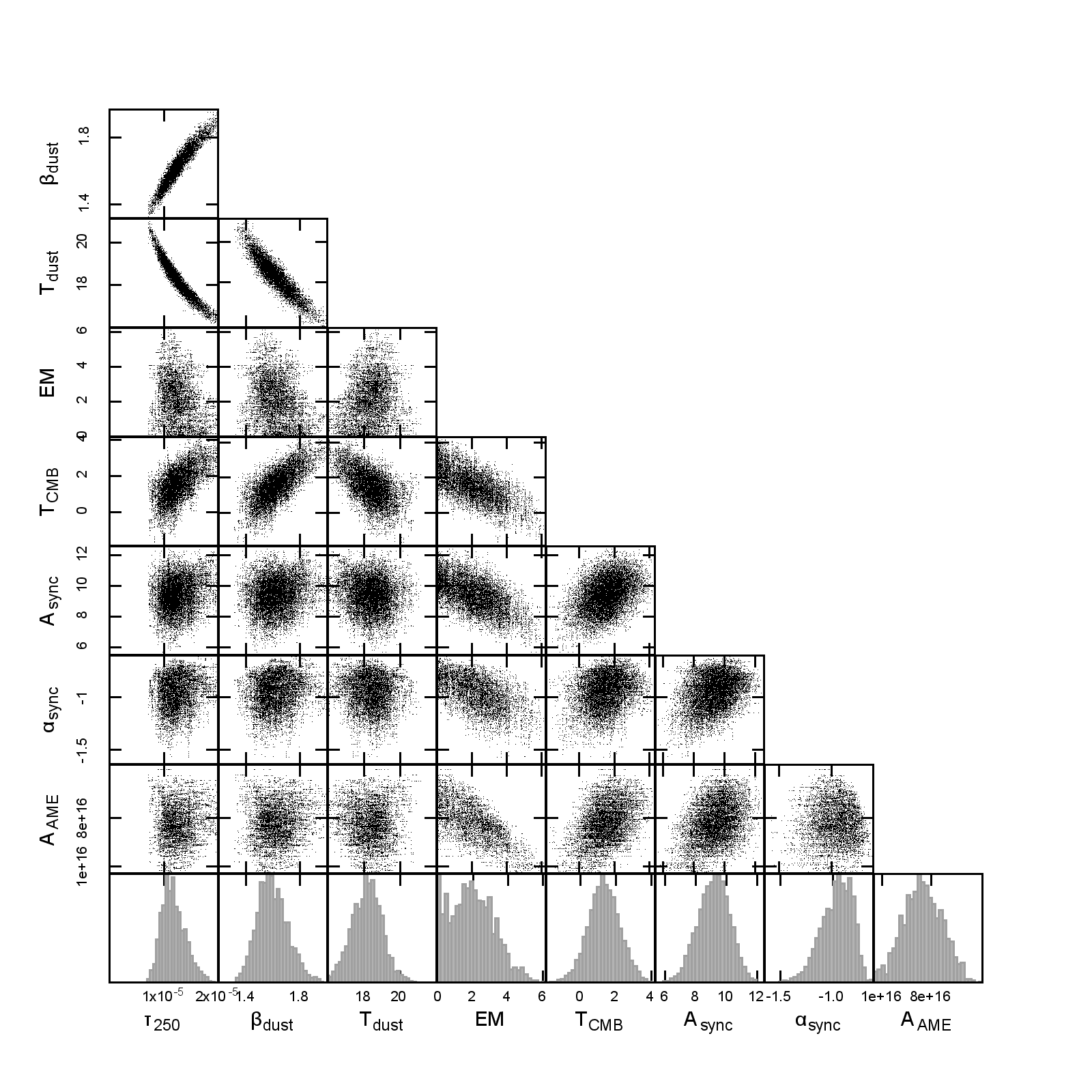}
\caption{The samples from the SED likelihood function for the Bayesian analysis. The bottom row shows the one-dimensional (marginalized) posterior for the parameters, with the other rows showing all of the two-dimensional marginal distributions.}
\label{fig:integratedspectrumbayesian_samples}
\end{center}
\end{figure*}

The best-fit model from the Bayesian analysis is shown in Fig. \ref{fig:integratedspectrumbayesian}, with the mean fit parameters given Table \ref{tab:fitparams}. The samples from the SED likelihood function are shown in Fig. \ref{fig:integratedspectrumbayesian_samples}. We find that there is a good fit across all of the frequencies ($\chi^2=$\getsymbol{chi2} with $N_\mathrm{dof}=$\getsymbol{ndof}), with steep spectrum synchrotron emission dominating the radio frequencies, a significant amount of free-free emission dominating the flux densities between 10 and 50\ghz, along with CMB at frequencies around 100\ghz, and cold thermal dust emission at higher frequencies. We report results here using the 2013 \Planck\ data set. We have repeated this analysis with the 2015 \Planck\ data \citep{planck2014-a01}, and find no significant differences.

\begin{table}
\begingroup
\newdimen\tblskip \tblskip=5pt
\caption{Best-fitting parameters for the integrated spectrum from the Bayesian analysis.}
\label{tab:fitparams}
\nointerlineskip
\vskip -3mm
\footnotesize
\setbox\tablebox=\vbox{
   \newdimen\digitwidth 
   \setbox0=\hbox{\rm 0} 
   \digitwidth=\wd0 
   \catcode`*=\active 
   \def*{\kern\digitwidth}
   \newdimen\signwidth 
   \setbox0=\hbox{+} 
   \signwidth=\wd0 
   \catcode`!=\active 
   \def!{\kern\signwidth}
    \halign{\hfil#\tabskip 1.0em&
            \hfil#\hfil\tabskip 0pt\cr
    \noalign{\doubleline\vskip 2pt}
    \omit\hfil Parameter & Value\cr
\noalign{\vskip 4pt\hrule\vskip 6pt}
$\tau_{250}$ & \getsymbol{dusttau}\cr
$\beta_\mathrm{dust}$ & \getsymbol{dustbeta}\cr
$T_\mathrm{dust}$ [K] & \getsymbol{tdust}\cr
EM [cm$^{-6}$\,pc] & \getsymbol{freefree}\cr
$\Delta T_\mathrm{CMB}$ [K] & \getsymbol{cmbamp}\cr
$A_\mathrm{synch}$ [Jy] & \getsymbol{Async}\cr
$\alpha_\mathrm{synch}$ & \getsymbol{alphasync}\cr
$A_\mathrm{AME}$ [Sr\,cm$^{-2}$] & \getsymbol{AMEamp}\cr
$\chi^2$ & \getsymbol{chi2}\cr
$N_\mathrm{dof}$ & \getsymbol{ndof}\cr
\noalign{\vskip 3pt\hrule\vskip 4pt}
}}
\endPlancktable
\endgroup
\end{table}

We find that the synchrotron emission in the aperture has an amplitude at 1\ghz\ of \getsymbol{Async}\,Jy and a steep spectrum with a power-law index of ($\alpha$\,$=$\,\getsymbol{alphasync}), which is in the range of typical synchrotron spectral indices for normal galaxies of about $-0.85$ (e.g., \citealp{Condon1992,Niklas1997}, but these may include a contribution from a thermal component, see \mbox{\citealp{Peel2011})}. This is poorly constrained as synchrotron is not a significant component in the fit at the \Planck\ and \WMAP\ frequencies.\footnote{Additional data at frequencies around 5--15\GHz, e.g., by C-BASS \citep{King2010}, would be required to see whether the synchrotron radiation is steepening or flattening at high frequencies, see e.g., \citet{Strong:2011} and \citet{Peel2012}.} This emission includes both compact and diffuse emission from M31 that is within the aperture, as well as foreground emission from our Galaxy and emission from bright quasars behind M31 (particularly at the lowest frequencies) -- although faint AGN, being more numerous, should statistically cancel out, as there should be as many of them in the background annulus as in the aperture. Bright compact sources are particularly visible in the high-resolution surveys of M31, as described by \citet{Graeve1981}. \citet{Berkhuijsen2003} use a compilation of source-subtracted measurements of M31 to measure the integrated flux densities out to 16\,kpc and find significantly lower flux densities than the fit here. For example, they quote ($11.20\pm0.81$)\,Jy of emission at 408\,MHz from \citet{Pooley1969} compared to the ($23.1\pm4.5$)\,Jy that we find here from the \citet{Haslam1981} 408\,MHz map; in contrast \citet{Graeve1981} find a much higher flux density of ($63.7\pm6.7$)\,Jy when the ``halo'' emission around M31 (which comes from point sources and a Galactic spur) is also considered. This difference is therefore presumably due to foreground emission and extragalactic sources in the aperture.

A significant amount of free-free emission is needed to account for the flux density detected in the lower \Planck\ and \WMAP\ bands; the model fit implies that a large fraction of the flux density seen at frequencies of 20--60\ghz\ is caused by free-free emission. Similar results have been seen in other nearby galaxies. For example, \citet{Peel2011} also found that a significant amount of free-free emission was needed to give realistic star formation rates for Messier\,82, NGC\,4945, and NGC\,253. Free-free emission scales linearly with the star-formation rate (see e.g., \citealp{Murphy2011}). \citet{Azimlu2011} have calculated the star formation rate based on H$\alpha$ imaging and find it to be $0.44$\mdotyr, which is higher than the ($0.25\pm0.05$)\mdotyr\ found by \citet{Ford2013} using ultraviolet and 24\um\ data. The SFR of ($0.36\pm0.14$)\mdotyr estimated by \citet{Xu1996} using \IRAS\ data of M31 lies in between these. Star formation rates can be calculated from the EM (and flux densities) that we are fitting for here using the formulae from \citet{Condon1992}. In our standard model, without applying any priors, we find an EM of (\getsymbol{freefree})\,cm$^{-2}$\,pc ($\approx$$0.7$\,Jy at 10\GHz), which equates to an SFR of $\approx0.12$\mdotyr; this is substantially lower than the SFR from \citet{Ford2013} and \citet{Xu1996}, and lower than the SFR from H$\alpha$.

The Bayesian analysis gives an AME amplitude of \mbox{\getsymbol{AMEamp}\,Sr\,cm$^{-2}$}, which is a $2.3\,\sigma$ marginal detection of the presence of AME in M31. We can estimate the amount of AME that we might expect based on observations of anomalous emission in our Galaxy \citep{Todorovic2010,Alves2012}. The ratio between AME at 30\ghz\ and thermal dust emission as seen by \IRAS\ at 100\um\ is expected to be about 1:3000 in flux density, although there is a considerable amount of scatter in this number (see \citealp{planck2013-XV}; additionally \citealp{planck2011-6.4b} found a ratio of approximately 1:2000 for the Small Magellanic Cloud) and it is sensitive to dust temperature \citep{Tibbs2012}. As we find ($3330\pm440$)\,Jy at 100\um, this implies that the 30\ghz\ AME emission should be ($1.11\pm0.15$)\,Jy (with the uncertainty only coming from the 100\um\ flux density, not from the ratio). At 30\ghz\ we find an AME flux of ($0.7\pm0.3$)\,Jy, which is comparable with the expectation based on the AME level seen in our Galaxy. The results are also compatible with those for AME in other nearby galaxies (M\,82, NGC\,253, and NGC\,4945)  presented by \citet{Peel2011} and the detection of AME within only one of ten star-forming clouds observed by \citet{Murphy2010} and \citet{Scaife2010} within NGC\,6946. The main source of uncertainty here is the level of the free-free emission. Observations at frequencies of 5--10\GHz\ (e.g., a 5\GHz\ measurement from C-BASS, \citealp{King2010}; \citealp{Irfan2015}) would improve these constraints.

Due to the correlations between $\alpha_\mathrm{sync}$ and EM, and EM and A$_\mathrm{AME}$, we have tried setting Gaussian priors on the $\alpha_\mathrm{sync}$ and EM parameters, with values of $\alpha_\mathrm{sync}$\,$=$\,$-0.90\pm0.3$, and EM\,$=$\,$5.0\pm1.0$. Applying both priors results in a slightly higher $\chi^2$\,$=$\,$18.6$, with $\alpha_\mathrm{sync}$\,$=$\,$-1.02\pm0.13$, EM\,$=$\,$4.0\pm0.9$ and \mbox{$A_\mathrm{AME}$\,$=$\,$(4\pm2)$\,$\times10^{16}$\,Sr\,cm$^{-6}$\,pc}, and there is a negligible effect on the mean values of the other fit parameters (changes are less than 1\,$\sigma$). As such, we have conservatively used wide uniform priors for our canonical model here.

We find that the thermal dust emission can be well characterized with a single modified blackbody spectrum, with an effective temperature of (\getsymbol{tdust})\,K, a spectral index \mbox{$\beta_\mathrm{dust}=$\getsymbol{dustbeta}} and an optical depth of \getsymbol{dusttau}. The overall cold temperature of the dust agrees with the \Herschel\ observations, which found an average temperature of \mbox{($18.1\pm0.2$)\,K} \citep{Fritz2012}. This is higher than the temperature of (16.96\,$\pm$\,0.13)\,K found in Sect. \ref{sec:dustparammaps}. The higher temperature obtained when allowing $\beta_\mathrm{dust}$ to vary is an expected consequence of the degeneracy between temperature and $\beta_\mathrm{dust}$ \citep{Shetty2009}. The spectral index is lower than would have been expected from the statistical $\beta_\mathrm{dust}$--$T_\mathrm{dust}$ relation seen in nearby galaxies in \citet{planck2011-6.4a}, where at this temperature $\beta_\mathrm{dust}$\,$>$\,$2$ would have been expected. It is, however, higher than the values seen in the LMC and SMC of around 1.5 and 1.2, respectively \citep{planck2011-6.4b}, and is comparable to the typical values of $\beta_\mathrm{dust}$ seen in our Galaxy \citet{planck2013-XVII}.  We note that it is surprising how well the single temperature modified blackbody fits the data. Even though this function does not replicate the emission at $<100$~$\mu$m from small dust grains that are stochastically heated by the diffuse ISRF, it still appears to be a good model for the overall emission from a galaxy and can provide a characteristic temperature for the large dust grains.

A positive differential CMB contribution is found, with an amplitude of \getsymbol{cmbamp}\,K. That this signal is positive is not too surprising, given the presence of the large positive CMB anisotropy seen in Fig. \ref{fig:m31cmb}. \citet{planck2011-6.4b} also found positive residual CMB contributions for the LMC and SMC, and \citet{planck2013-XV} found that AME sources preferentially had larger CMB contributions to their spectra. The CMB amplitude is degenerate with the dust spectral index, so a flattening of this spectral index (e.g., as seen in our Galaxy by \citealp{planck2013-XIV}) would reduce the contribution from the CMB in this spectrum; the positive CMB fluctuation may be accounting for a small excess of emission in the range 100--300\GHz. Further investigation of the CMB contribution would be needed to identify whether that is the case here or not.

The 100 and 217\GHz\ bands of \Planck\ are particularly contaminated with $^{12}$CO,\footnote{$^{13}$CO also emits in these frequency bands with a similar unit conversion coefficient, but the $^{13}$CO line amplitudes are a factor of around 10 fainter than the $^{12}$CO amplitudes (e.g., \citealp{Allen1995}); $^{12}$CO/$^{13}$CO ratios are expected to be similar in M31 to Galactic values, since $^{13}$CO is produced through hot-bottom burning in asymptotic giant branch stars \citep{Herwig2005}.} and consequently these bands were not included in the fit. Using the unit conversion coefficient estimates of the $^{12}$CO contamination of the \Planck\ maps from \citet{planck2013-p28} of ($14.78\pm0.21$) and \mbox{($45.85\pm0.11$)\,$\upmu\mathrm{K_{CMB}}$ [K\,km\,s$^{-1}$]$^{-1}$} at 100 and 217\GHz, respectively, and the sum of the emission in the \citet{Nieten2006} CO map of $109\,481$\,K\,km\,s$^{-1}$, we estimate that within the elliptical aperture centred on M31, the \Planck\ maps should contain a total of 0.58\,Jy of CO emission at 100\ghz\ and 3.6\,Jy at 217\GHz. These estimates agree within a factor of two with measurements of the CO contamination given by the residuals after subtracting the thermal dust model. At 100\ghz, there is a difference of (\getsymbol{residual100})\,Jy between the data and the model fit, corresponding to \getsymbol{residual100percent} of the flux density measured at that frequency. At 217\ghz\ it is (\getsymbol{residual217})\,Jy, or \getsymbol{residual217percent}. In both cases, the uncertainties in the measurement are larger than the expected level of CO from M31.

\section{Conclusions} \label{sec:conclusions}

\Planck\ has clearly detected emission from the Andromeda galaxy across all of its frequency bands, and the High Frequency Instrument has mapped out the emission at 5\arcm\ scales. Multiple spiral arm features are seen to both the north and south of the galaxy, with the 10\,kpc ring clearly detected down to 100\ghz, and additional spiral arm structures extending out to 26\,kpc from the nucleus. Knots of emission at various points within M31 can also seen in the \Planck\ data.

Smoothing the \Planck\ data to a common resolution and combining it with ancillary data, has provided a powerful probe of the dust properties.  By looking at colour ratio plots, it can be seen that the dust observed at the longest wavelengths ($\gtrsim 0.3$\,mm) is heated by the diffuse stellar population (as traced by the stellar emission at 3.6\um), whilst the dust observed at higher frequencies is heated by more local features (as traced by the star formation emission at 24\um). Throughout the \Planck\ frequency range the dust emission is primarily from dust heated by the diffuse stellar population rather than star formation.

By fitting a single modified thermal dust spectrum to individual pixels, the variations in dust mass and temperature across M31 can be seen. When a fixed $\beta_\mathrm{dust}=2$ is used and the SED is rescaled to include only the component that is attributable to the dust heated by the old stellar population, then the dust temperature is inversely proportional to the galactocentric distance. When the spectral index is allowed to vary, then we recover values of $\beta_\mathrm{dust}\approx2$ when only fitting the rescaled pixel SEDs. When we fit the unscaled pixel SEDs using a variable $\beta_\mathrm{dust}$ (see \mbox{Sect.~\ref{sec:dustparammaps}}), we find that \changed{the variations in $\beta_\mathrm{dust}$ could be due to variations in the relative amplitudes of emission from hotter dust heated by star formation and cooler dust heated by the old stellar population rather than variations in the physical emissivity of the dust grains.}

Using aperture photometry, the integrated spectrum of all of the emission from M31 has been measured. This spectrum is well fitted with a combination of free-free, CMB and cold thermal dust emission. Overall, the fitted dust temperature is reasonably cold, at (\getsymbol{tdust})\,K, with a spectral index of \getsymbol{dustbeta}. The significant amount of free-free emission \changed{corresponds to a star formation rate of around 0.12\mdotyr}. We find a $2.3\,\sigma$ marginal detection of AME in M31, with an amplitude of ($0.7\pm0.3$)\,Jy at 30\GHz, which is comparable to the expected level of around 1\,Jy from measurements of the ratio of AME to thermal dust emission in our Galaxy.

\begin{acknowledgements}
We thank M. Haas for providing a copy of the reduced 170\um\ data from \ISO, P. Barmby for providing the \Spitzer\ IRAC 3.6\um\ data, L. Chemin for providing the \hi\ map from DRAO, J. Fritz, M. Smith and others in the HELGA collaboration for providing a copy of the \Herschel\ data, \changed{S. Viaene for providing a copy of the foreground star-subtracted GALEX map,} and A. Richards for assistance with the CO data. We acknowledge the use of the Legacy Archive for Microwave Background Data Analysis (LAMBDA); support for LAMBDA is provided by the NASA Office of Space Science. This research has made use of the NASA/IPAC Extragalactic Database (NED) which is operated by the Jet Propulsion Laboratory, California Institute of Technology, under contract with the National Aeronautics and Space Administration. Some of the results in this paper have been derived using the \healpix\ package. \changed{The research leading to these results has received funding from the European Research Council under the European Union's Seventh Framework Programme (FP7/2007-2013) / ERC grant agreement no.~307209, as well as funding from an STFC Consolidated Grant (no.~ST/L000768/1).}

The development of \Planck\ has been supported by: ESA; CNES and CNRS/INSU-IN2P3-INP (France); ASI, CNR, and INAF (Italy); NASA and DoE (USA); STFC and UKSA (UK); CSIC, MICINN, JA, and RES (Spain); Tekes, AoF, and CSC (Finland); DLR and MPG (Germany); CSA (Canada); DTU Space (Denmark); SER/SSO (Switzerland); RCN (Norway); SFI (Ireland); FCT/MCTES (Portugal); and PRACE (EU). A description of the Planck Collaboration and a list of its members, including the technical or scientific activities in which they have been involved, can be found at \url{http://www.sciops.esa.int/index.php?project=planck&page=Planck_Collaboration}.

\end{acknowledgements}

\bibliographystyle{aat}

\bibliography{Planck_bib,m31_refs}

\appendix
\section{\changed{Comparison data sets}}\label{sec:data}
\changed{As described in Sect. \ref{sec:highres}, we make use of the following data sets for comparison purposes.}

\changed{We use the} \IRAS\ data at 12, 25, 60, and \changed{100\um\ in} the form of \changed{the improved r}eprocessing of the \IRAS\ \changed{s}urvey (IRIS) \citep{Miville2005},\footnote{\url{http://www.cita.utoronto.ca/~mamd/IRIS/}} which is based on the \IRAS\ Sky Survey Atlas \citep{Wheelock1994}, but has improved processing to achieve a higher resolution of 3\parcm8--4\parcm3, depending on the \changed{wavelength.}

\changed{We make} use of the \ISO\ 175\um\ data from \citet[][private communication]{Haas1998}, which has $1\parcm3$ resolution. This data set (along with the low-resolution \DIRBE\ data) sits in the the frequency gap between \Planck\ and \Spitzer. As the version of these data that we use has not been pre-calibrated, we calibrate them such that the sum of the pixels adds up to the overall flux density of 7900\,Jy within the area surveyed by \ISO, as reported in \citet{Haas1998}. We caution that the \ISO\ data are not well sampled, meaning that some of the large-scale emission will be missing from these data, which may result in underestimates of the flux densities in the following analysis. We do not fit to this data; instead we include it for comparison purposes.

Several maps of \hi\ emission from M31 exist \citep[e.g.,][]{Unwin1983,Chemin2009,Braun2009}. We make use of the recent \hi\ emission map from \citet{Chemin2009} to trace the gas within M31. This map was constructed using a combination of interferometric and single dish data using the Synthesis Telescope and the 26-m Telescope at the Dominion Radio Astrophysical Observatory (DRAO). This combination means that the map has 22\arcs\ resolution, whilst not resolving out extended emission, and it also covers the full extent of M31, including the outermost spiral arms. Although the original map is contaminated by \hi\ emission from our Galaxy at the same frequencies as that from part of M31, this has been removed by \citet{Chemin2009}. We flatten the image cube to an intensity map (by summing up the brightness temperature over the frequency channels within a spatial pixel), convolve it to 5\arcm, and resample it to match the \Planck\ data set using {\tt Aladin} \footnote{\url{http://aladin.u-strasbg.fr/}} \citep{Bonnarel2000} to take into account differences in the coordinate projection systems used.

Finally, we also make use of the $^{12}$CO $J$=1$\rightarrow$0 map of M31 from \citet{Nieten2006}, which is constructed from 23\arcs\ resolution observations with the IRAM 30-m telescope. This map covers the 10\,kpc ring structure, but does not include the outermost spiral arm structures. We use this map to compare with the \Planck\ data in the bands that have been contaminated with CO emission, particularly for the integrated spectrum in Sect. \ref{sec:lowressed}. The data (downloaded from NED) were reprojected using {\tt AIPS} prior to being smoothed and repixelised.

\section{Compact sources in and around M31}\label{sec:sources}

Following from Sect. \ref{sec:morphology}, we provide some comments about objects in the field around M31 that are visible in Figs. \ref{fig:m31planck} and \ref{fig:m31planck857}, and features of M31 that are included in the ERCSC and PCCS catalogues.

M110/NGC205 is clearly detected at 857\ghz\ (350\um; position: $00^{\rm h}40^{\rm m}22.1^{\rm s}$, +41\deg41\arcm7\arcs\ J2000, distance ($809\pm24$)\,kpc, \citealp{McConnachie2005}, ``F6'' in Fig. \ref{fig:m31planck857}). This dwarf galaxy has also been seen in thermal dust emission by \Spitzer\ and \Herschel\ \citep{DeLooze2012}, and it is known to be surprisingly dust-rich. Messier 32, which is at position $00^{\rm h}42^{\rm m}41.8^{\rm s}$, +40\deg51\arcm55\arcs, is not detected. This is because it is faint at these frequencies (it has been measured by \ISO\ to be less than 0.27\,Jy at 200\um, \citealp{Temi2004}), and it is also very close to the bright 10\,kpc ring, which means that it is not in a clean environment to be detected with the spatial resolution of the \Planck\ data.

There is also a detection of the blazar \object{B3\,0035+413} as a compact source at 28.4, 100, 143, and 217\ghz\ (13, 3.0, 2.1, and 1.38\,mm; position $00^{\rm h}38^{\rm m}30^{\rm s}$, +41\deg34\arcm40\arcs; marked in the left-hand \changed{panels} of Fig. \ref{fig:m31planck857} as ``B3''). The high-frequency flux density of this flat spectrum source, as measured by the One Centimetre Receiver Array at 30\GHz\ (10\,mm), is ($395\pm20$)\,mJy \citep{Lowe2007}, so it is unsurprising that it is seen in the \Planck\ maps at these frequencies. It is not included in the \Planck\ Early Release Compact Source Catalogue (ERCSC) \citep{planck2011-1.10}, since the CMB signal is much stronger than the source flux at these frequencies. It is, however, in the second \Planck\ catalogue, the \Planck\ Catalogue of Compact Sources \citep[PCCS; ][]{planck2013-p05}, due to the filtering carried out prior to source extraction; it is listed there as PCCS1 100 G120.34$-$21.17 (DETFLUX of $350\pm80$\,mJy), PCCS1 143 G120.32$-$21.16 ($260\pm40$\,mJy), and PCCS1 217 G120.28$-$21.16 ($220\pm40$\,mJy). At a redshift of 1.35, the source is not physically associated with M31, but is just a line-of-sight coincidence.

Several bright segments of the M31 ring structure were detected in the ERCSC at the 143 to 857\ghz\ (2.1 to 0.35\,mm) channels. Most of these sources are flagged as EXTENDED (EXT) in the ERCSC, which indicates that under a Gaussian fit, the square root of the product of the fitted major axis and minor axis is larger than 1.5 times the beam FWHM at a given frequency. The ERCSC sources are marked in the \changed{top-right} panel of Fig. \ref{fig:m31planck857}, where they are circled at their 857\GHz\ (350\um) positions, with the circles being 5\arcm\ in radius, such that they have a diameter approximately equal to twice the FWHM at the 857\GHz\ (350\um) channel. Sources 2 to 5 are clearly within the star-forming ring $\sim$10\,kpc offset from the nucleus, while source 1 belongs to the outer ring that is seen both by \ISO\ and \Spitzer\ MIPS. A much larger number of segments of the ring are included by the PCCS, with 21--32 sources detected in PCCS 217--857\GHz\ (1.38-0.35\,mm) bands on the inner ring alone, which are marked as extended. Due to the large number of segments included in the PCCS, we do not circle them in Figure \ref{fig:m31planck857}.

\raggedright 
\end{document}